Waterford Institute *of* Technology
INSTITIÚID TEICNEOLAÍOCHTA PHORT LÁIRGE

**MSc. Innovative Technology Engineering**

# Impact of power supply noise on image sensor performance in automotive applications

by

**Shane Gilroy**

Dissertation Submitted in Partial Fulfilment of the Requirements for the Degree
of
Master of Science
in
Innovative Technology Engineering

Submitted to the School of Engineering
Waterford Institute of Technology

September 2016

# Abstract

Vision Systems are quickly becoming a large component of Active Automotive Safety Systems. In order to be effective in critical safety applications these systems must produce high quality images in both daytime and night-time scenarios in order to produce the large informational content required for software analysis in applications such as lane departure, pedestrian detection and collision detection. The challenge in producing high quality images in low light scenarios is that the signal to noise ratio is greatly reduced, which can result in noise becoming the dominant factor in a captured image, thereby making these safety systems less effective at night.

A project has been undertaken to develop a systematic method of characterising image sensor performance in response to electrical noise in order to improve the design and performance of automotive cameras in low light scenarios.

The root cause of image row noise has been established and a mathematical algorithm for determining the magnitude of row noise in an image has been devised. An automated characterisation method has been developed to allow performance characterisation in response to a large frequency spectrum of electrical noise on the image sensor power supply. Various strategies of improving image sensor performance for low light applications have also been proposed from the research outcomes.

# Declaration

I certify that this dissertation is all my own work and contains no plagiarism. By submitting this dissertation, I agree to the following terms:

Any text, diagrams or other material copied from other sources (including, but not limited to, books, journals and the internet) have been clearly acknowledged and referenced as such in the text by the use of 'quotation marks' (or indented *italics* for longer quotations) followed by the author's name and date [eg (Byrne, 2008)] either in the text or in a footnote/endnote. These details are then confirmed by a fuller reference in the bibliography.

I have read the sections on referencing and plagiarism in the handbook or in the WIT plagiarism policy and I understand that only submissions which are free of plagiarism will be awarded marks. By submitting this dissertation, I agree to the following terms. I further understand that WIT has a plagiarism policy which can lead to the suspension or permanent expulsion of students in serious cases. (WIT, 2008).

Signed: ______________________Shane Gilroy___________________________

Date: _____________________5th September 2016_________________________

# Acknowledgements

I would like to thank Valeo Vision Systems for the facilitation of this research project, particularly Stephen O'Donnell and Enda Ward for defining the research problem and coordinating the project activities from within Valeo Vision Systems. I would also like to thank the VVS Image Quality department, particularly Brian Deegan, Thomas Devaney, Peadar Conneely, Patrick Denny and Mícheál Garvey for the access to the knowledge and resources required for the project, Gerry Conway, Margaret Glavin and everyone in the VVS Hardware group for their contributions and insight and Aidan Walsh, Robert Carroll and Robert Melody of the VVS Test department for their interaction and advice at various stages of the project.

I would also like to thank my supervisor Dr John O'Dwyer for the academic support throughout my time with Waterford Institute of Technology.

# Table of Contents

# List of Abbreviations and Acronyms

| | |
|---|---|
| ABS | Antilock Braking Systems |
| ADC | Analogue to Digital Conversion |
| ADAS | Advanced Driver Assistance Systems |
| APS | Active Pixel Sensor |
| AWB | Auto White Balance |
| BLC | Black Level Calibration |
| CDS | Correlated Double Sampling |
| CFA | Colour Filter Array |
| DPS | Digital Pixel Sensor |
| DSP | Digital Signal Processor |
| DUT | Device Under Test |
| FPN | Fixed Pattern Noise |
| FPS | Frames per Second |
| MP | Megapixel |
| PCB | Printed Circuit Board |
| PPD | Pinned Photodiode |
| PPS | Passive Pixel Sensor |
| PRNU | Photo-response Non-uniformity |
| RGB | Red-Green-Blue |
| SNR | Signal to Noise Ratio |
| VI | Virtual Instrument |
| VVS | Valeo Vision Systems |
| WIT | Waterford Institute of Technology |

# List of Tables

# List of Figures

# Chapter 1 Introduction

There were 472 road deaths in Ireland in 1997 according to the Road Safety Authority (1). There were 166 road deaths in 2015 and the number of fatalities has decreased in an almost linear fashion in the intervening years as seen in Figure 1. This reduction is due in part to the advancement and standardization of automotive safety systems. Automotive safety systems fall into two main categories, passive safety systems which protect the driver and passengers in the event of a collision such as seat belts, airbags etc. and active safety systems which prevent the occurrence of a collision such as traction control, ABS Brakes and blind spot monitoring.

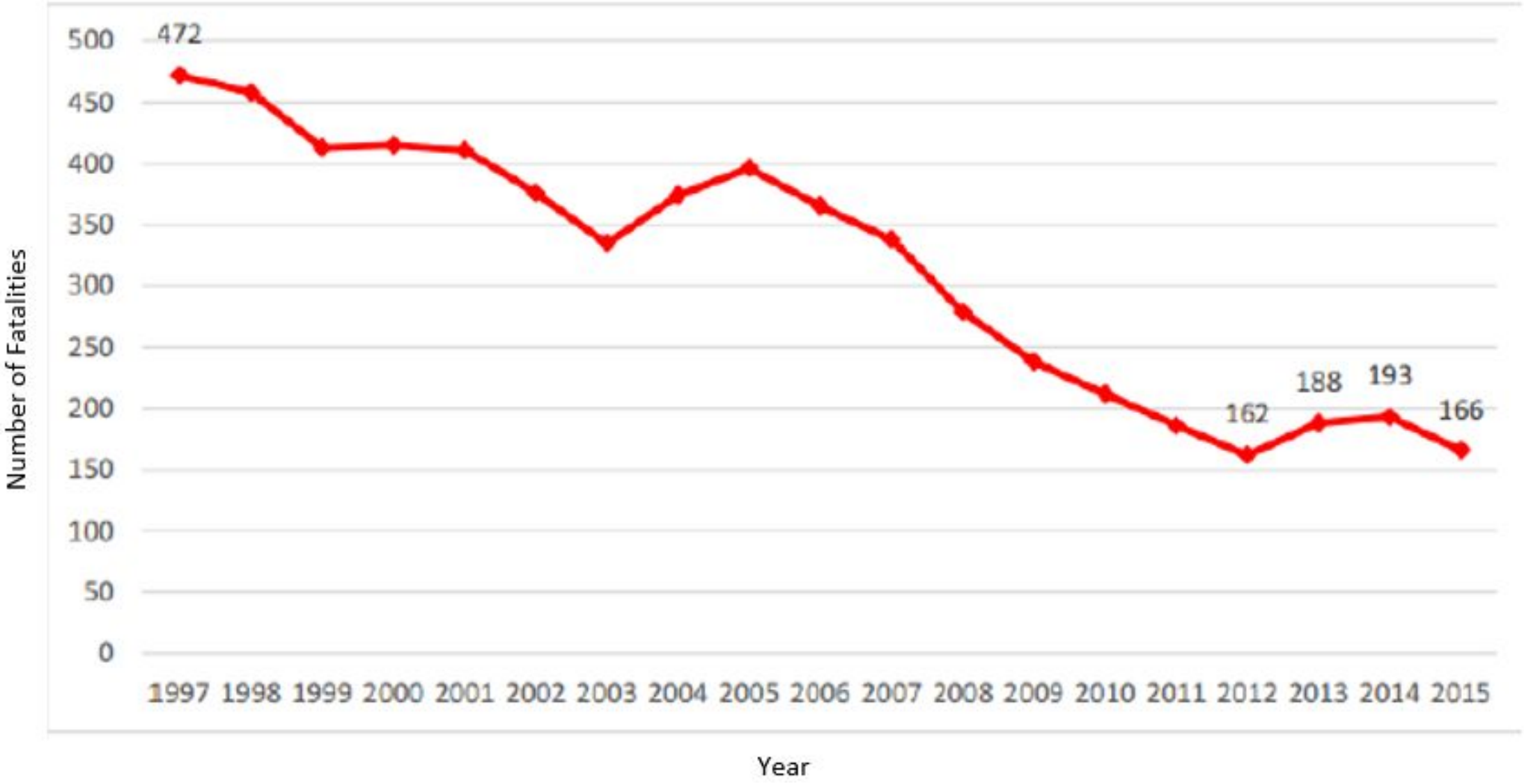

**Figure 1: Road Deaths in Ireland 1997-2015**

Modern active safety systems known as Advanced Driver Assistance Systems (ADAS) incorporate the convergence of a range of Radar, LIDAR, Camera and Ultrasonic technology (2) in order to increase the safety of drivers and passengers through many applications as shown in Figure 2. Intelligent vision systems are quickly becoming a larger component of active automotive safety systems for a wide range of applications as manufacturers strive to provide 360° protection for their customers.

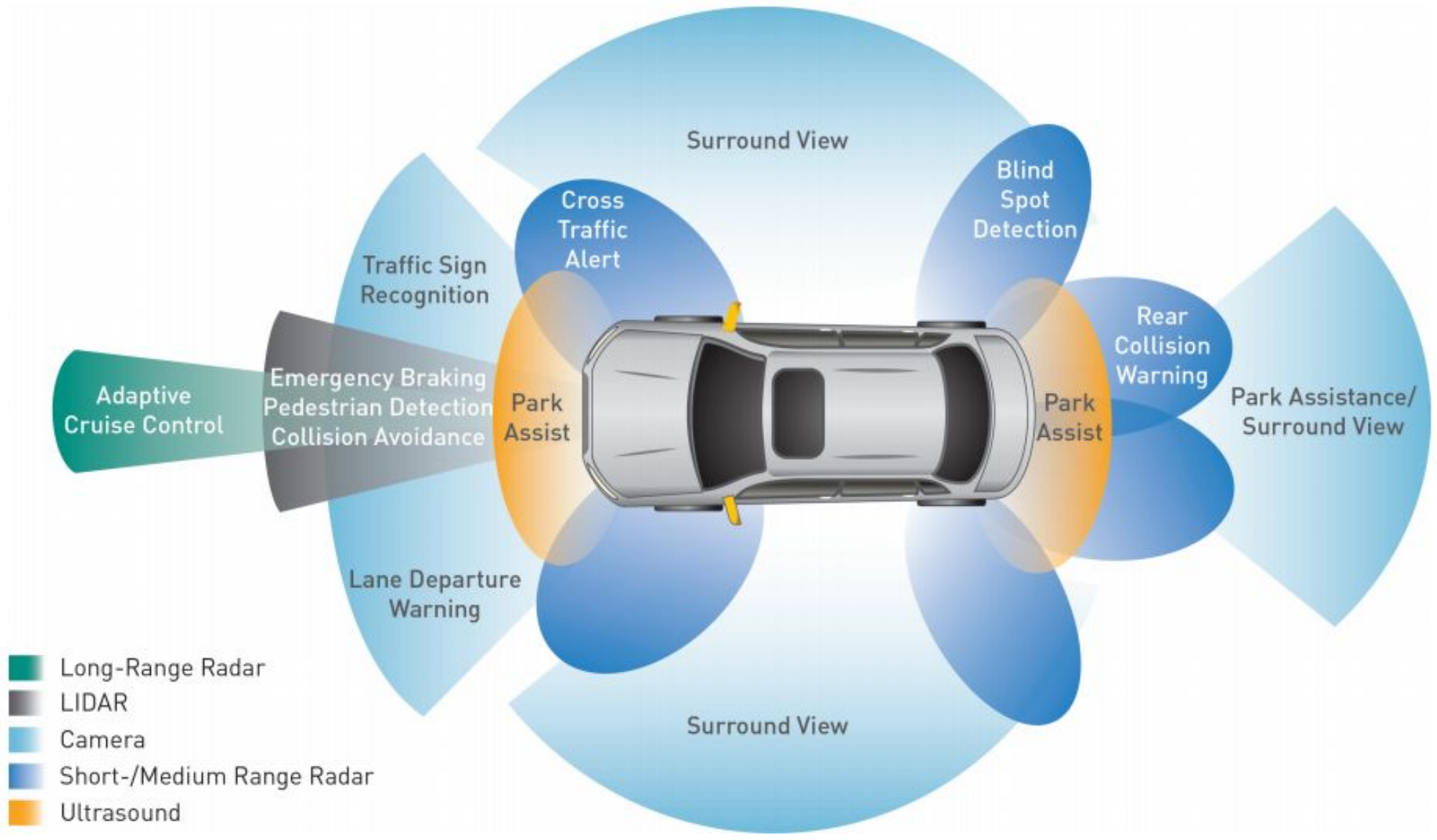

**Figure 2: Growing Applications of Advanced Driver Assistance Systems (3)**

One example of a vision based safety system is lane departure. Failure to stay in the correct lane was the largest single factor in fatal collisions in the US in 2002 (4), playing a role in 32.8% of all cases. Vision Systems can be used to detect lane departure before a collision occurs and mitigate the risk by alerting the driver, adjusting the steering angle or applying the brakes on the opposite side of the vehicle in order to prevent lane departure.

Another example of a vision based safety system is backover protection. There is an average of 232 fatalities and over 13,000 injuries each year in the US as a result of backovers (5, 6). The victims are primarily children under the age of 5 year old and the elderly (6, 7). As a result, new legislation will be enforced from May 2018 requiring the mandatory installation of rear view camera on all new vehicles weighing under 10,000lb or 4,500kg (6). This law will apply to all cars, SUVs, buses and light trucks. A minimum field of view of a 10ft by 20ft (3m x 3.5m) zone directly behind the vehicle and a minimum image size is specified. These cameras will not only reduce the risk of backovers by allowing the user to see behind the vehicle, but also through the use of advanced software algorithms such as Object Detection which allows the vehicle to autonomously mitigate the risk of backovers by alerting the user or applying emergency braking etc.

## 1.1 Signal to Noise Ratio

In order to be effective in safety applications, these systems must perform without error in both daytime and night time scenarios. In order to operate successfully in both scenarios vision systems must have a high Signal to Noise Ratio. Signal to Nosie Ratio (SNR) is an important metric which defines the quality of any electronic device which transmits or receives a signal.

The various noise sources within an image sensor and camera accumulate and create a noise floor which imposes limitations on the performance capability of the image sensor. The Absolute Sensitivity Threshold defines the number of captured photons required to get a signal equal to the noise level captured by the image sensor. Therefore, the absolute sensitivity threshold dictates the minimum amount of light required by an image sensor in order to obtain any kind of valid signal (8). The greater the ratio of signal to noise in the system, the more valid useable data is captured.

In low light situations, the Signal to Noise Ratio of a captured image can be greatly reduced. This can lead to noise becoming the dominant signal as shown in Figure 3, thus reducing the amount of useful information available for software analysis and thereby impacting the performance of the safety system or in extreme cases making the safety system defunct.

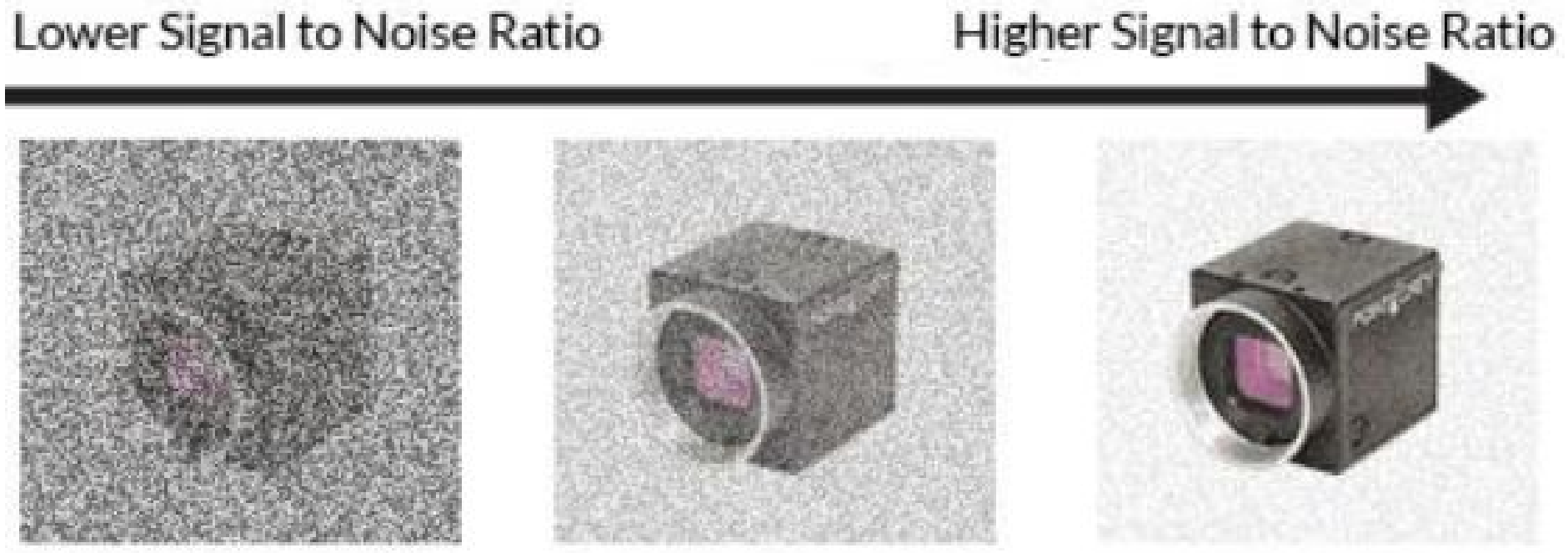

**Figure 3: Effect of Signal to Noise Ratio (SNR) on Image Quality**

There are two methods by which the signal to noise ratio of an image sensor can be increased, increase the photon to electron conversion rate of each pixel, known as the Quantum Efficiency or decrease the noise floor. As a result, pixel technology has progressed greatly in recent times and with developments in Back Side Illuminated CMOS technology (BSI) manufacturers are now commonly experiencing quantum efficiency rates of over 70% with some even as high as 90% at particular wavelengths (9). Therefore, attention has now been diverted towards reducing the noise floor in order to obtain any further improvements in signal to noise ratio.

## 1.2 CMOS Image Sensor Evolution

Technological trends have shown a demand for increased spatial resolution in image sensing applications. In order to obtain increased resolution or total number of pixels without increasing the physical size of the sensor, pixel size must be reduced (10, 11). There are also many other driving factors for the reduction of pixel size in CMOS image sensors while maintaining the pixel count, such as a reduction of chip cost, energy to read the sensor, camera volume and camera weight (12). Reducing pixel size has a negative impact on an image sensors optical and electrical performance by reducing the amount of light photons that can be captured by each individual pixel as demonstrated by Figure 4. This contributes further to the reduction of Signal to Noise Ratio in low light applications.

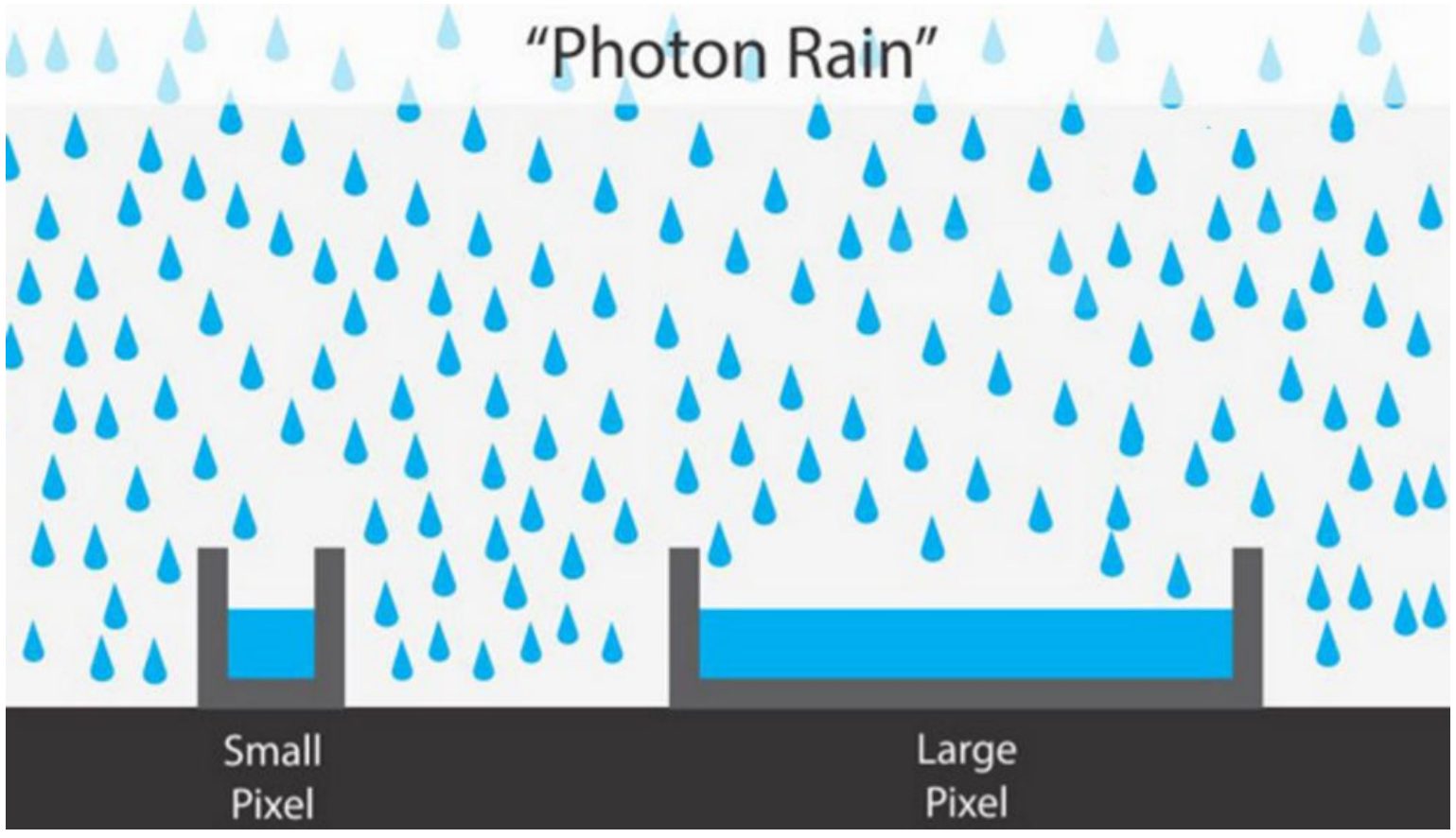

**Figure 4: Small Pixel vs. Large Pixel Photon Capture**

Current research in CMOS image sensor design such as Feruglio *et al* 2006 (13), Seo *et al* 2013 (14) and Chen *et al* 2012 (15) attempt to address this issue at image sensor chip level, with a view to reducing internal noise sources within an image sensing device such as fixed pattern noise, shot noise and read noise. These chip level improvements of SNR can be difficult to observe in practical automotive applications as external sources of noise created by the surrounded circuitry and the automotive environment can dominate the noise floor and negatively impact image sensor performance. This can prevent camera designers from achieving the full capability of a modern image sensors low light performance.

## 1.3 Knowledge Gap

A knowledge gap exists for a systematic method of characterising image sensor performance in response to the external noise sources which occur in vision systems in practical applications that can be used in order to improve the design of automotive cameras for low light applications.

## 1.4 Research Objectives

A research project has been undertaken in order to satisfy this knowledge gap. In order to further define the scope of the project to the suit allocated time scale, the focus has been placed on image sensor performance characterisation in response to electrical noise on the power supply lines. This project has been carried out in association with Valeo Vision Systems (VVS) on site at Dunmore Road, Tuam, Co. Galway. Valeo Vision Systems, founded as Connaught Electronics Ltd in 1982, were the first to market in Europe with automotive rear view and surround view camera systems. Connaught Electronics was acquired by the world's leading provider of driving assistance systems, Valeo in 2007 and has produced over 5.2 million cameras in 2015 for automotive manufacturers such as BMW, Daimler, VW Group, Volvo, Ford and Nissan.

The primary objectives of the research project are as follows:

1. Design and develop a structured methodology for image sensor performance characterisation in response to electrical noise on the power supply lines.

2. Perform characterisation on a number of current state-of-the-art image sensors used in automotive applications.
3. Review cost and size effective solutions to improve image sensor performance in response to electrical noise on the power supply lines based on research outcomes.

## 1.5 Dissertation Overview

The structure of this dissertation is as follows:

**Chapter 1 Introduction and Context** provides an overview of the motivation and key objectives of the research project.

**Chapter 2 Literature Review** provides an in-depth overview of pixel electronics, CMOS image sensor readout architecture, noise sources which occur in CMOS image sensors and image row noise.

**Chapter 3 Methodology** provides an overview of the steps taken in order to achieve the objectives of the research project.

**Chapter 4 Results** presents the results obtained throughout the project. The results section describes the root cause of an image sensors reaction to power supply noise, outlines the proposed characterisation method used to quantify this noise over a frequency spectrum, presents validation results of the proposed methodology and provides the results of image sensor characterisation of two state-of-the-art image sensors currently used in automotive applications.

**Chapter 5 Analysis and Discussion** provides analysis of the image sensor characterisation method and results as well as a more detailed discussion of the image sensor settings used for characterisation and options for improving image sensor performance in response to electrical noise on the power supply lines in automotive applications.

**Chapter 6 Conclusion and Recommendations for Future Work**

# Chapter 2 Literature Review

A detailed literature review of CMOS pixel electronics, CMOS image sensor readout architecture, the noise sources which occur in CMOS image sensors and image row noise is provided in this section of the dissertation.

## 2.1. CMOS Pixel Electronics

The photosensitive component in a CMOS imager sensor pixel is known as a Photodiode. A photodiode is a P-N junction semiconductor diode that generates an electrical signal in response to light intensity through a principle known as the Photoelectric Effect.

When a photodiode is exposed to light as in Figure 5, the energy of the photons which strike the diode can be transferred into the individual silicon atoms. If this transferred energy is larger than the band gap of the material, the electrons in the valance band of the material can jump into the conduction band and generate an electron-hole pair.

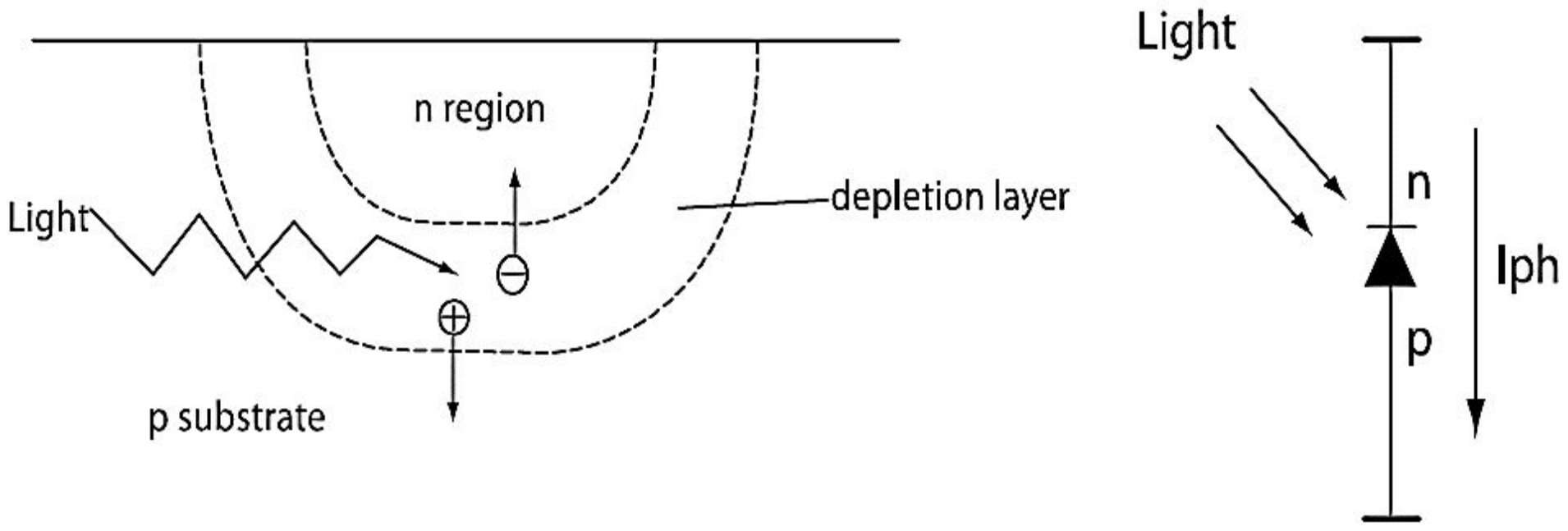

**Figure 5: Photodiode Exposed to Light**

The energy of a photon is determined by its wavelength. Equation 2.1 below can be used to determine if a photon has a sufficient wavelength to excite electrons within a semiconductor:

$$E_{Photon} = h \times f = \frac{h \times c}{\lambda} \geq E_g \quad (2.1)$$

Where $E_{Photon}$ is the energy of the photon, $h$ is Planck's Constant, $f$ is the frequency of light, $c$ is the speed of light, $\lambda$ is the wavelength of light and $E_g$ is the band gap of the material.

The band gap energy of silicon is 1.1 eV, therefore only light photons with a wavelength shorter than 1100nm can excite electrons in a silicon photodiode (16). When this occurs an electrical field is created within the depletion layer of the p-n junction causing electrons to travel towards the n-doped region and electron holes to move towards the p-doped region generating a photocurrent across the terminals of the photodiode. The magnitude of this photocurrent is proportional to the number of photons captured by the photodiode per unit time, which in turn corresponds to the intensity of light (17).

During its operation a photodiode is reverse biased. Therefore, after reset occurs a voltage, Vbias is applied to the photodiode. As light photons strike the diode during exposure, a charge corresponding to light intensity is generated which accumulates on to the junction capacitance and the voltage across the diode begins to decrease. The difference between the reset voltage Vbias and the final voltage (Vd) after exposure represents the light intensity captured by the photodiode.

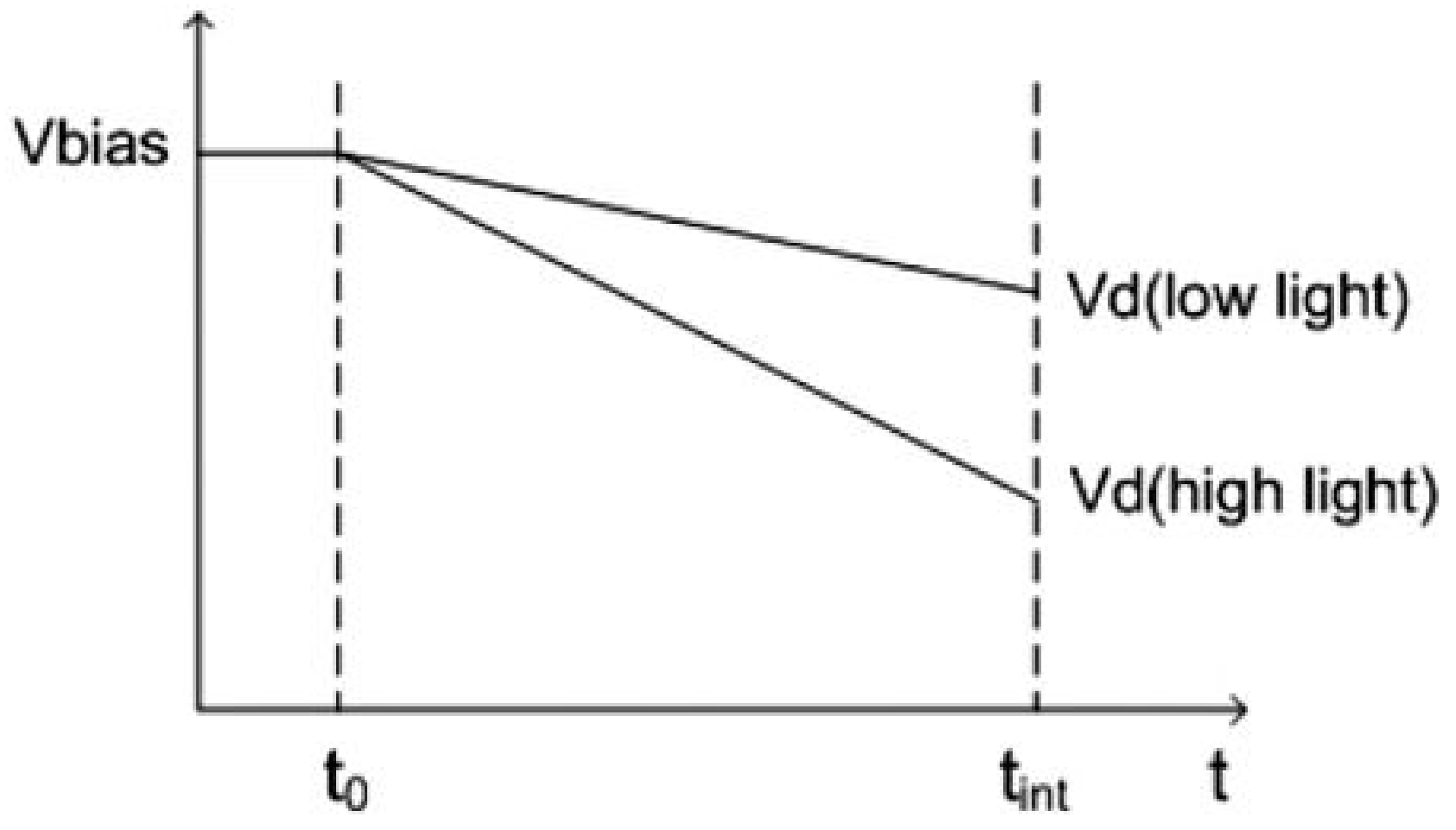

**Figure 6: Photodiode Response to High and Low Light**

## 2.2. Passive Pixel Sensor

The first generation of CMOS image sensors are referred to as Passive Pixel Sensors (PPS) and consist of one photodiode and one row select transistor which acts as a switch as seen in Figure 7 below (12).

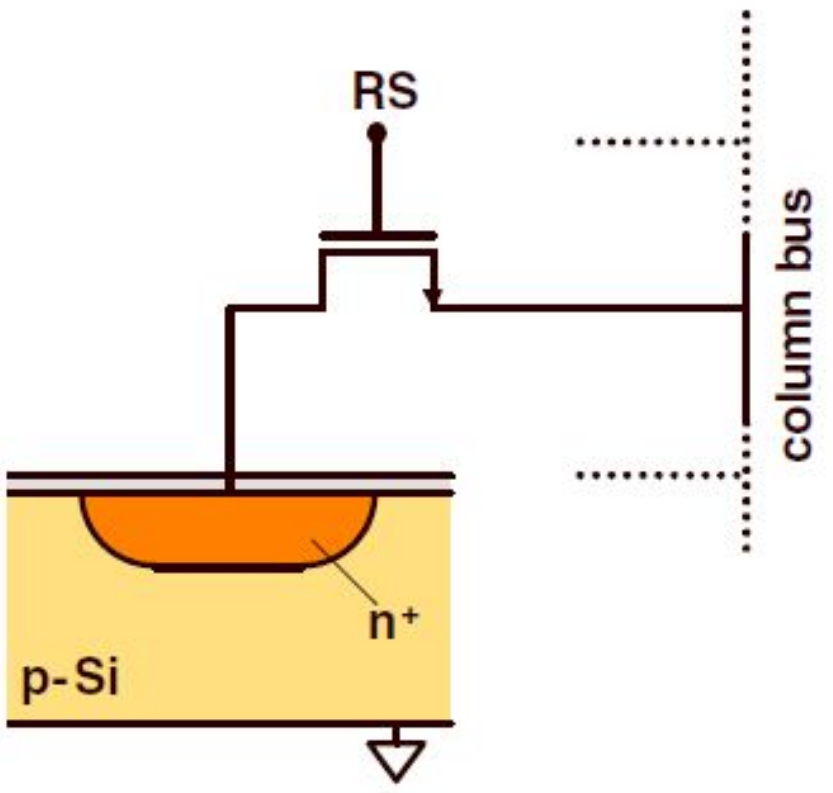

**Figure 7: Passive Pixel Sensor**

Passive Pixel Sensor Readout operates as follows:

- At the beginning of the exposure the photodiode is reverse biased to a high voltage, typically 3.3V etc.
- During the exposure time, photons strike the photodiode decreasing the reverse voltage across the photodiode.
- At the end of the exposure period, the remaining voltage across the diode is measured. The drop in voltage from the pre-exposure voltage quantifies the amount of photons captured by the photodiode during the exposure time.
- The photodiode is then reset in order to allow for a new exposure cycle

Passive pixel sensors have a large Fill Factor or ratio of photosensitive area to total pixel area, due to the minimal number of components within the pixel. Fill Factor can be expressed as:

$$FF = \frac{A_{ps}}{A_{pix}} x\ 100\% \quad (2.2)$$

Where $FF$ is the Fill Factor, $A_{ps}$ is the photosensitive area of the pixel and $A_{pix}$ is the total pixel area (18).

The drawback from passive pixel sensors is that they incur a large amount of noise due to the mismatch in capacitance between the small pixel and the large vertical bus capacitance.

## 2.3. Three Transistor Active Pixel Sensor

Active Pixel Sensors (APS) provide a significant improvement in noise performance through the addition of an amplifier in the form of a source follower transistor. This is known as a Three Transistor Active Pixel Sensor (3T APS) and consists of a photodiode, a reset transistor, a source follower transistor and a row select or addressing transistor as shown in Figure 8 below (12).

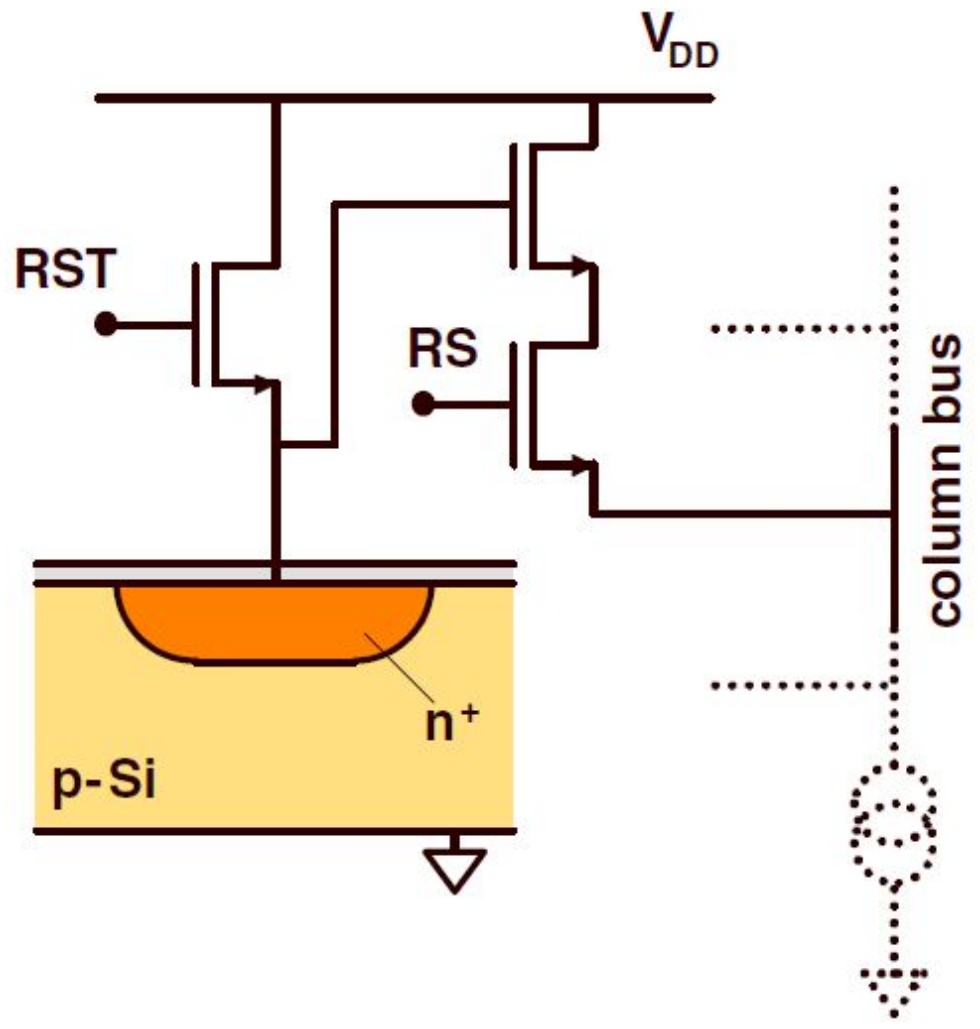

**Figure 8: 3T Active Pixel Sensor**

The principle of operation of a 3T Active Pixel Sensor is similar to a Passive Pixel Sensor:

- At the beginning of the exposure the photodiode is reverse biased or reset.
- During the exposure time, photons strike the photodiode decreasing the reverse voltage across the photodiode.
- At the end of the exposure, the pixel is addressed and the voltage across the photo diode is transferred outside of the pixel by the source follower.
- The photodiode is then reset for a new exposure cycle

Although the extra circuitry in a 3T APS reduces the Fill Factor of the pixel, it solves a significant amount of the pixel sensor noise issues of a PPS as the in-pixel amplifier isolates the photodiode capacitance from the large column capacitance. The Reset / kTC noise created during the resetting of the photodiode remains prominent in 3T APS and in order to improve this a 4T Active Pixel Sensor was developed.

## 2.4. Four Transistor Active Pixel Sensor

The primary technology used in modern automotive image sensors is Four Transistor Active Pixel Sensors also known as a Pinned Photodiode (PPD). 4T Active Pixel Sensors introduced a Floating Diffusion (FD) and a Transfer Gate transistor (TG) to the traditional 3T APS design, separating the photodiode from the readout circuit as illustrated in Figure 9 below (19).

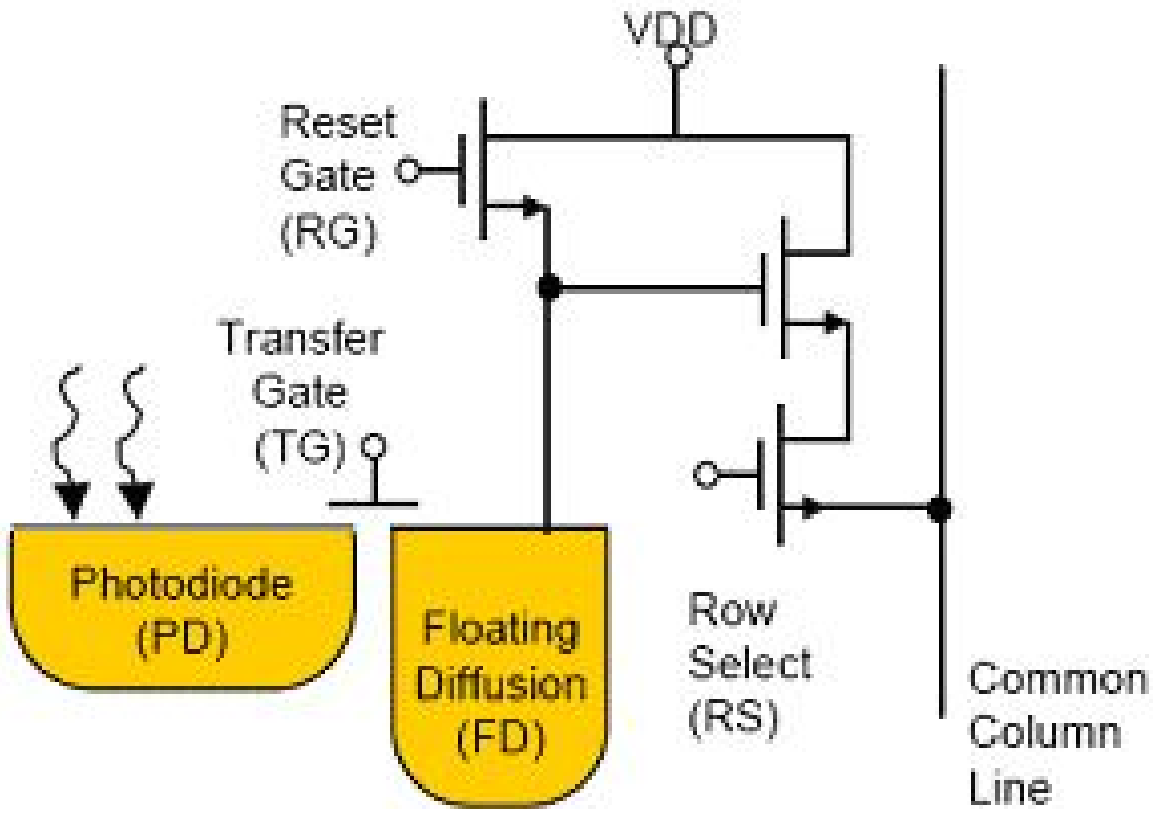

**Figure 9: Pinned Photodiode 4T Active Pixel Sensor**

This pinned photodiode structure makes it possible to carry out Correlated Double Sampling in order to cancel out a significant portion of the noise generated within pixels such as the reset noise, 1/f noise and the dc offset introduced by the source follower, thereby dramatically improving CMOS image sensor performance for mainstream applications. Correlated Double Sampling (CDS) is the process of reading the output of the sensor twice, once after reset when the signal is zero and once with the full signal present. This allows the offset voltage and the read noise etc. to be subtracted from the signal, reducing noise and improving sensor performance. 4T Active Pixel Sensor operation is as follows:

- Incoming photons are converted to electrons in the photodiode for the duration of the exposure time.
- At the end of the exposure, the floating diffusion is reset by the reset transistor.
- A measurement is taken of the output voltage after reset and stored on the column circuit.
- The transfer gate is then activated and the voltage in the photodiode is transferred to the floating diffusion.
- Once the transfer is complete the charge is readout and also stored on the column circuit.
- The two stored voltages are then subtracted in order to remove the noise measurement from the signal in a process called Correlated Double Sampling (CDS).

## 2.5. CMOS Pixel Readout Architecture

Once the exposure cycle in a 4T Active Pixel Sensor is complete, the charge collected in the photodiode is buffered by the source follower and is transferred towards the column signal line. Analogue processing such as amplification and analogue to digital conversion can occur before or after the signal reaches the column signal line depending on the type of readout architecture used in the image sensor design.

There are three main types of readout architecture used in CMOS image sensor design, each of which is characterised by whether analogue processing is carried out at chip level, column level or pixel level (17).

### 2.5.1. Chip Level Processing

Chip level processing architecture carries out post exposure analogue processing consisting of noise suppression, amplification and analogue to digital conversion outside of the pixel column array as illustrated in Figure 10 below.

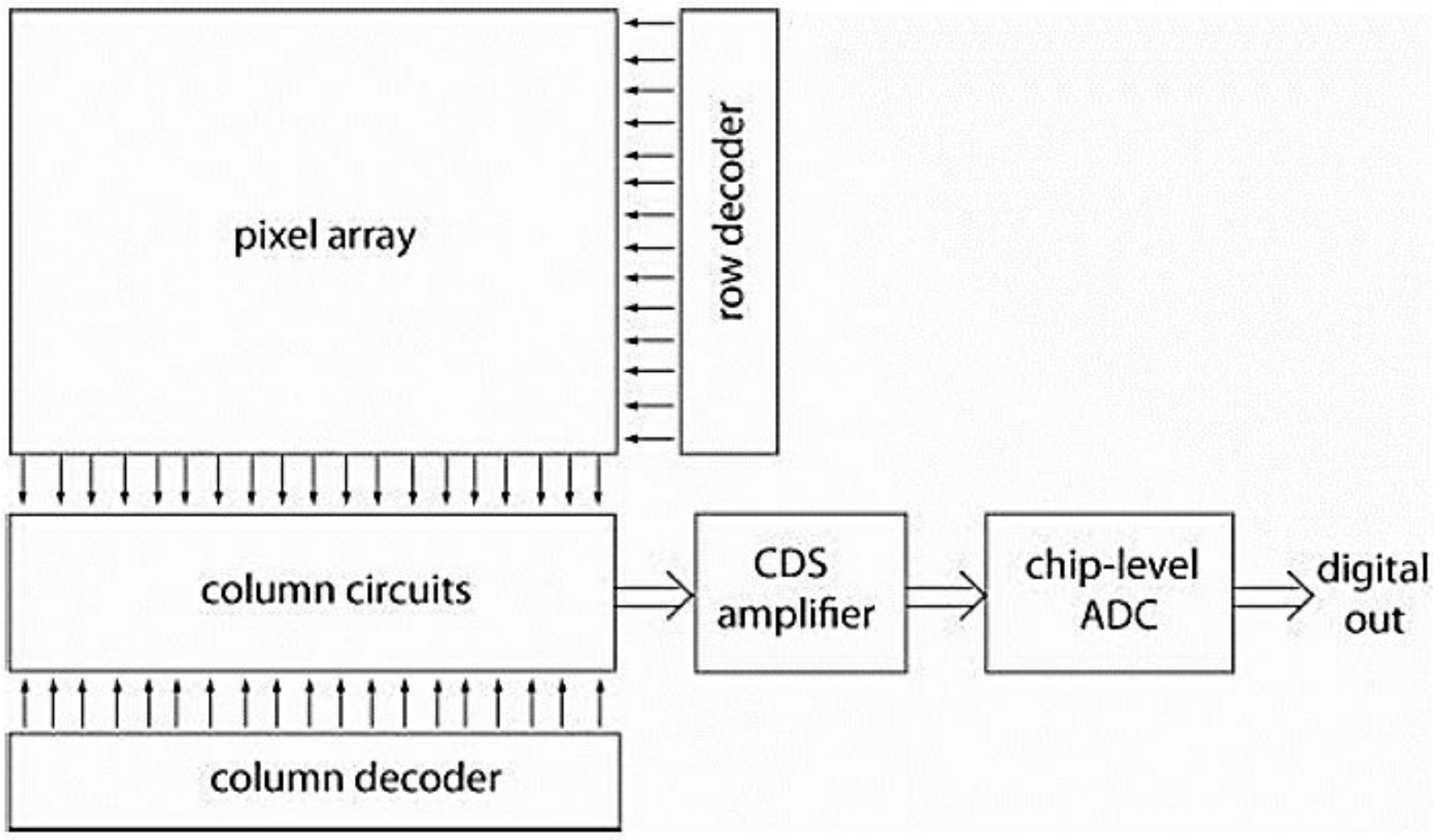

**Figure 10: Chip Level Processing Architecture**

All pixels on one row are simultaneously reset and moved to a sample and hold circuit in a process called column parallel readout represented by the "column circuits" block in Figure 10. Each pixel value is then amplified and converted into a digital signal one by one in sequence. When this is complete the next row is read out into the sample and hold circuit and the process is repeated. This linear process of analogue processing is simple from a circuit design point of view and maintains good uniformity as the amplifier and ADC are identical for each pixel. Chip level processing is restricted to low speeds however as a single ADC and amplifier must process each pixel which can create a bottle neck and restrict the maximum achievable frame rate. This approach also results in a longer analogue signal chain which can result in an increase of overall noise generation on the analogue signal path compared to column level or pixel level architectures (17, 18).

### 2.5.2. Column Level Processing

Column level processing is the most commonly used architecture in modern CMOS image sensors as it provides a good compromise between frame rate, fill factor, power consumption and speed (17). Column level processing solves chip level speed and long analogue path issues by using a separate analogue processing block for each column. A block diagram of column level architecture can be seen in Figure 11 below.

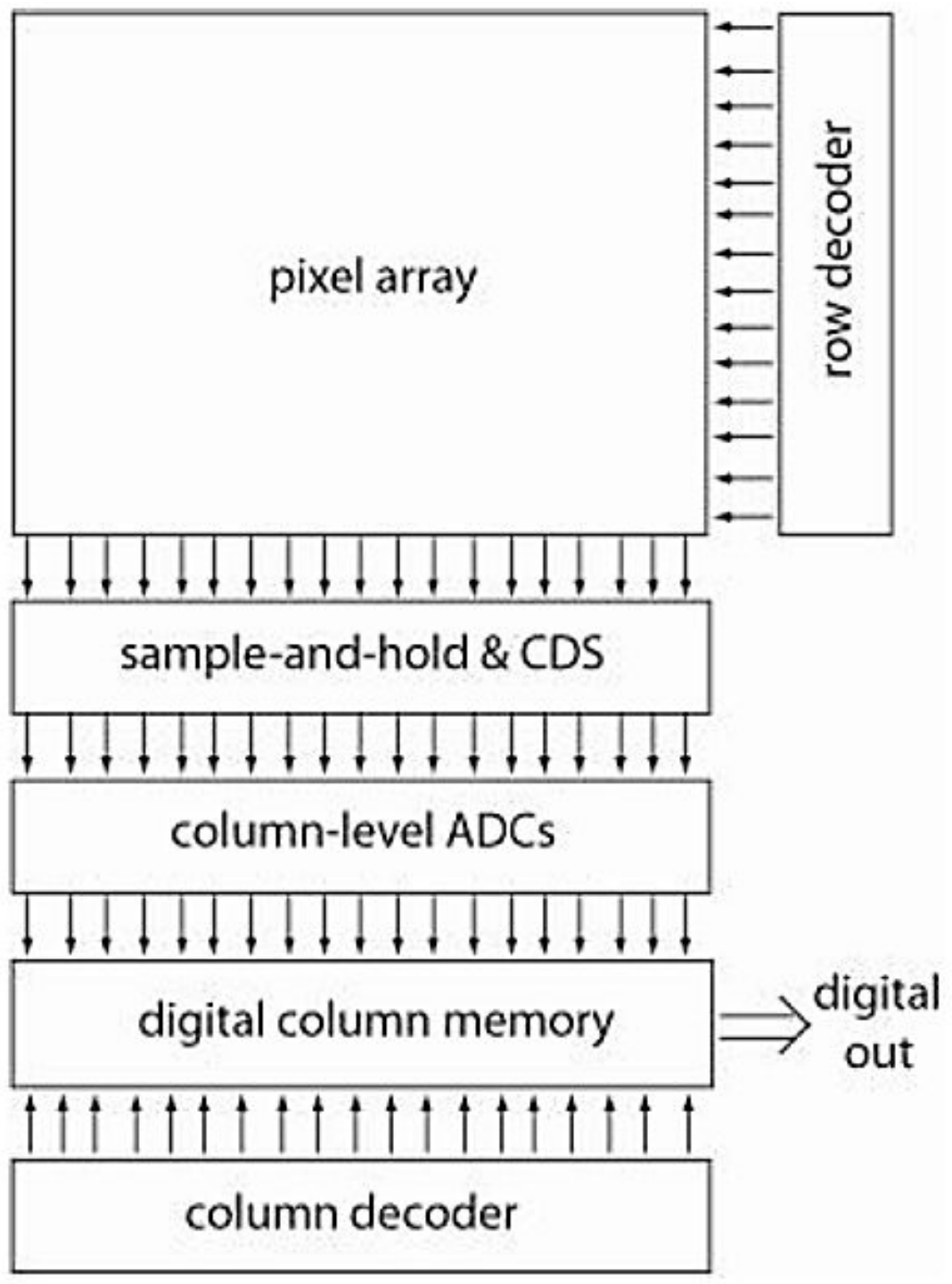

**Figure 11: Column Level Processing Architecture**

Every pixel on a signal row is sampled simultaneously in parallel and transferred to the sample and hold circuitry in a process known as column parallel readout. Each column has its own amplifier and analogue to digital converter in order to digitise the data in parallel before transferring the signal out. In order to achieve this column level processing, these devices require hundreds of thousands of amplifiers and ADCs that operate in parallel in each column. Tiny variations between these amplifiers and ADCs as a result of offset and gain mismatches create column level Fixed Pattern Noise (FPN) on the image which is highly visible to the naked eye.

Although the extra analogue processing circuitry increases the speed, efficiency and noise performance of chip level readout architecture it also reduces the fill factor of the pixel as the photosensitive area to total pixel area is reduced in order to fit the extra hardware electronic components. This can lead to larger required pixel sizes or reduced low light performance.

### 2.5.3. Column Parallel Readout Architecture

CMOS image sensors with chip level and column level architectures can also be particularly susceptible to horizontal row noise as a result of fluctuations on the power lines. This is due to the column parallel readout mechanism used in chip level and column level architectures as all pixels on one row are reset and read out simultaneously, all pixels on that row are therefore uniformly affected by the noise level on the power lines at that time. As each row is read out in sequence at different times variations can be seen in light intensity vertically between rows (16).

## 2.5.4. Pixel Level Readout Architecture

Pixel level readout or Digital Pixel Sensor (DPS) is the most recently developed readout architecture used in CMOS image sensor design. DPS brings the analogue processing down to pixel level as illustrated in Figure 12.

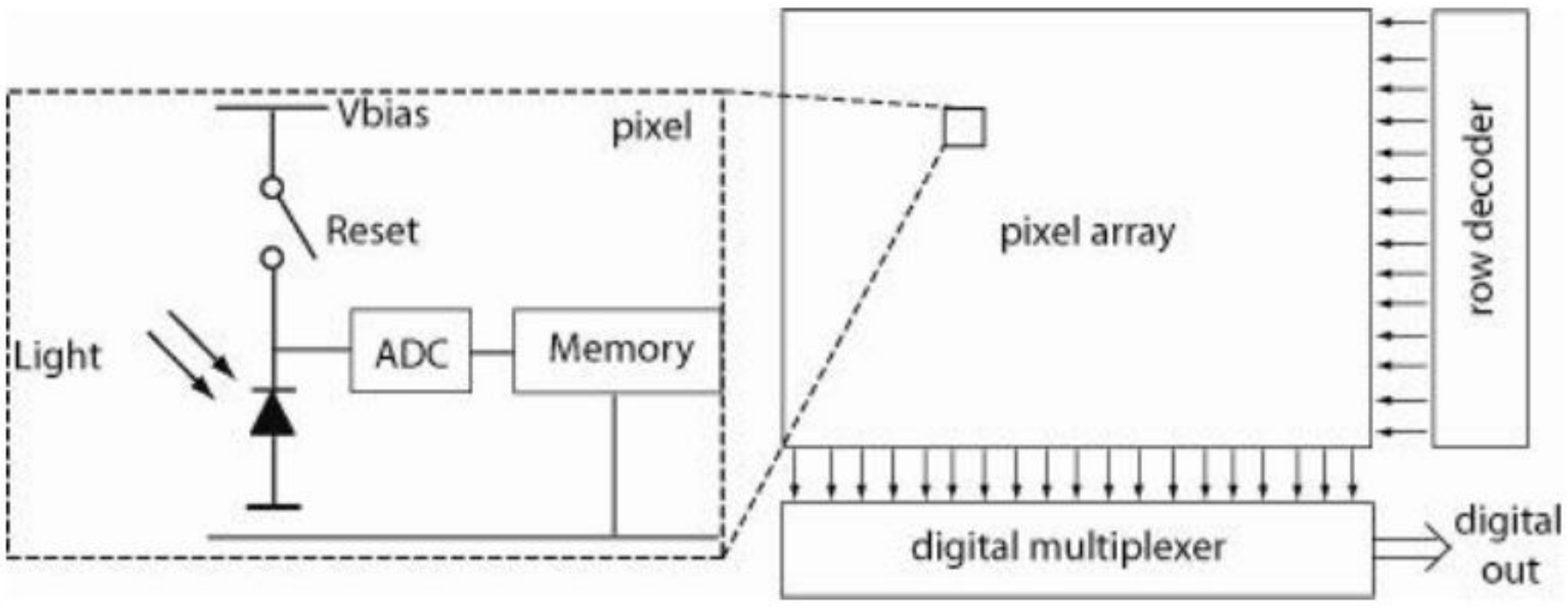

**Figure 12: Pixel Level Readout Architecture**

DPS image sensors are easily scalable and allow for very high readout speeds as the entire pixel array can be read out simultaneously. Although Digital Pixel Sensors also suffer from Fixed Pattern Noise as a result of ADC offset and gain mismatches, pixel level FPN is four to five times less visible to the naked eye that column level FPN (16). The row noise phenomenon can also be avoided by using this pixel architecture as all pixels in all rows can be reset and read simultaneously and therefore equally impacted by any transitions in power supply noise and image intensity.

The primary disadvantage of Digital Pixel Sensor technology is that very large pixel sizes are necessary as all analogue processing blocks are placed within the pixel and the fill factor is greatly reduced compared to column level and chip level readout technologies (17).

### 2.5.5. Sensing Colour

An image sensor is made up of millions of electrically identical pixels in an array, each of which is unable to distinguish between the colour of light absorbed during its operation. This colour separation is carried out by the addition of a Colour Filter Array (CFA) which is placed over the image sensor such as the Bayer colour filter array shown in Figure 13 below.

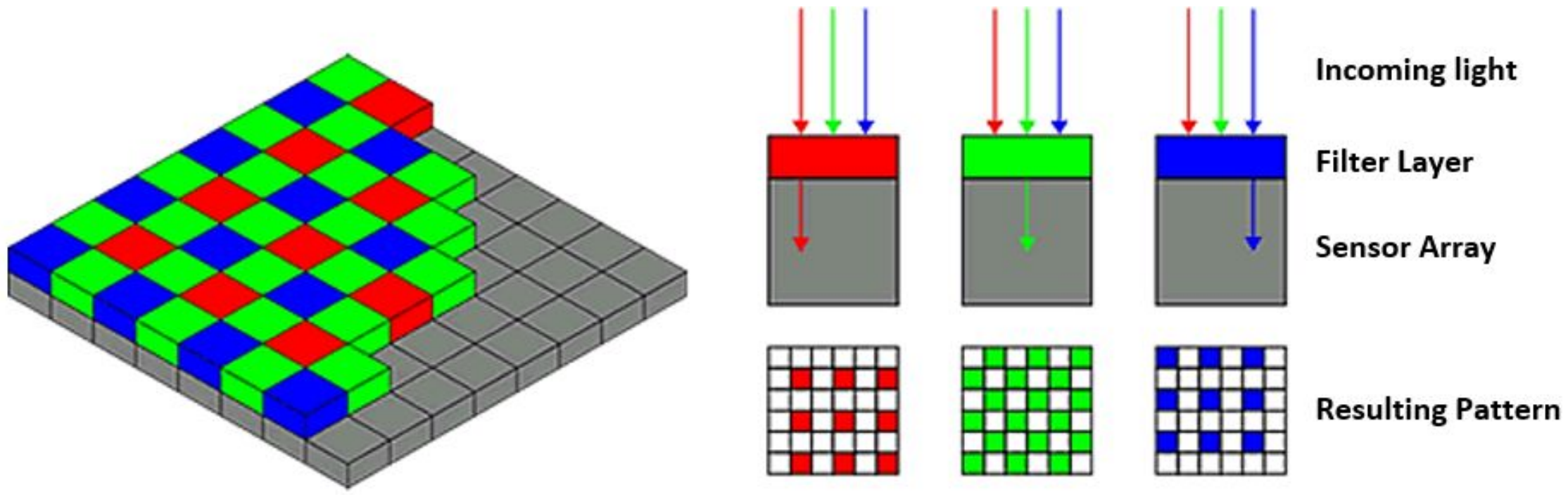

**Figure 13: Bayer Colour Filter Array**

The Bayer CFA is an array of red, green and blue colour filters arranged in a mosaic pattern. Each of the individual filters allows through only light of particular wavelengths representing either red, green or blue colour so that each pixel can perceive only one colour. As the human eye is more sensitive to green light than blue or red light, the Bayer colour filter contains twice as many green pixels as blue or red pixels (18). This raw image information is then used to inform the creation of a full colour image by using a demosaicing algorithm in order to calculate the red, green and blue component of each pixel. This process is also known as colour reconstruction or CFA interpolation. Several techniques can be used for colour reconstruction such as "nearest neighbour interpolation" where the colour values for the closest red, blue and green pixels are used to create an RGB pixel in each pixel location. Another example is "bilinear interpolation" where the red value of a non-red pixel is calculated as an average of two or four of its nearest red pixels and the same process is carried out for green and blue (16).

## 2.6. Noise Sources in CMOS Image Sensors

CMOS image sensors suffer from several different noise sources which set fundamental limits on image sensor performance particularly in video and low light applications. These noise sources can be grouped into two main categories, ***Temporal*** noise which varies with each frame captured and ***Spatial*** noise which does not vary from frame to frame. Table 1 below lists the noise sources which occur in each category.

**Table 1: Noise Sources in CMOS Image Sensors**

| Temporal Noise | Spatial Noise |
|---|---|
| Photon Shot Noise | Fixed Pattern Noise |
| Reset / kTC Noise | Column Fixed Pattern Noise |
| Thermal / Johnson Noise | |
| Flicker / $\frac{1}{f}$ Noise | |
| Quantization Noise | |

Although the origins of each noise source and the stage at which they are introduced differs as illustrated in Figure 14 below (20), they all appear as variations in the image intensity (21).

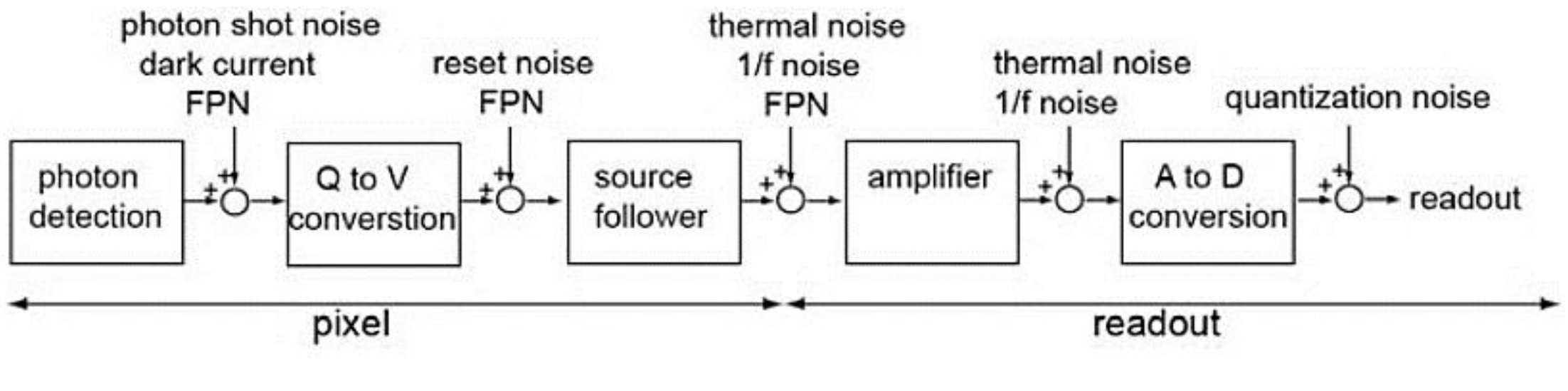

**Figure 14: Introduction of Noise Sources in CMOS Image Sensors**

This chapter provides an overview of the noise sources which occur within CMOS image sensors.

## 2.6.1. Photon Shot Noise

Photon Shot Noise occurs as a result of the fundamental physics of light rather than the nature of the image sensor technology or design characteristics. Shot noise is caused by the particle nature of photons and the statistical variation in the amount of photons which strike the sensor at any point during the exposure time (12). This process follows a Poisson distribution. Photon shot noise of a pixel, $\sigma_{ph}$ is equal to the square root of the mean number of photons which strike the pixel throughout the exposure time, $\mu_{ph}$.

$$\sigma_{ph} = \sqrt{\mu_{ph}} \tag{2.2}$$

Once the photons are absorbed by the silicon and converted to electrons this relationship remains constant in that the shot noise in electrons, $\sigma_e$ is also equal to the square root of the mean number of electrons captured by the sensor $\mu_e$.

$$\sigma_e = \sqrt{\mu_e} \tag{2.3}$$

In the case of an ideal noise-free sensor in a noise-free camera, the performance of the camera would be restricted by the photon shot noise and therefore the maximum Signal to Noise Ratio, $SNR_{Max}$ is equal to the square root of the signal value (12).

$$SNR_{Max} = \frac{\mu_e}{\sigma_e} = \frac{\mu_e}{\sqrt{\mu_e}} = \sqrt{\mu_e} \tag{2.4}$$

## 2.6.2. Reset / kTC Noise

Reset noise, also referred to as kTC noise occurs when an Active Pixel Sensor (APS) pixel is reset. Reset noise, $\sigma_{Reset}$ is a form of thermal noise generated by the resistance within the reset transistor and is expressed by:

$$\sigma_{Reset} = \sqrt{\frac{kT}{C}} \tag{2.5}$$

Where $k$ is Boltzmann's constant, $T$ is the temperature in degrees Kelvin and $C$ is the junction capacitance of the photodiode (22). Reset noise can be much larger than other temporal noise sources such as Johnson noise or Flicker noise however it is significantly reduced via correlated double sampling and therefore typically neglected in most analyses (21).

### 2.6.3. Thermal / Johnson Noise

Thermal noise, also known as Johnson noise is the electronic noise that is generated by the random thermal motion of the charge carriers in an electrical conductor. In a CMOS image sensor this noise is generated by the source follower transistor (23). The open-circuit Thermal noise voltage, $\sigma_{Thermal}$ is given by:

$$\sigma_{Thermal} = \sqrt{4kTRB} \tag{2.6}$$

The power spectral density, $\overline{V_T^2}$ of the noise can be given by:

$$\overline{V_T^2} = 4kTR \tag{2.7}$$

Where $k$ is Boltzmann's constant, $T$ is the temperature in degrees Kelvin, $R$ is the resistance of the electrical conductor and $B$ is the bandwidth in Hertz (24). Thermal noise has a flat frequency spectrum, i.e. white noise, therefore the equivalent noise amplitude is proportional to its bandwidth. Thermal noise bandwidth and amplitude can therefore be decreased in circuit design through the use of a large sampling capacitor (23).

### 2.6.4. Flicker / 1/f Noise

Flicker noise is a common noise source in transistors where the power spectral density is inversely proportional to the frequency and therefore it is also referred to as 1/f noise or pink noise. Flicker noise occurs in CMOS image sensors as a result of lattice defects and contamination at the interface of the silicon channel in CMOS transistors and also as a result of the gate oxide trapping charge carriers causing carrier number fluctuation (23, 25). Flicker noise voltage, $\sigma_{\frac{1}{f}}$ can be quantified by:

$$\sigma_{\frac{1}{f}} \approx \frac{1}{WL} \tag{2.8}$$

Where $W$ is the transistor width in $\mu m$ and $L$ is the transistor length in $\mu m$. Improvements in process technology to reduce defects and contaminants and correlated double sampling can minimise flicker noise as can increasing the transistor area in circuit design as can be seen by the equation below:

$$\overline{V_F}^2 = \frac{K_{\frac{1}{f}}}{C_{OX}WL} \cdot \frac{1}{f} \qquad (2.9)$$

Where $\overline{V_F}^2$ is Flicker noise power spectral density, $K_{\frac{1}{f}}$ is the flicker noise coefficient, $C_{OX}$ is the gate capacitance per unit area, $W$ is transistor width, $L$ is the transistor length and $f$ is the frequency.

## 2.6.5. Quantization Noise

Quantization noise can occur as a result of the CMOS image sensors analogue to digital (ADC) conversion. As an ADC produces discrete output levels, a range of analogue inputs may result in the same output as determined by the resolution of the ADC. This uncertainty or error produces a quantization noise, $\sigma_{ADC}$ that can be calculated by:

$$\sigma_{ADC} = \frac{V_{LSB}}{\sqrt{12}} \qquad (2.10)$$

Where $V_{LSB}$ is the analogue voltage in the least significant bit (12, 21). Quantization noise can be reduced by increasing the resolution of the analogue to digital convertor used in the image sensor design.

## 2.6.6. Fixed Pattern Noise

Fixed Pattern Noise (FPN) refers to pixel to pixel variation in the sensor array which is consistent from frame to frame. In an ideal image sensor, each pixel would produce an identical output in response to an identical input, however this is not the case in current technology and slight variations exist from pixel to pixel. Fixed pattern noise is referred to spatial noise as it does not change with time and it can be divided into two categories, Dark Fixed Pattern Noise and Light Fixed Pattern Noise known as Photo-response Non-uniformity (PRNU).

Dark Fixed Pattern Noise is signal independent, is primarily due to dark current generation from pixel to pixel and occurs when the sensor array is in the dark. PRNU in contrast is due to variations in pixel responsivity when light is applied and is the dominant pattern in most arrays as dark current and therefore dark FPN becomes negligible by cooling the array (21). PRNU is proportional to the signal and can be defined by the relation:

$$\sigma_{PRNU} = P_N S \tag{2.11}$$

Where $\sigma_{PRNU}$ is the PRNU in $rms\ e^-$, $P_N$ is the FPN quality factor (Generally < 1%) and $S$ is the Signal (26).

There are two primary sources for Fixed Pattern Noise in CMOS image sensors with a column level ADC, the offset of the source follower transistor in the pixel which contributes to pixel Fixed Pattern Noise and the column level ADC offset which can result in column Fixed Pattern Noise (23). Prevention of column FPN is of particular importance in image sensor design as striped noise is four to five times more visible to the human eye than pixel level FPN (16).

Figure 15 below demonstrates the impact that Fixed Pattern Noise has on image quality by displaying an identical image with (a) no FPN, (b) pixel level FPN and (c) column level FPN.

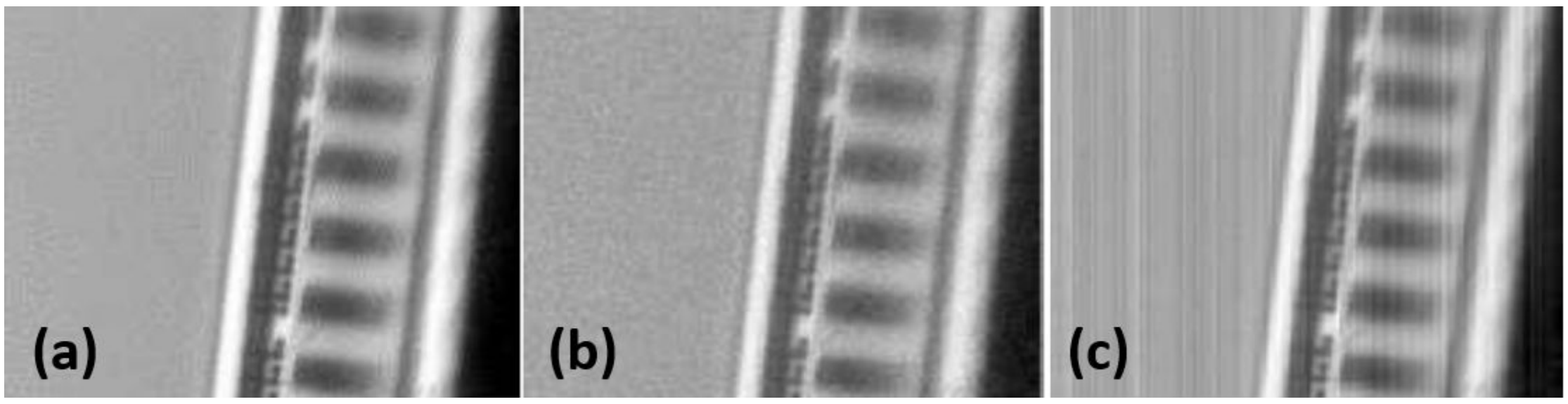

**Figure 15: Impact of Fixed Pattern Noise on Image Quality**

## 2.7. Image Row Noise

Once an image sensor has been integrated into a camera design, external noise forces can interfere with the image sensor and impact image quality. Electrical noise on the image sensor power supply lines can result in distortion observed on images in the form of row noise.

The column parallel readout architecture commonly used in CMOS image sensors can make images susceptible to row noise as a result of fluctuations of the power supply lines during pixel reset, amplification and analogue to digital conversion (18). As every pixel in a single row is reset and read in parallel, power fluctuations impact each pixel on a specific row equally. However, as each row is then read in sequence, power supply fluctuations can impact the image intensity on a row by row basis rather than pixel by pixel, thereby causing a banding effect or row noise on the image as demonstrated by Figure 16 below.

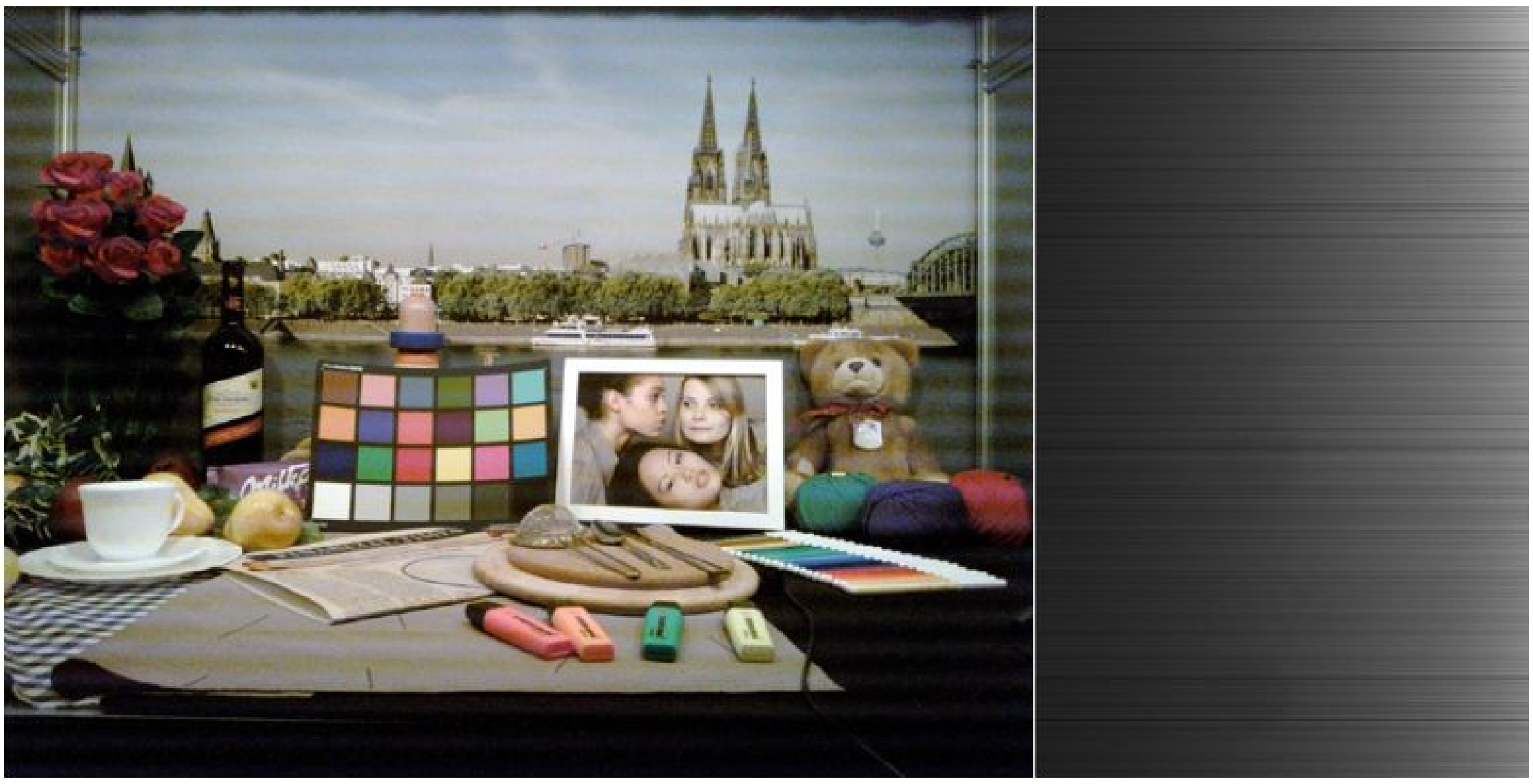

**Figure 16: Examples of Row Noise on Captured Images**

The impact of row noise on an image can vary according to the relationship between the noise signal frequency and the image line frequency. Image line frequency can be determined by the following equation:

$$Line\ frequency = frames\ per\ second\ x\ frame\ length \tag{2.12}$$

If the noise frequency is identical to the line frequency or harmonics of the line frequency, no row noise will be observed on the image however the noise may still uniformly impact the overall image intensity, or brightness of the image. Row noise will not occur in this case as the noise frequency is identical during readout for every row. This scenario is illustrated by Figure 17 below (18). The critical moment or readout is signified by the blue line and the noise level in red is identical when each row is read.

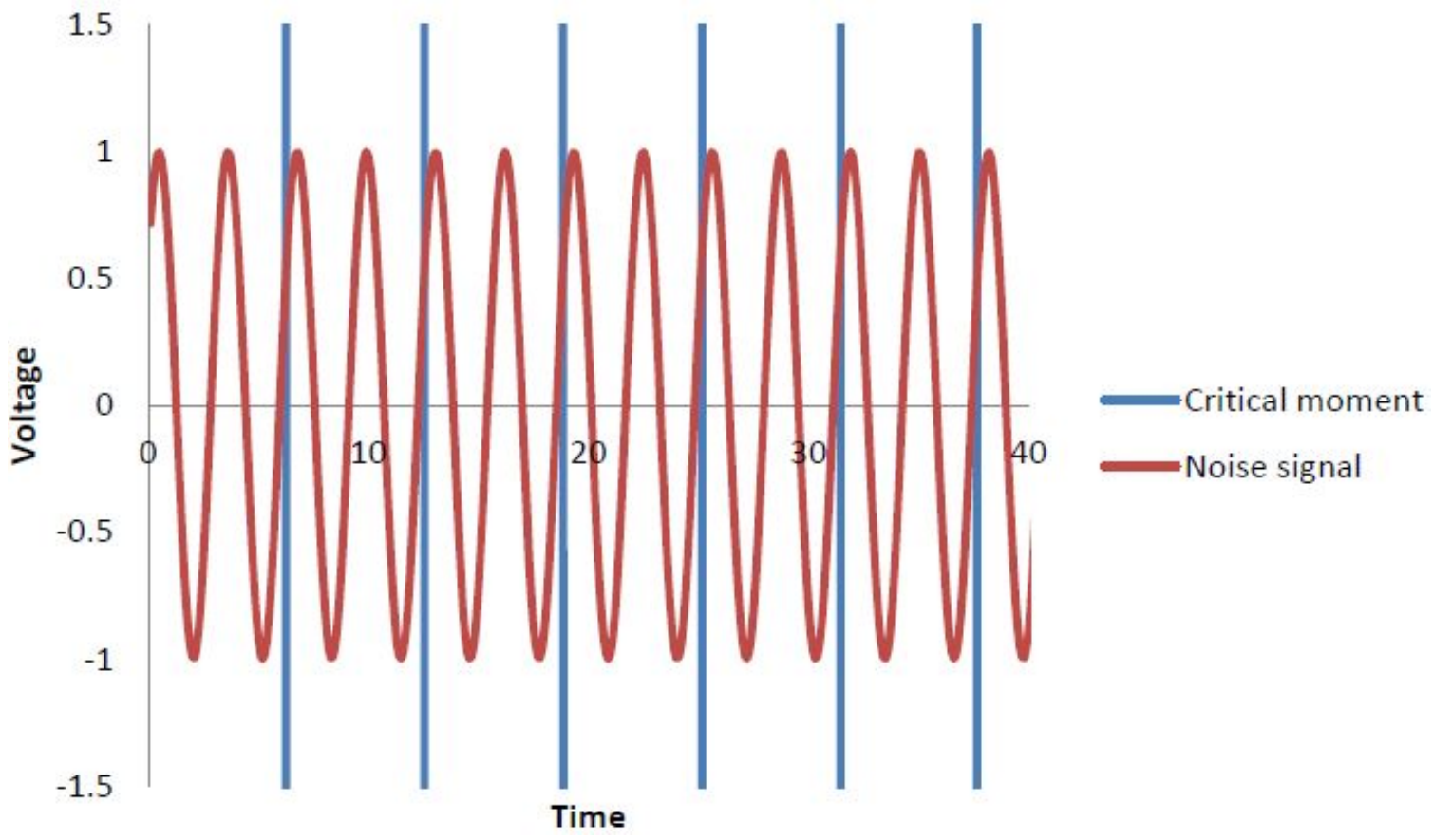

**Figure 17: Synchronised Noise Frequency and Line Frequency Harmonics**

In the case where noise frequency differs from the line frequency and its harmonics, image intensity or brightness may vary from row to row as a result of power supply fluctuations creating row noise or banding on the image. Small variations from the line frequency produces wide bands on the image and the larger the noise frequency differs from the line frequency the narrower the observed bands on the image. When the noise frequency reaches the midpoint between harmonics of the line frequency the observed bands will be just one-pixel high as the signal will be in phase on every second row readout {Mikkonen, 2014 #14}(18). Figure 18 below illustrates the scenario where the noise frequency is at the midpoint between line frequency harmonics and the noise signal alternates between two levels on sequential rows.

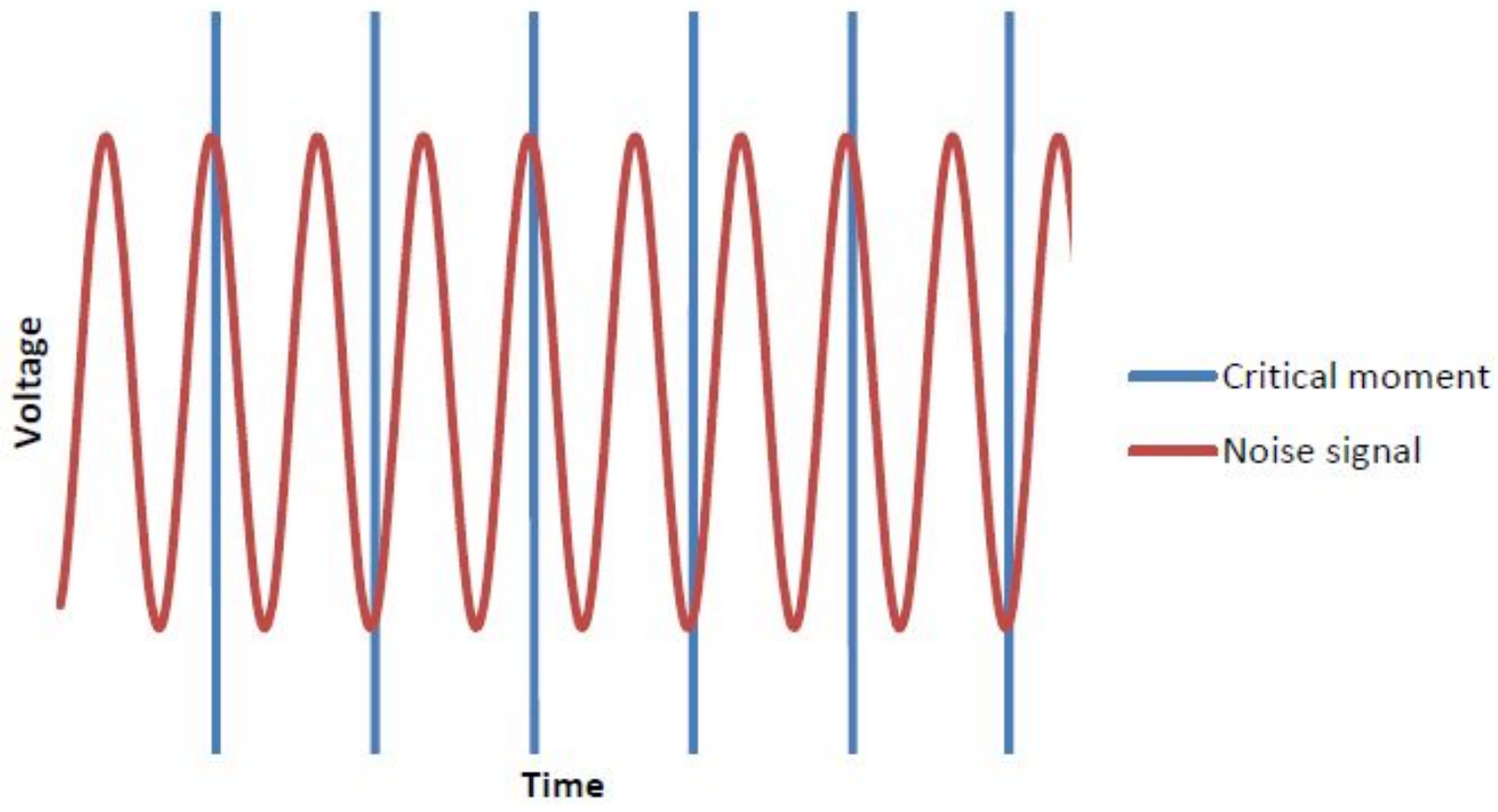

**Figure 18: Noise Frequency at Midpoint of Line Frequency Harmonics**

Row noise is perceived at its strongest when the noise frequency is close to but not in sync with the line frequency as this is the point at which the banding is at its largest row height (18).

# Chapter 3 Methodology

A methodology has been developed in order to achieve the objectives of the research project identified in Section 1.4. A flowchart of the methodology is shown in Figure 19.

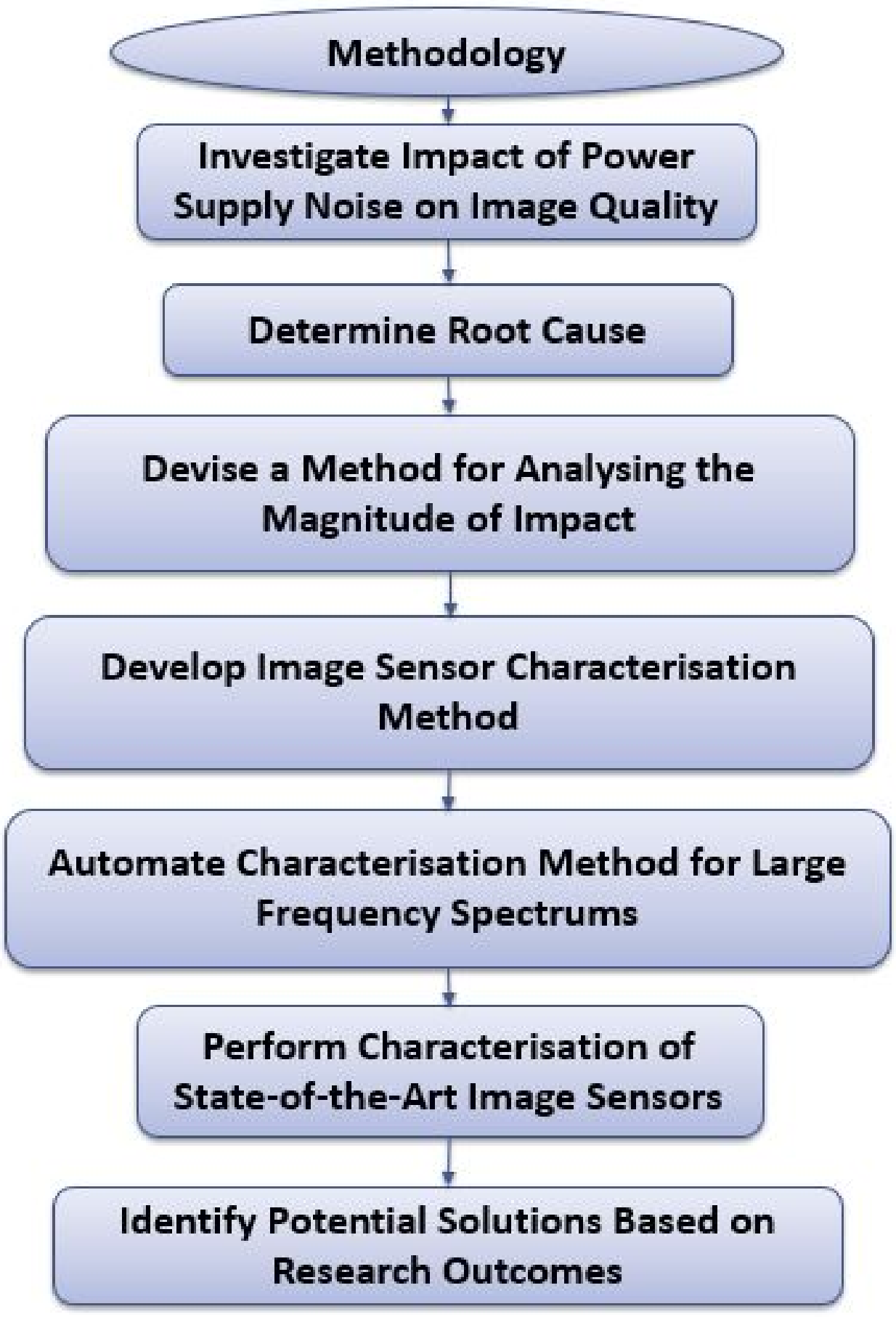

**Figure 19: Methodology Flowchart**

In order to adequately develop a characterisation method for CMOS image sensors in response to electrical noise, the impact of electrical noise on image quality must first be understood and the root cause in terms of hardware electronics must be identified. In order to assess and understand this impact a detailed literature review must be completed on CMOS pixel electronics, hardware architecture and the creation of noise sources within CMOS image sensors in order to sufficiently comprehend the formation of image distortion from root cause through to realisation.

Once the root cause is established a suitable analysis method for measuring and quantifying the magnitude and extent of any distortion in a thorough, scientific and repeatable manner must be developed. A characterisation method can then be built around this analysis in order to provide controlled frequencies of electrical noise as an input and to conduct analysis to provide the characteristics in terms of immunity or susceptibility to power supply noise as an output, unique to the specific model of image sensor under test. Once a characterisation method is established, this method can be automated in order to facilitate its use over a wide frequency spectrum. This will allow thorough characterisation to be carried out on a number of state-of-the-art image sensors currently used in automotive applications. The characterisation results in conjunction with the detailed literature review and research outcomes can then be used in order to identify potential solutions for mitigating the susceptibility of a particular image sensor in response to electrical noise on the power supply lines.

# Chapter 4 Results

This chapter presents the results of the activities outlined in the methodology shown in Chapter 3. The impact of power supply noise has been assessed and the root cause has been established. A mathematical algorithm to quantify the magnitude of row noise in an image has then been developed and an appropriate characterisation method over a large frequency spectrum has been implemented. Algorithm validation, frequency spectrum investigation, and repeatability testing were carried out in order to set the initial parameters for image sensor performance characterisation. Performance characterisation was then carried out on two state of the art CMOS image sensors for automotive applications, Omnivision OV10635 and Omnivision OV7955. The results of these activities are displayed in this section of the dissertation.

## 4.1. Impact of Power Supply Noise on Image Quality

Row noise is a type of image noise created as a result of electrical noise on the image sensor power supply line interacting with the design characteristics of a CMOS image sensor to produce a horizontal banding on images as displayed in Figure 20 below.

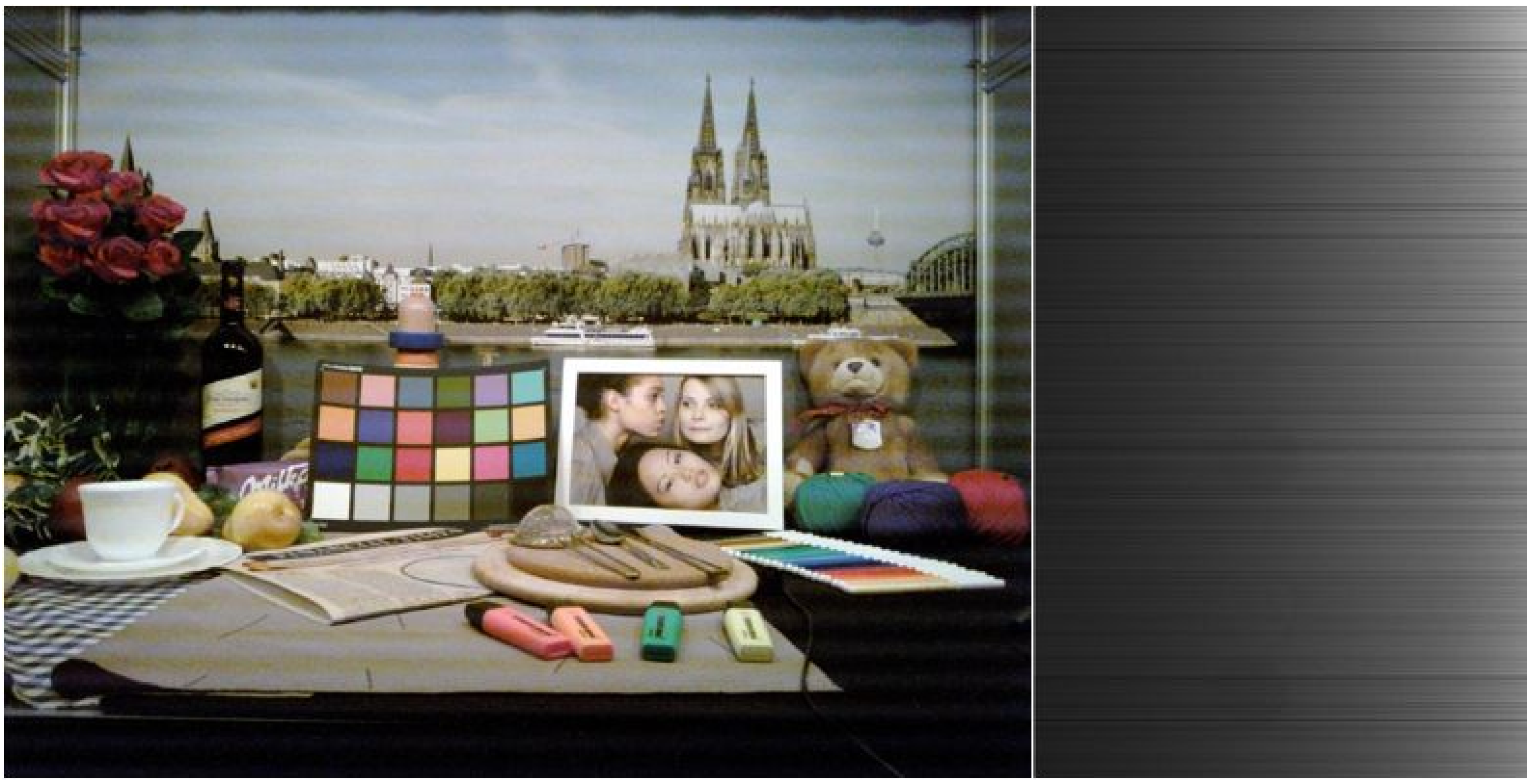

**Figure 20: Examples of Image Row Noise**

Row noise is a form of spatial noise in a similar manner to Fixed Pattern Noise (FPN) and therefore does not vary between frames like the temporal noise sources that occur within CMOS image sensor technology. Row noise is not signal dependent and therefore can be observed equally not only on a pixel array with a varying light input, but also on the Optical Black Rows (OPR) of an image sensor, designated pixels with an identical electrical make up as all other pixels however are not light sensitive. This allows row noise analysis to be conducted at 0 lux when all rows should have a uniformed 0 light intensity. The key characteristics of row noise are shown in Figure 21.

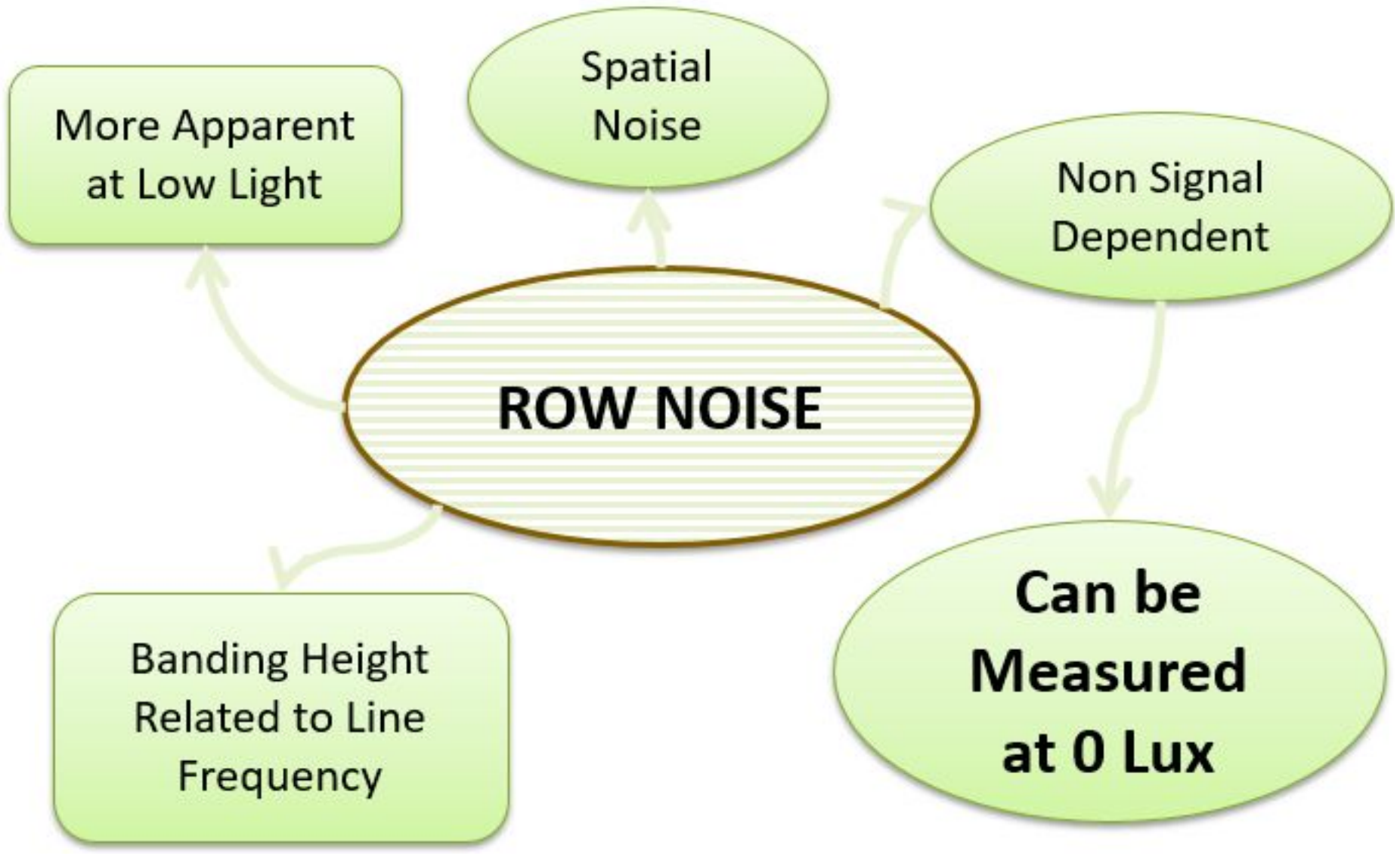

**Figure 21: Row Noise Characteristics**

Although row noise is non signal dependent it is found to be more apparent in low light applications where the Signal to Noise Ratio (SNR) of the device is greatly reduced and image noise becomes more apparent. The banding height of row noise on an image can be related to the Line Frequency which is defined as:

$$Line\ frequency = frames\ per\ second\ x\ frame\ length \tag{2.12}$$

Therefore, changes to the frame length or the number of frames per second can manipulate the visibility of row noise on an image.

## 4.2. Root Cause of Row Noise in a CMOS Image Sensor

Row noise occurs as a result of the fusion of electrical noise on the image sensor power supply lines, the operating principle of a photodiode and the column parallel readout structure commonly used in modern CMOS image sensors.

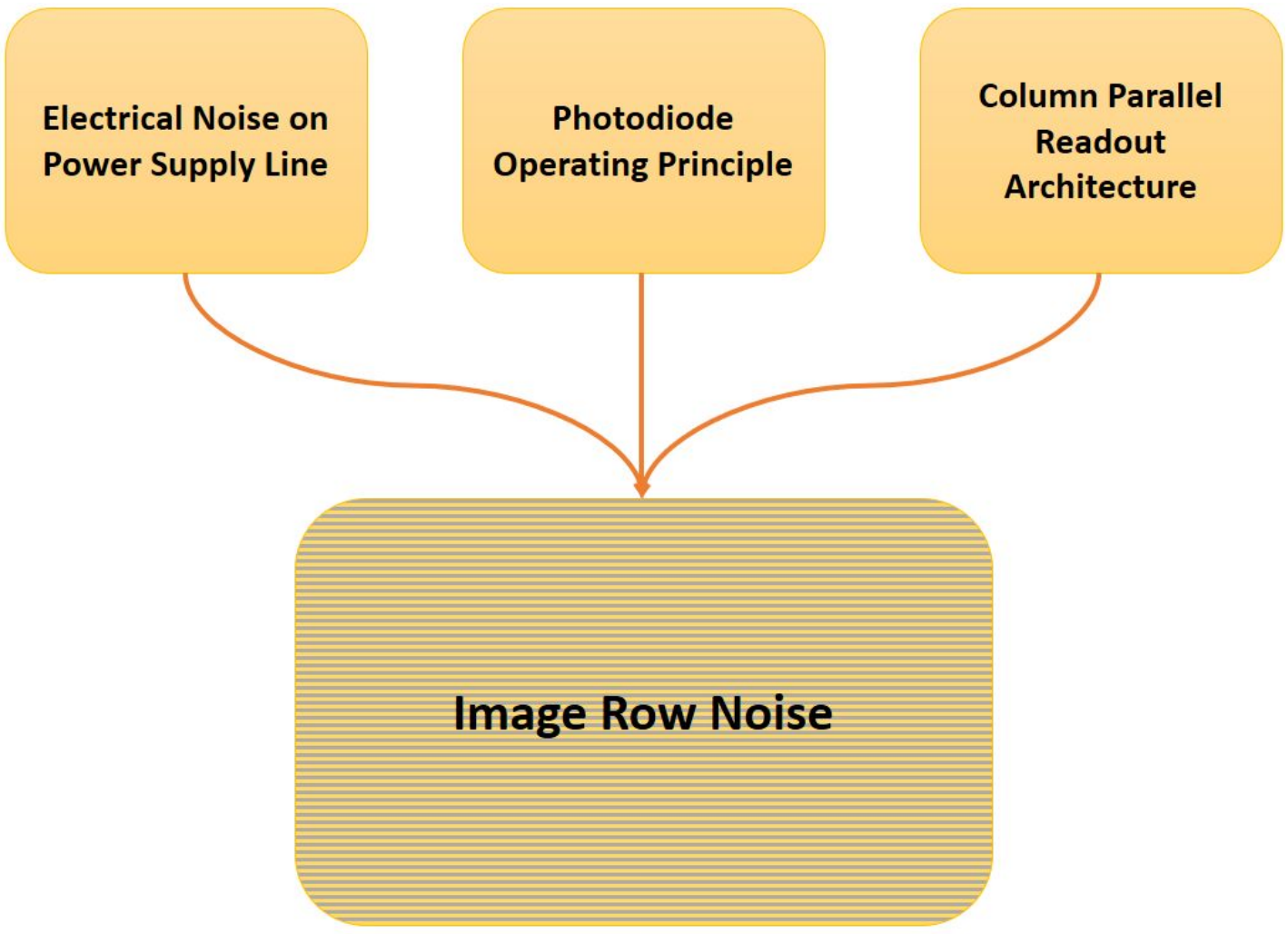

**Figure 22: Elements of Row Noise**

The current flow in a photodiode as a result of the photoelectric effect is extremely low and therefore the easiest way to measure incident light is to apply a reverse bias voltage to the photodiode and calculate the voltage drop which occurs as a result of incident light as shown in Figure 23.

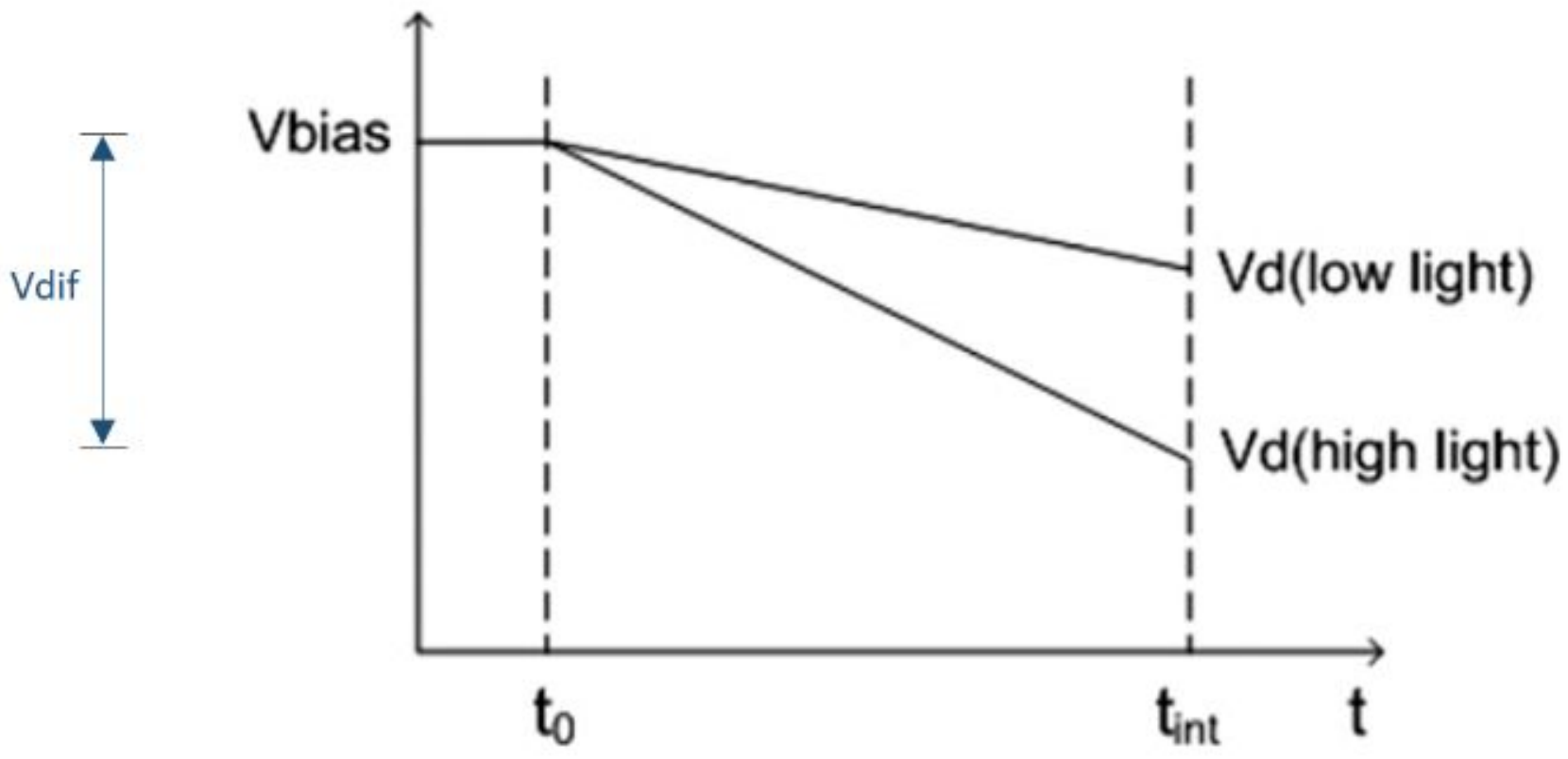

**Figure 23: Photodiode Operation**

Every pixel in an image sensor pixel array contains its own photodiode which is first reset by placing a charge across the photodiode known as Vbias. The photodiode is then exposed to incident light which causes the charge stored on the photodiode to be dissipated in proportion to the intensity of the incident light. The difference between the voltage after reset, Vbias and the voltage after exposure, Vdif is equivalent to the amount of light captured by the photodiode during exposure.

Electrical noise on the image sensor power supply voltage can in turn lead to fluctuations in Vbias, the reverse bias voltage of the photodiode. This can effectively change the starting point of the light intensity measurement and result in different light intensity values for identical inputs over time as illustrated in Figure 24.

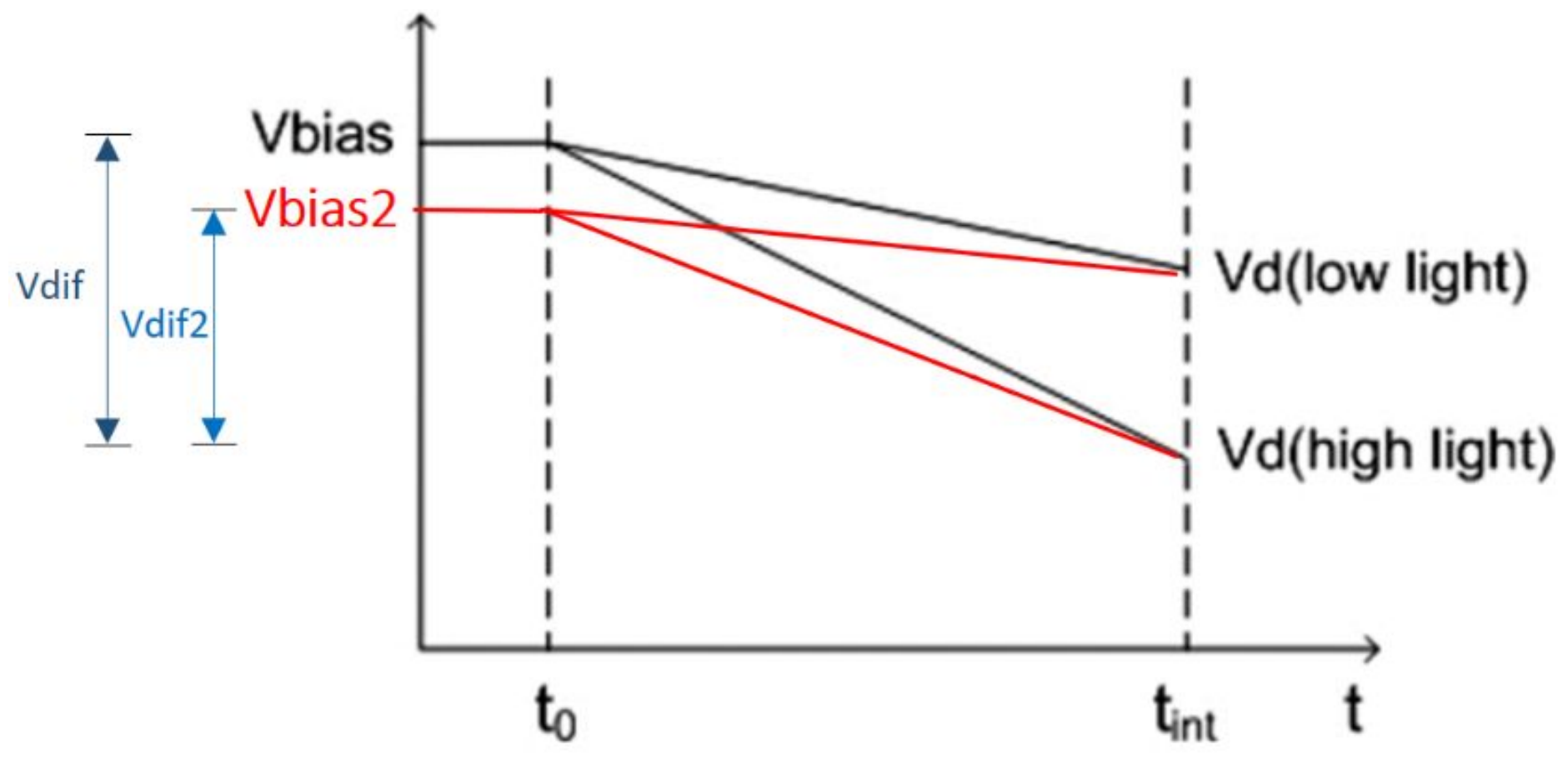

**Figure 24: Power Supply Noise Impact on Light Intensity Measurement**

Modern image sensor technologies which use chip level or column level analogue processing incorporate a column parallel readout architecture as illustrated in Figure 25.

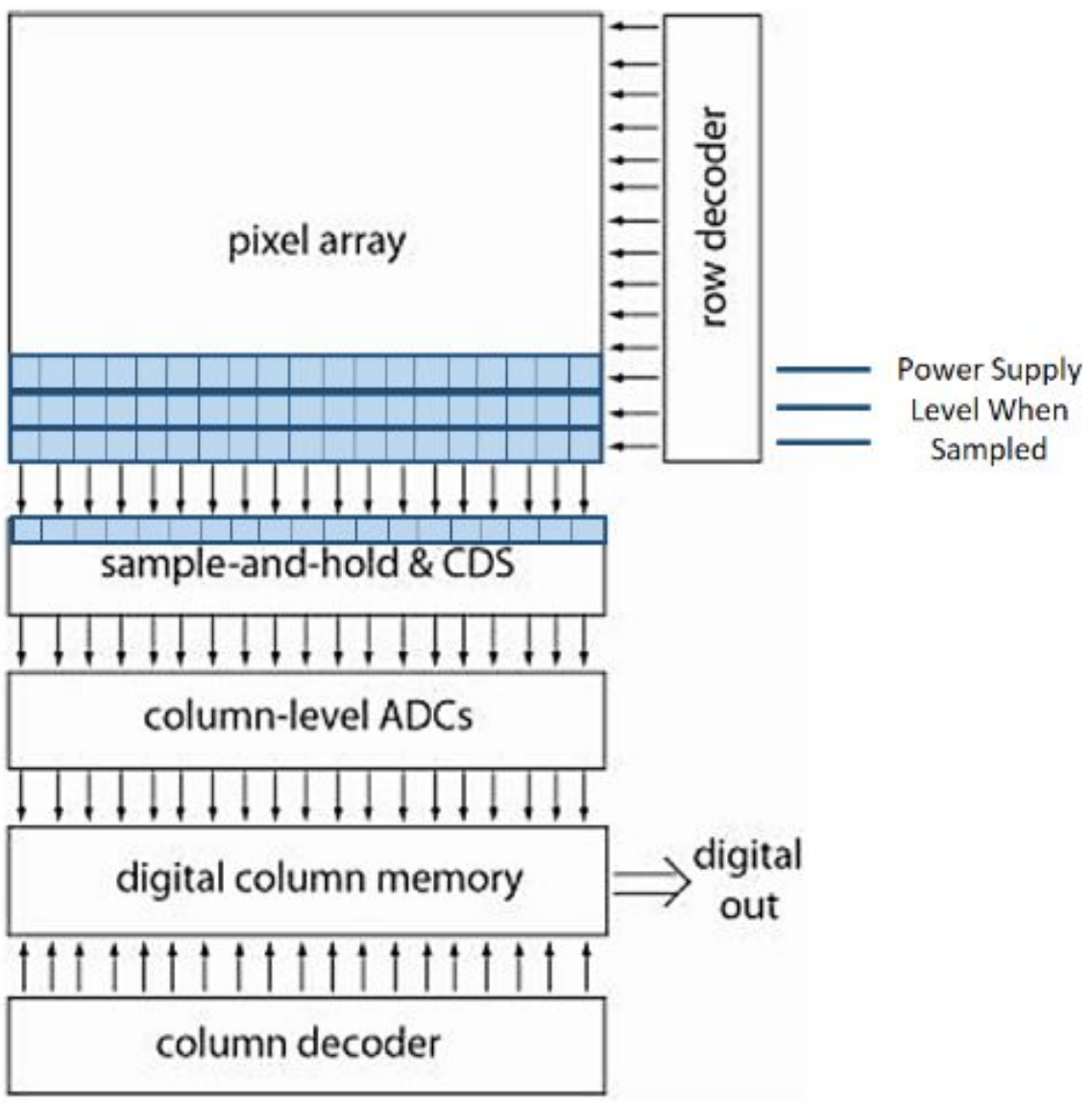

**Figure 25: Column Parallel Readout Operation**

Column parallel readout resets and samples all pixels on one row at a single point in time in a column parallel operation. The pixels on the sampled row are then digitized and the signal is transferred out. The next row in sequence is then sampled and the operation is repeated.

As discussed, in instances where electrical noise is present on the image sensor power supply lines, pixel photodiodes may produce a different light intensity value for the same input at a different point in time. This can result in light intensity variation from row to row due to the simultaneous nature by which rows are sampled in devices with a column parallel readout architecture, as demonstrated in Figure 26.

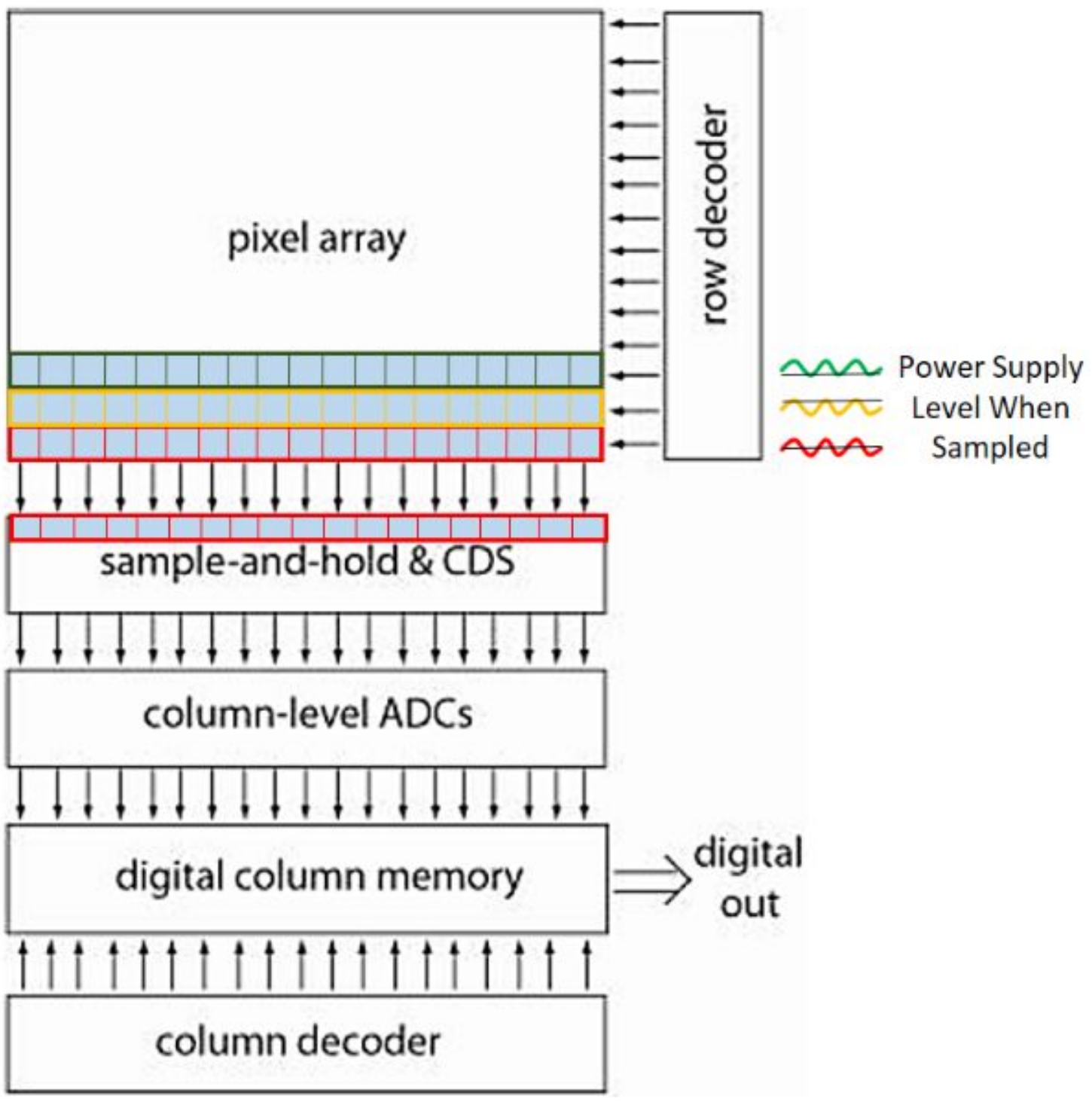

**Figure 26: How Column Parallel Readout Produces Row Noise**

These light intensity variations between rows result in horizontal bands of distortion on the output image known as row noise.

In summary, the creation of row noise in response to power supply fluctuations as a result of column parallel readout occurs as follows:

1. Electrical noise present on the image sensor power supply lines cause image intensity or brightness variations on the signal due to variation of the pixel photodiode reverse bias voltage.
2. All pixels in one row are sampled simultaneously in a column parallel operation leading to every pixel on the sampled row having an identical bias voltage when readout occurs.
3. Pixels on the sampled row are digitized and the signal is transferred out.
4. The varying nature of the noise level results in a different power supply voltage when the next complete row is read out leading to a uniformed variation in image intensity or brightness between rows known as row noise.

## 4.3. Row Noise Algorithm

A flowchart of the algorithm used to calculate the magnitude of row noise in an image is provided in Figure 27.

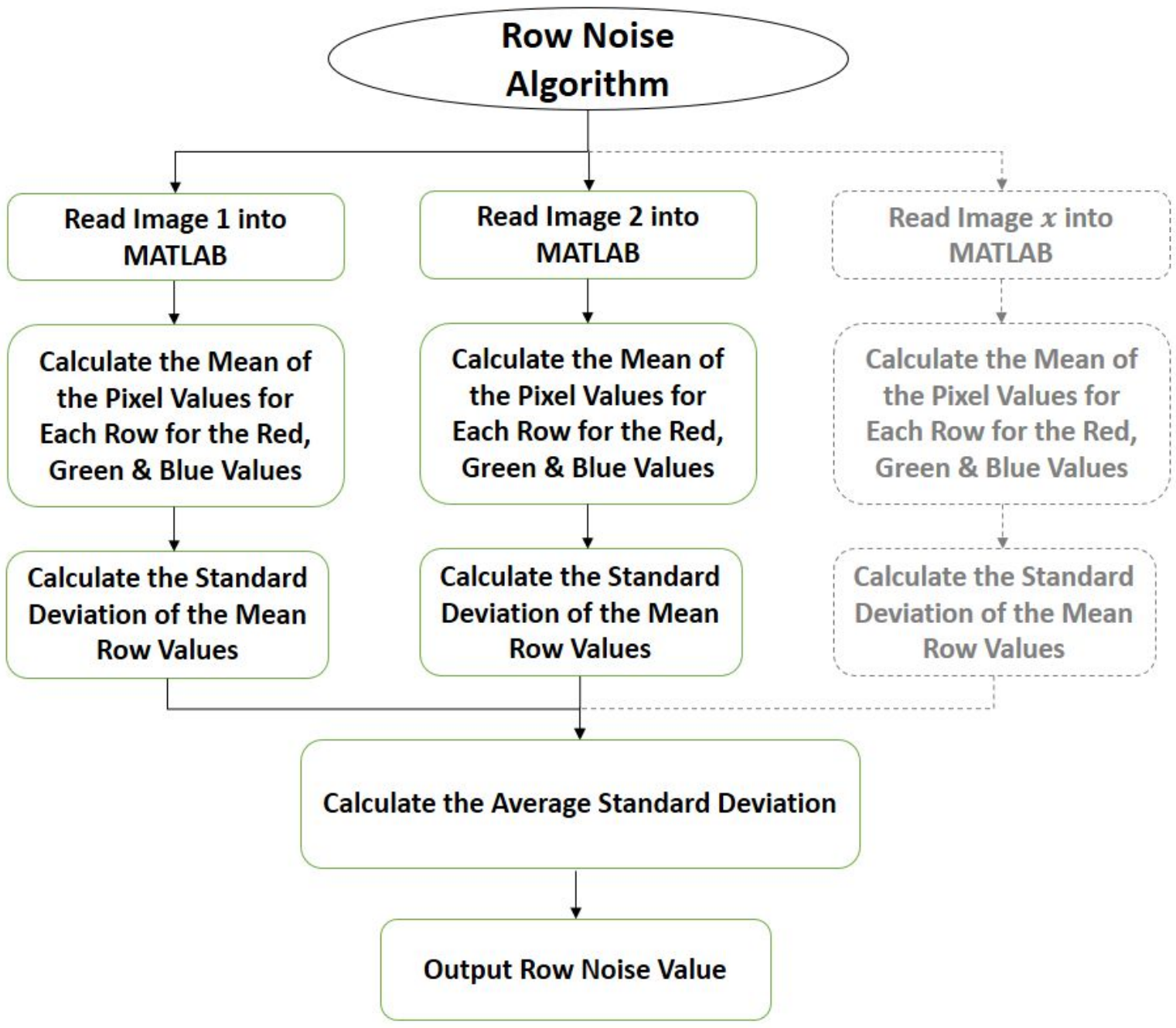

**Figure 27: Row Noise Algorithm Flowchart**

- The RGB image data for multiple images are read into MATLAB as three dimensional arrays of a size automatically determined by the image aspect ratio in order to preform mathematical analysis.
- The pixel values for each row are analysed and the mean pixel value for each row is calculated.
- This process is repeated for the red, blue and green pixel data for each row.
- The standard deviation of the mean pixel value for each row is calculated. This represents the variation between rows, identifying the presence and magnitude of row noise in the image.
- This process is repeated for multiple images at the same noise frequency step in order to reduce the effects of temporal noise on the results.
- The mean standard deviation is calculated representing the magnitude of row noise present at the specific frequency step accounting for the effects of temporal noise. The row noise value is recorded and output from the algorithm to a .csv file.

## 4.4. Image Sensor Characterisation in Response to Power Supply Noise

A method has been developed to characterise image sensor performance in response to electrical noise on the power supply lines, Figure 28.

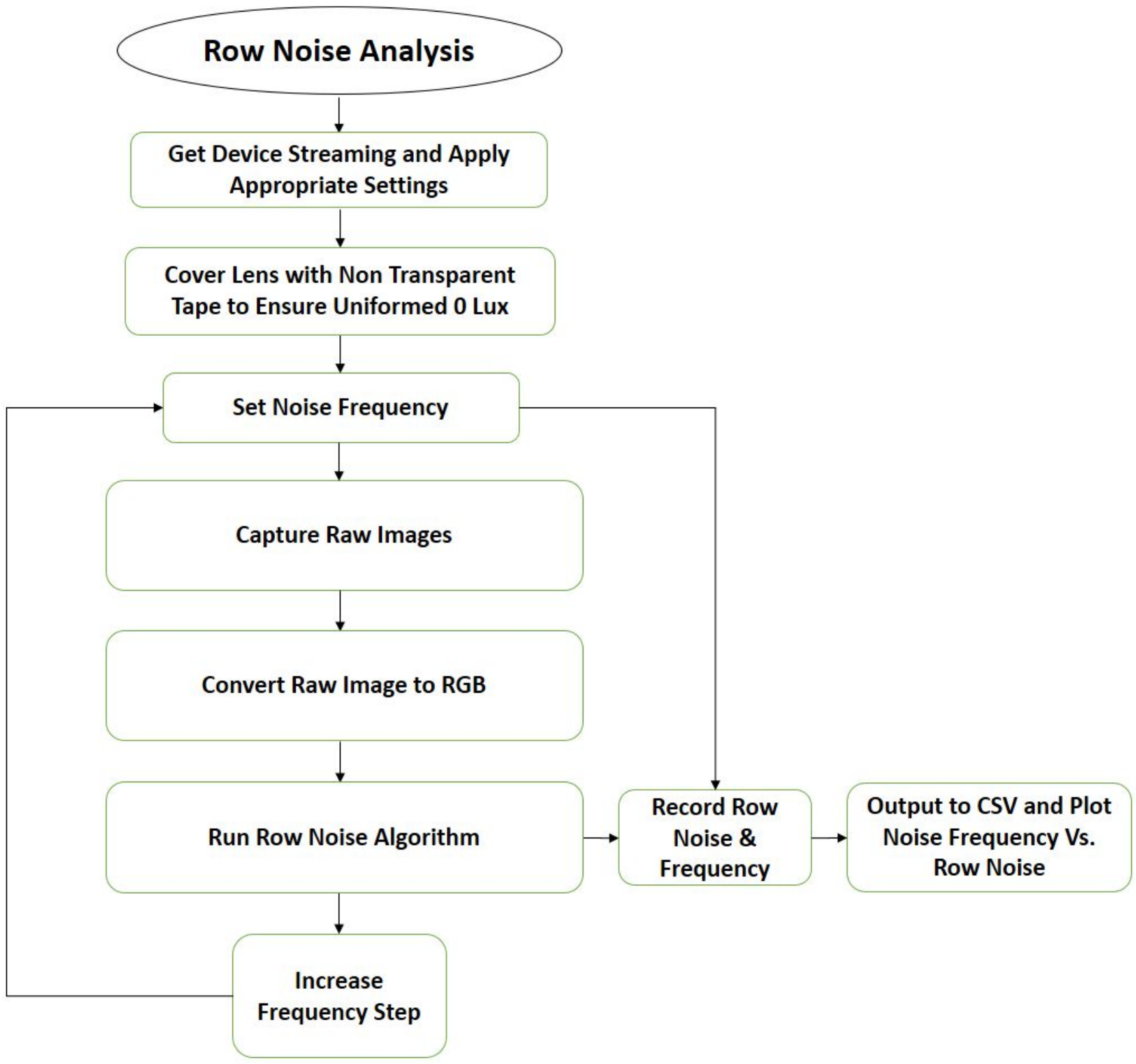

**Figure 28: Image Sensor Characterisation Flowchart**

- The appropriate device settings are applied and the device is set to streaming using proprietary software for the image sensor under test available from the manufacturer.
- The lens is covered with a non-transparent material in order to ensure no light strikes any part of the image sensor throughout the characterisation process.
- Controlled noise of a specific frequency is applied by a function generator controlled via LabVIEW.
- Raw images are captured by the proprietary software for the image sensor under test from the manufacturer.
- The raw images are converted to RGB format using proprietary or Imatest software in order to obtain the colour data of the red, green and blue component for each pixel in the pixel array.

- This RGB data is then analysed using a custom algorithm implemented in Matlab in order to determine the magnitude of row noise present in the image.
- The row noise value for the specific noise input frequency is recorded.
- The noise frequency step is then increased by the desired frequency determined by the user specified sample rate i.e. 1 kHz.
- The process is repeated until the user specified frequency range is complete.
- The row noise values for each frequency step are recorded in a .csv file.

## 4.5. Equipment Required

The following equipment was utilised in order to carry out image sensor characterisation in response to electrical noise on the power supply lines, Figure 29.

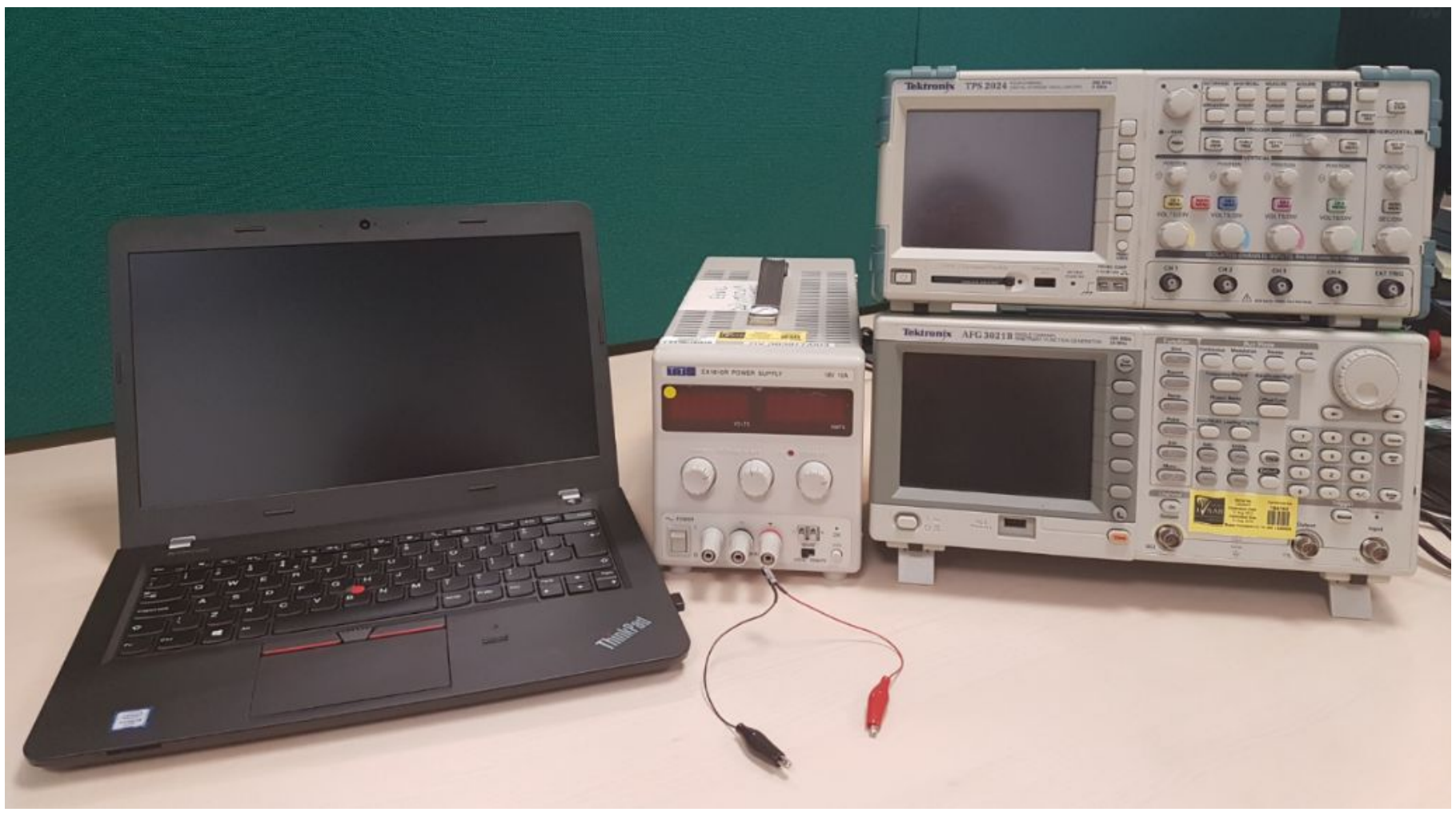

**Figure 29: Equipment Required for Image Sensor Performance Characterisation**

- Function Generator – LabVIEW Compatible i.e. Tektronix AFG 3021B
- Variable Power Supply i.e. TTi EX1810R
- Oscilloscope i.e. Tektronix TPS2024
- Coupling Decoupling Network (CDN) or 15 uF Electrolytic Capacitor
- Copper / Non-transparent tape
- PC with the following software:
    - LabVIEW
    - NI TestStand
    - Image Sensor Proprietary Software
    - IMATEST (*Optional)

## 4.6. Test Setup

The test setup for image sensor characterisation is shown in Figure 30.

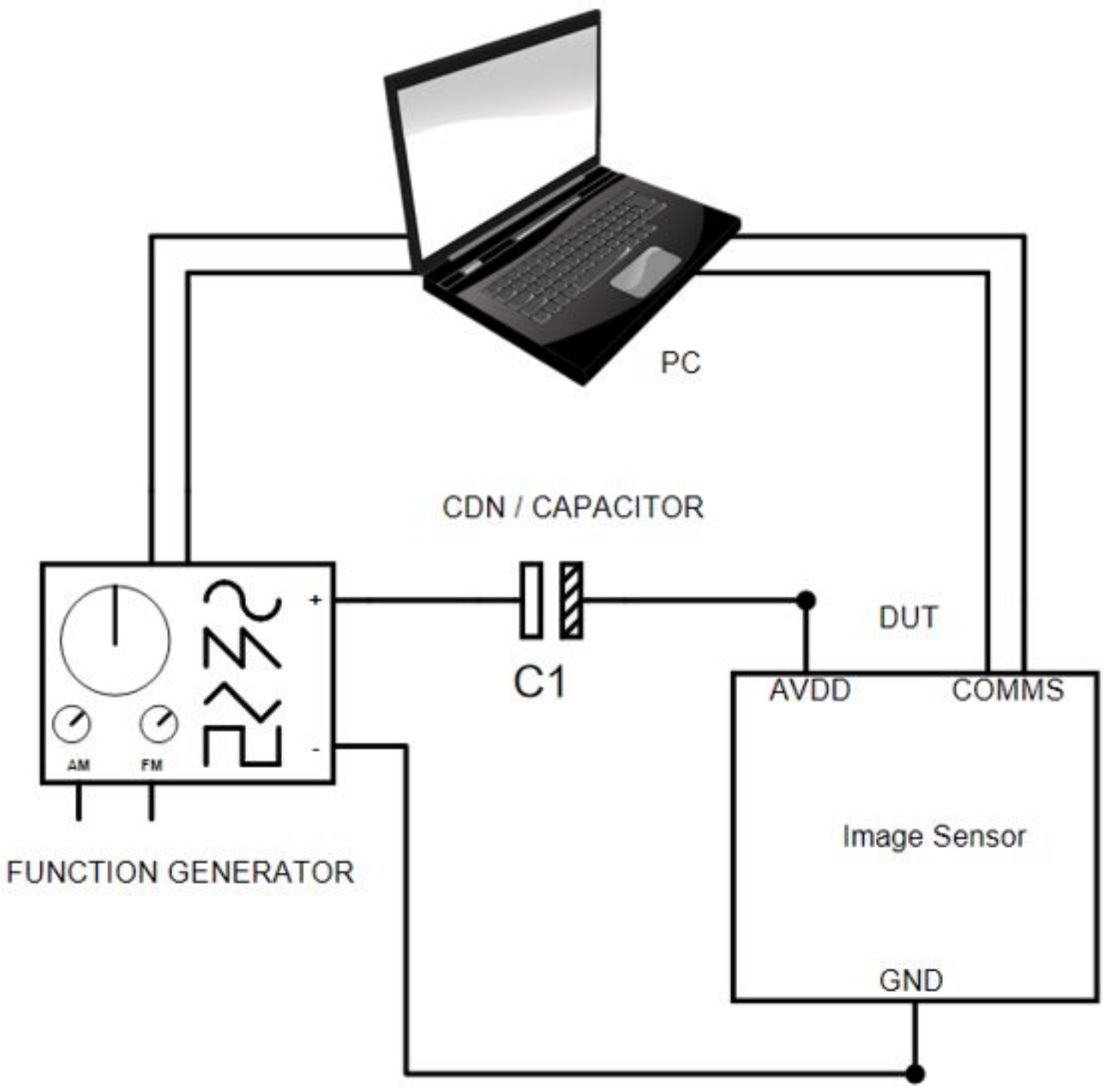

**Figure 30: Set Up Required for Image Sensor Performance Characterisation**

The procedure was employed as follows:

1. Connect Function Generator positive output to a Coupling/Decoupling Network or to the positive terminal of 15uF Electrolytic Capacitor.
2. Connect Coupling/Decoupling Network or the negative terminal of 15uF Electrolytic Capacitor to the desired voltage line on the image sensor evaluation PCB i.e. AVDD, DVDD etc. See evaluation board schematic for reference.
3. Connect the negative terminal of the Function Generator to a suitable Ground line on the evaluation PCB.
4. Connect the Function Generator to the PC via USB or GPIB interface.
5. Connect the device under test to the PC through the standard interface for power and communication i.e. USB.
6. Determine the normal level of the power supply the noise is to be coupled on to i.e. 2.8V, 3.3V or 5V etc.

7. In order to avoid image sensor damage, perform manual test of noise coupling using a bench power supply, function generator and oscilloscope. Couple the desired amplitude of electrical noise on to power supply output of appropriate voltage level and confirm correct operation using oscilloscope.
8. Find appropriate point on image sensor evaluation PCB schematic to apply injected electrical noise on desired power line and ground. *NB ensure noise is applied downstream of power regulator circuits as injected electrical noise may be suppressed by power regulators.
9. Confirm that "RowNoiseAnalyser" folder is saved at location "C:\RowNoiseAnalyser" directly on the C:\ Drive of host PC.
10. Open image sensor proprietary software and apply the following image sensor settings:
    - Output Single Pixel Raw Image 8-bit Data Format
    - Disable Black Level Calibration (BLC)
11. Confirm Device is streaming appropriately and then cover the lens with copper or non-transparent tape in order to block the image sensor from all light sources.
12. Apply the following capture settings:
    - 8-Bit Data format
    - Save files to location "C:\RowNoiseAnalyser"
    - Save as **.bmp** format
13. If image sensor proprietary software does not support saving Raw images as RGB .bmp files, IMATEST or similar software may need to be initialised also at this point.

## 4.7. Characterisation Procedure

Once setup is complete the following procedure can be used to conduct image sensor performance characterisation in response to electrical noise on the power supply lines.

1) Open “RowNoiseAnalysis” test sequence within NI TestStand as seen in Figure 31 below:

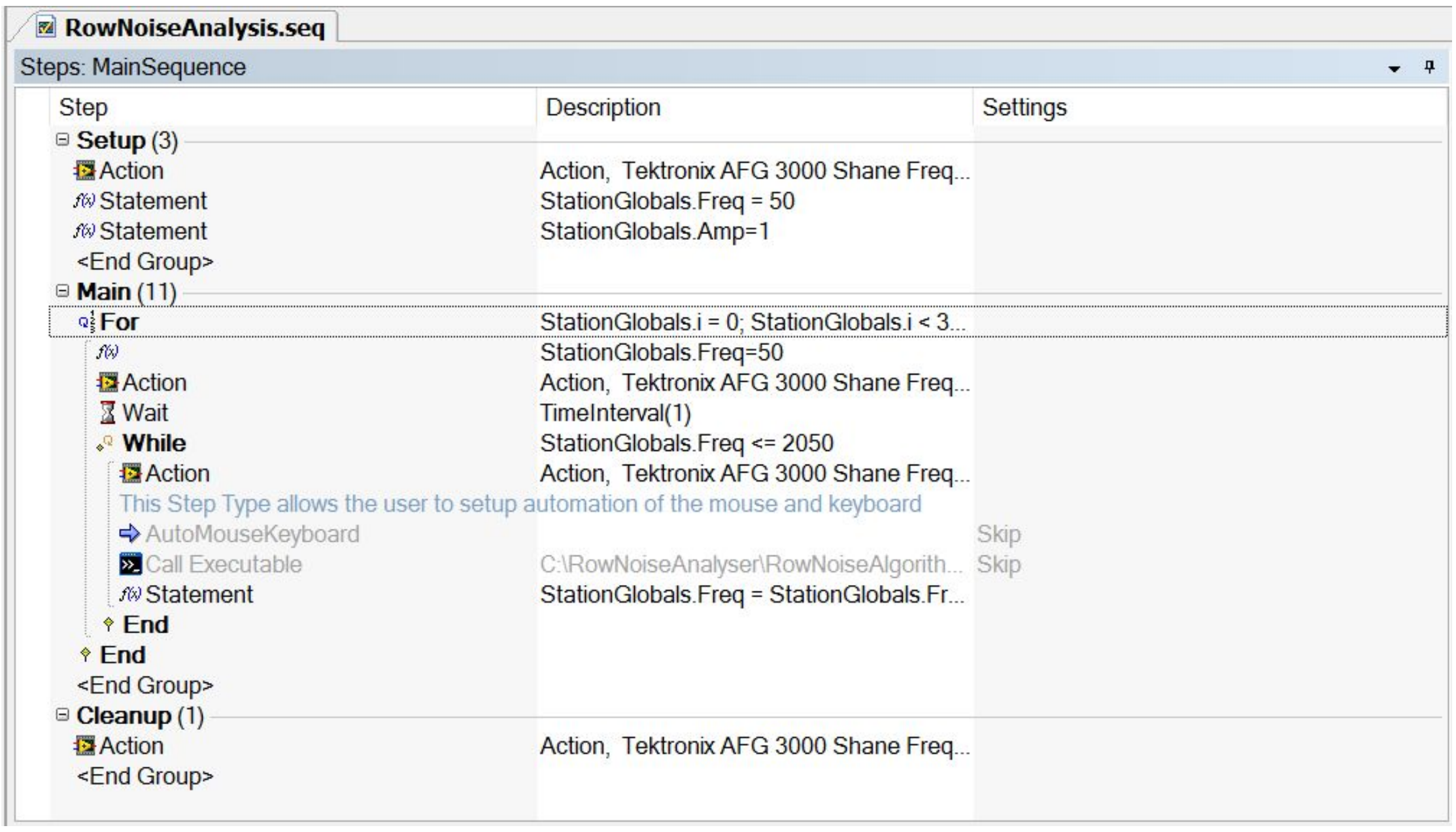

**Figure 31: “RowNoiseAnalysis” TestStand Sequence**

2) Specify starting frequency in Hertz (Hz) and amplitude in Volts peak to peak (Vpp) by editing the variables “Freq” and “Amp” respectively in the **Setup** section of the TestStand sequence.
3) Specify the end value of the frequency sweep (Hz) by editing the “Freq” variable of the **While** statement in the **Main** section of the TestStand sequence.
4) Specify the frequency step size by editing the increment statement at the end of the **While** loop as follows: “Freq = Freq + *Step Size (Hz)*”
5) Set up Image Capture using the AutoMouseKeyboard application as follows:
    a. Right Click “AutoMouseKeyboard” and click the **Edit** option to open the following dialogue box:

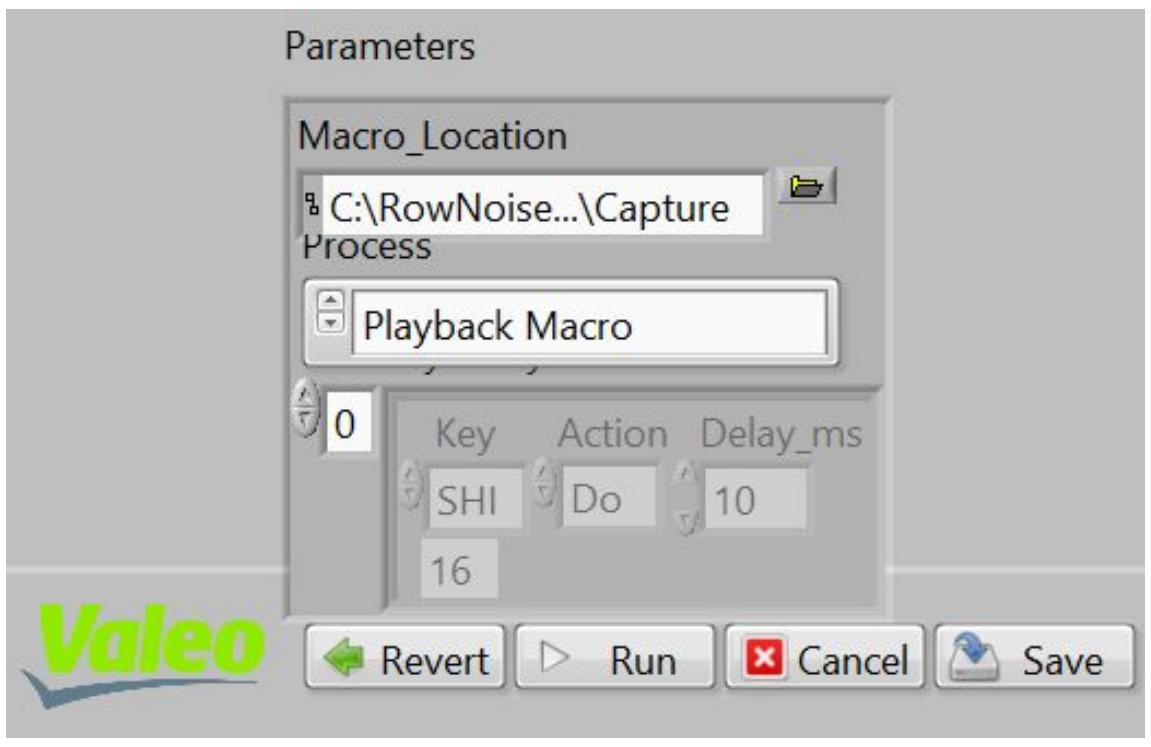

**Figure 32: AutoMouseKeyboard User Interface**

b. Change process from **Playback Macro** to **Record Macro**.

c. Click **Run** then proceed to use image sensor proprietary software to capture 3 images, convert to RGB if necessary and save images as "**im1**", "**im2**" & "**im3**" in **.bmp** format.

d. Click

to stop recording macro and save macro.

6) Click **Execute** – **Test UUTs** as demonstrated below in order to run image sensor characterisation.

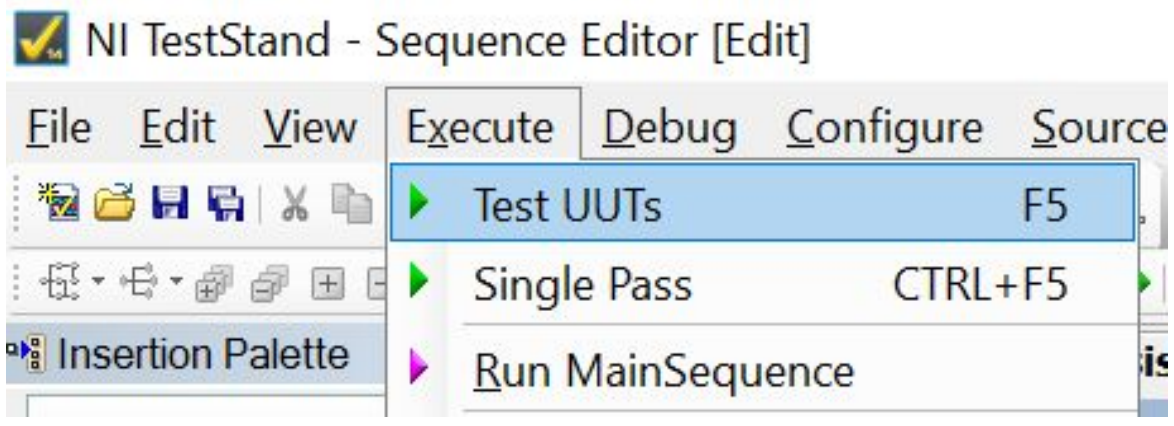

**Figure 33: Run Image Sensor Characterisation**

7) The program will output a .csv file "RowNoiseOut" 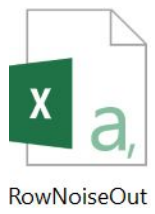

in the application folder upon completion.

## 4.8. Validation

Validation of the image sensor characterisation method was carried in order to assess its ability to systematically produce an appropriate image row noise value based on the magnitude of row to row variation in an image in a structured repeatable manner. Frequency spectrum analysis was also carried out in order to determine a suitable frequency sweep size for initial image sensor characterisation in response to electrical noise on the power supply lines.

### 4.8.1. Row Noise Algorithm Validation

The theory behind the row noise algorithm has been validated by analysing the row noise in images with increasing degrees of visible row noise using a foundational algorithm in order to calculate a row noise value for each image, Table 2.

It can be visually confirmed that the row noise algorithm successfully assigns an appropriate relative magnitude value based on the amount of visual row noise present in an image.

**Table 2: Row Noise Algorithm Validation**

| Input Noise Frequency | Captured Image | Row Noise |
|---|---|---|
| No Noise | 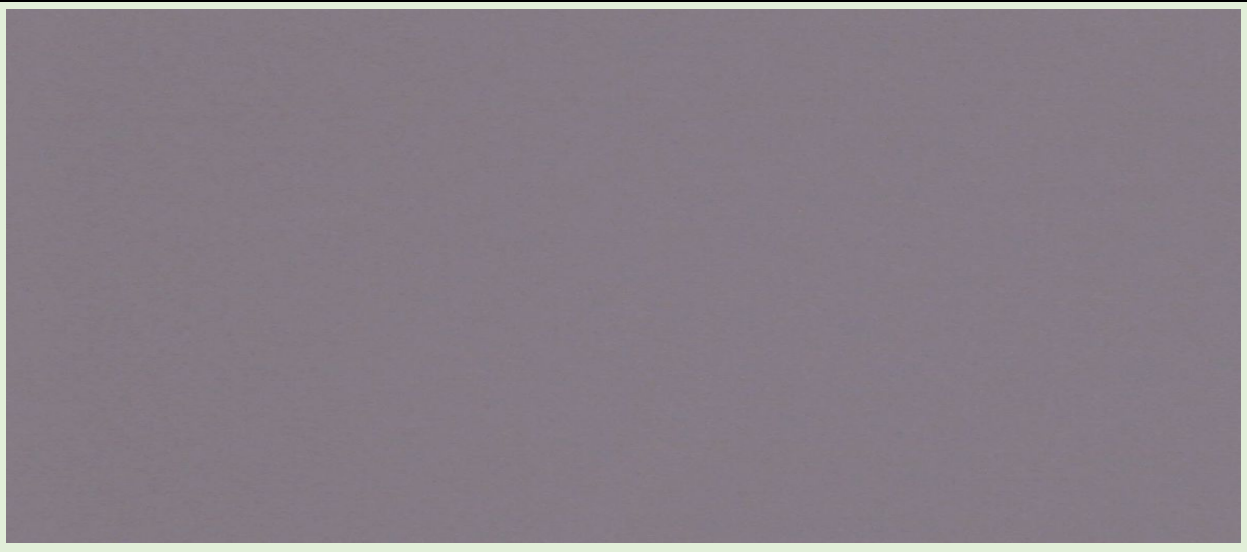 | **0.3805** |
| 25 kHz | 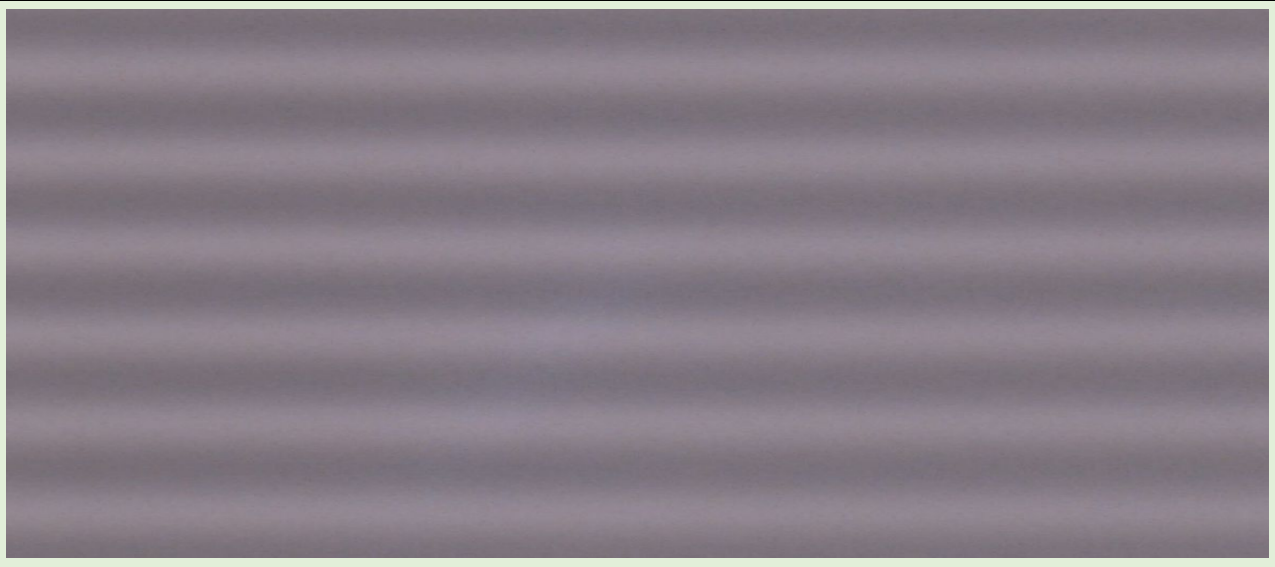 | **8.8069** |
| 50 kHz | 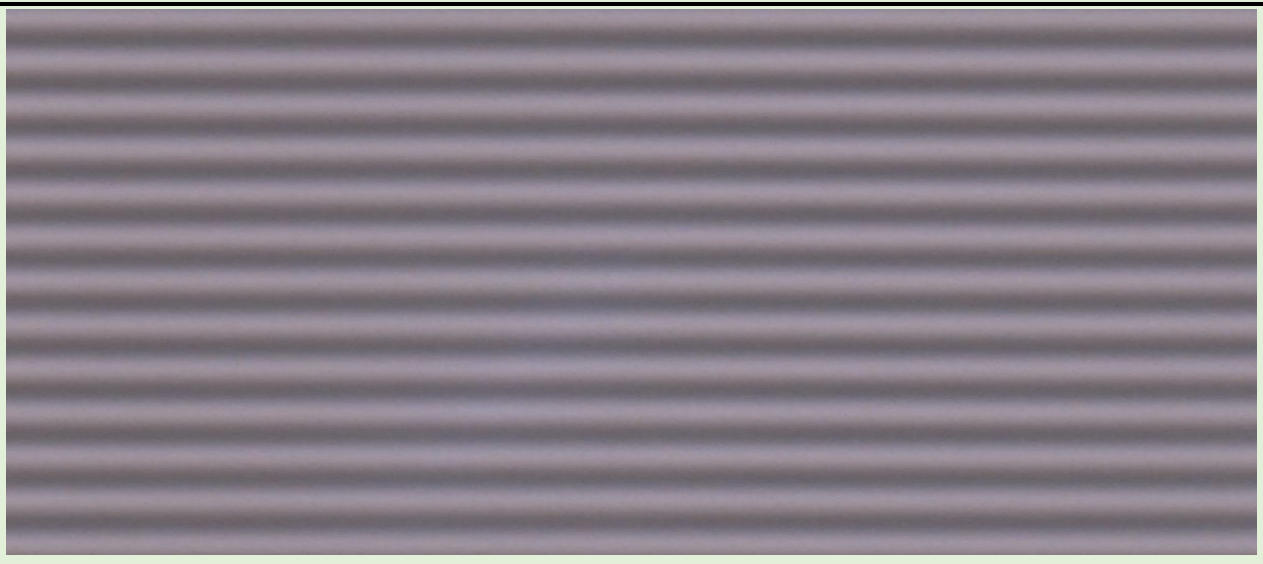 | **18.2273** |
| 75 kHz | 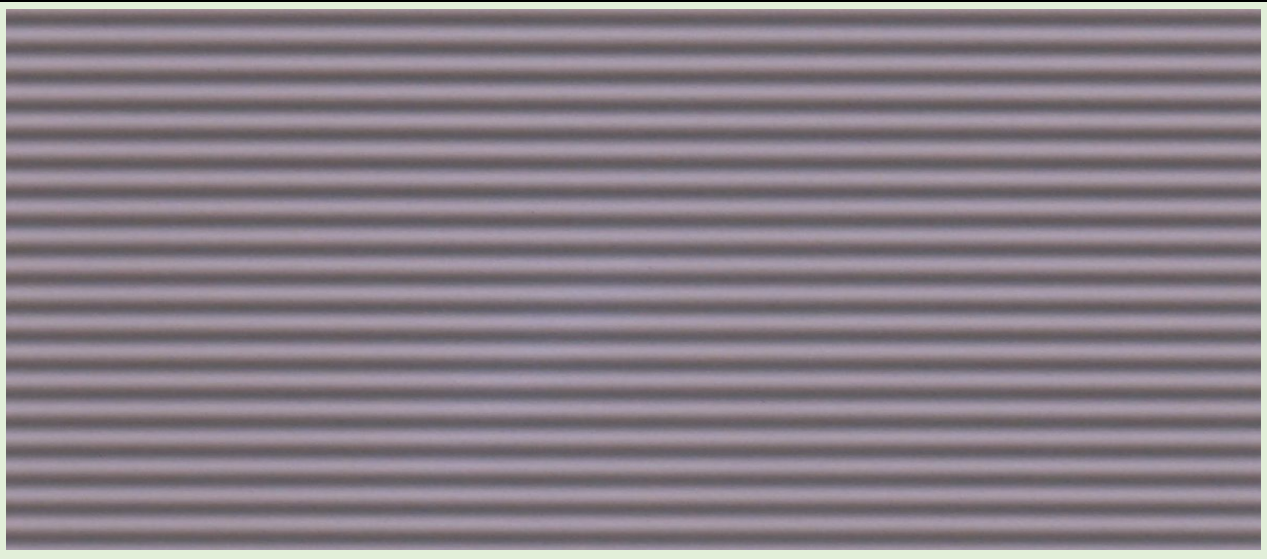 | **24.2720** |
| 100 kHz | 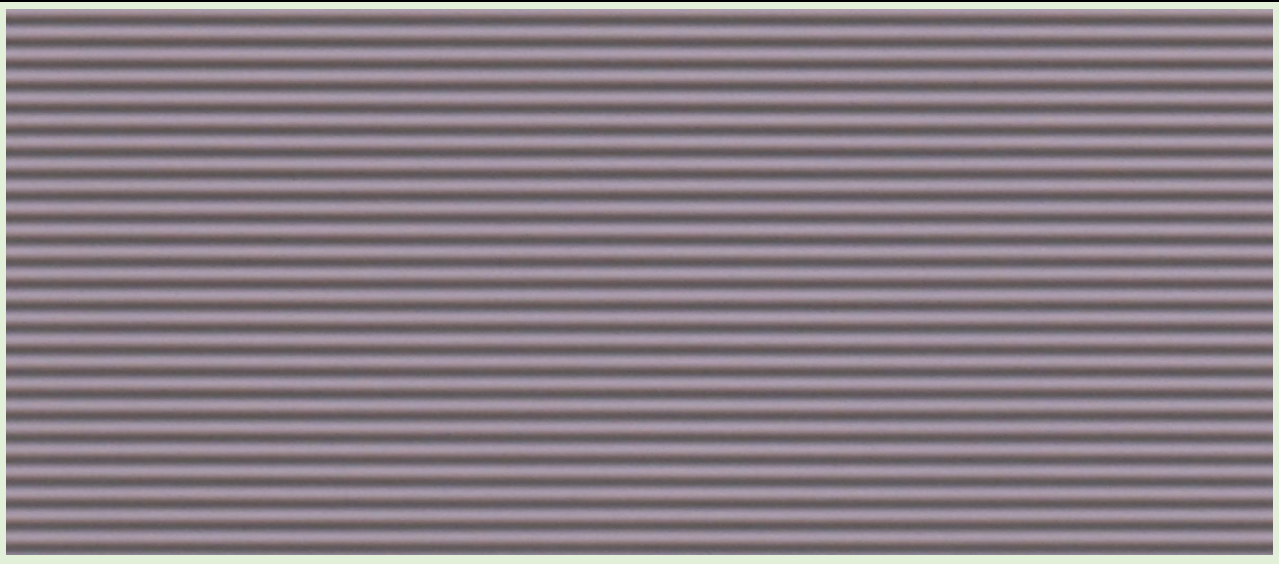 | **27.1401** |

*Input noise amplitude 1Vpp, Images taken from OV10635 sensor, Short Pixel Raw 30fps, No BLC

### 4.8.2. Repeatability

Repeatability of the characterisation method has been confirmed using an Omnivision OV7955 image sensor. Image sensor characterisation was repeated three times in succession for a frequency input range of 50Hz to 300kHz, Figure 34. The test setup has been dismantled and reinstalled between the second and the third test run in order to rule out any dependency on setup.

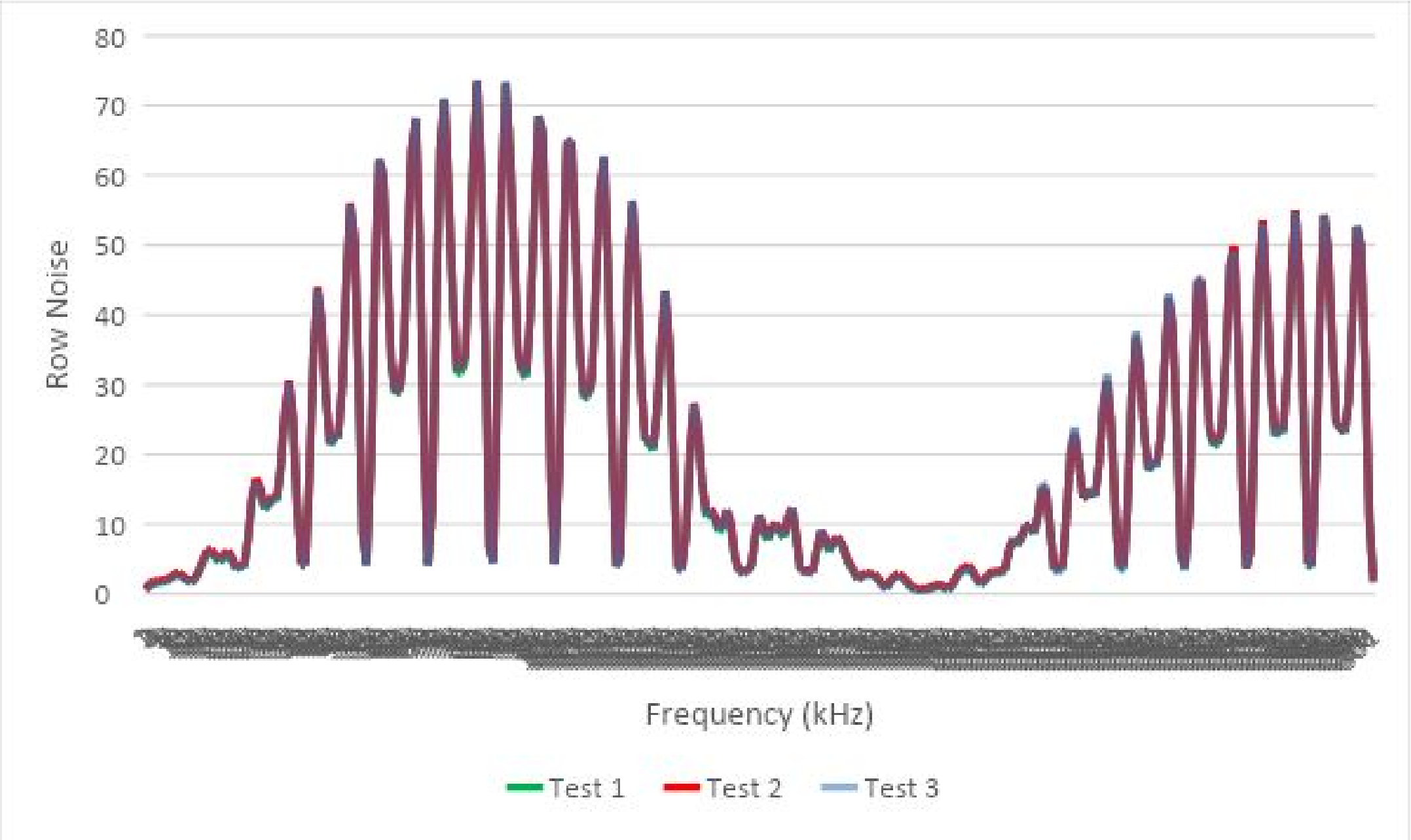

**Figure 34: Characterisation Method Repeatability**

The results indicate that the characterisation method is highly repeatable.

### 4.8.3. Frequency Spectrum

In order to carry out adequate characterisation of an image sensor, the devices response to electrical noise on the power supply lines across a wide frequency spectrum must be assessed. Image capture and analysis at each frequency step takes approximately 20 seconds including the 7 seconds required in order to conduct the row noise algorithm. This number can vary significantly depending on the format and ease of use of the proprietary software for the image sensor under test. Approximate characterisation time for various frequency spectrums can be seen in Table 3.

**Table 3: Frequency Spectrum vs. Characterisation Time, 1kHz Step Size**

| Frequency Spectrum | Approximate Characterisation time |
|---|---|
| 0-100kHz | 33 Minutes |
| 0-200kHz | 1 Hour 7 Minutes |
| 0-500kHz | 2 Hours 47 Minutes |
| 0-1Mhz | 5 Hours 33 Minutes |
| 0-2Mhz | 11 Hours 7 Minutes |
| 0-4Mhz | 22 Hours 13 Minutes |

In order to determine a suitable frequency sweep size for initial characterisation a 2 MHz sweep was conducted to identify specific areas that could be concentrated on with a view to reducing the overall characterisation time.

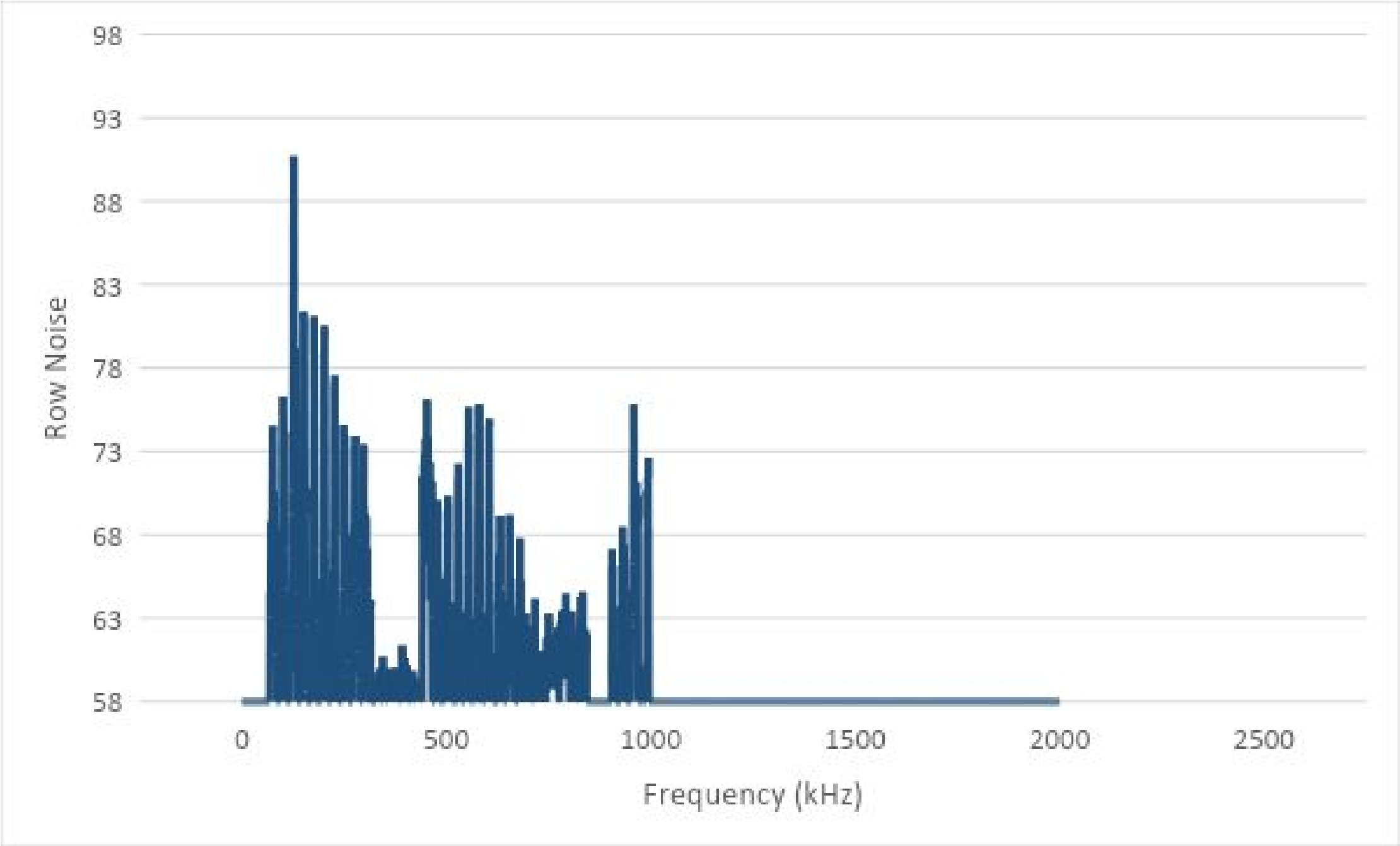

**Figure 35: Frequency Spectrum Investigation, 0Hz – 2MHz**

The results of this investigation indicate that the image sensor is most vulnerable to electrical noise of frequencies up to 1MHz on the power supply lines and thus a frequency sweep of 50Hz to 1MHz will be used for initial characterisation results.

## 4.9. Characterisation of State of the Art Image Sensors

Performance characterisation in response to electrical noise on the power supply lines has been carried out on two state-of-the-art image sensor models in order to display the use of the test methodology proposed in Chapter 3. The results of this characterisation can be seen in sections 4.9.1 and 4.9.2.

### 4.9.1. Omnivision OV10635 Performance Characterisation

Performance characterisation of the Omnivision OV10635 image sensor in response to electrical noise on the power supply lines using the methodology outlined in Chapter 3 is provided in Figure 36.

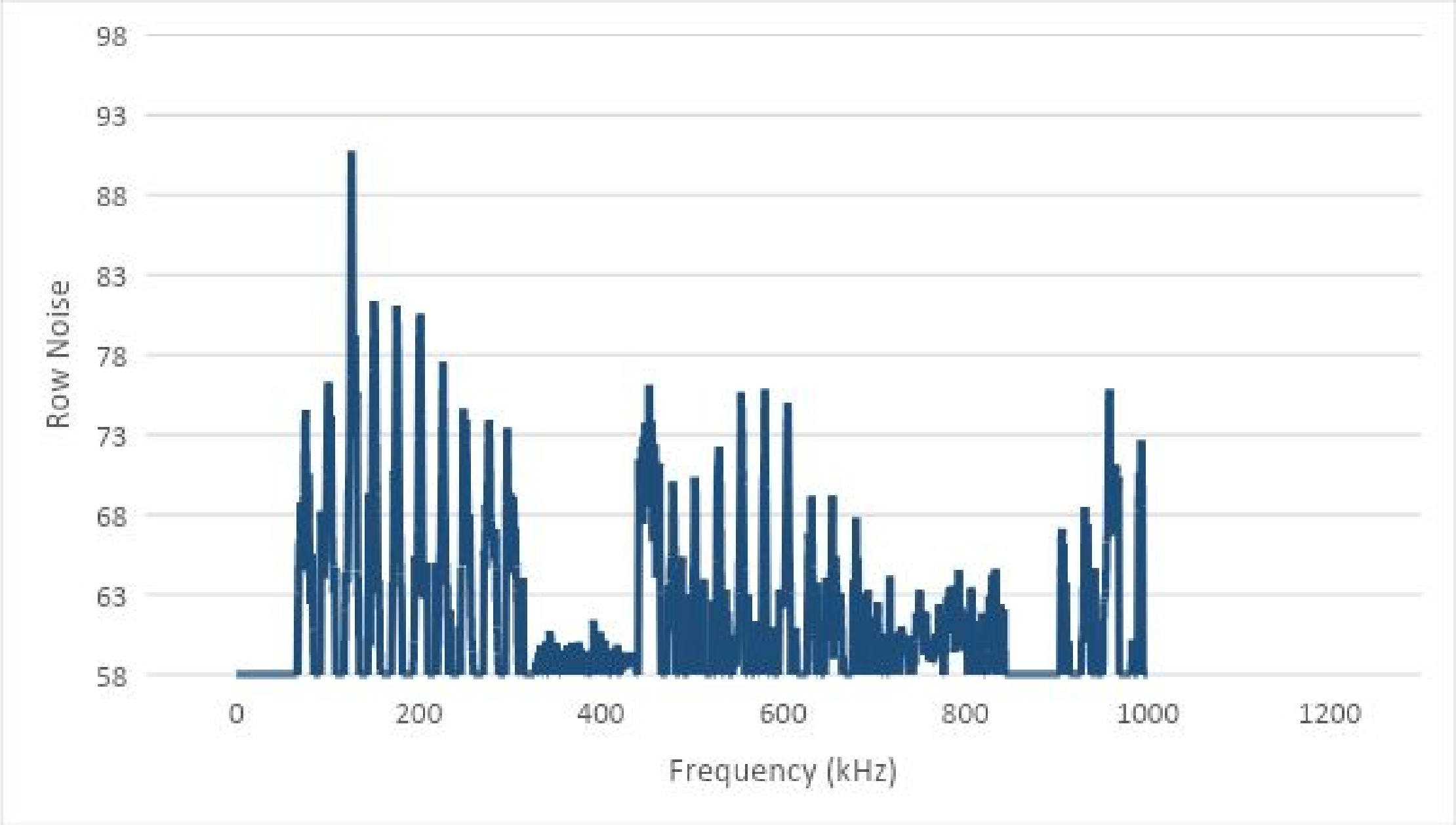

**Figure 36: Omnivision OV10635 Performance Characterisation**

The settings applied for the performance characterisation of the Omnivision OV10635 can be seen in Table 4.

**Table 4: Omnivision OV10635 Image Sensor Characterisation Settings**

| OV10635 Characterisation Settings | |
|---|---|
| **Format** | Raw |
| **Pixel** | Short Channel |
| **Aspect Ratio** | 1280x800 |
| **Frame Rate** | 30fps |
| **BLC** | Disabled – 4000 – Bit 0 |
| **PSU/Voltage** | AVDD/3.3V |
| **Noise Frequency** | 50Hz - 1MHz |
| **Noise Amplitude** | 1Vpp |

## 4.9.2. Omnivision OV7955 Performance Characterisation

Performance characterisation of the Omnivision OV7955 image sensor in response to electrical noise on the power supply lines using the methodology outlined in Chapter 3 is provided in Figure 37.

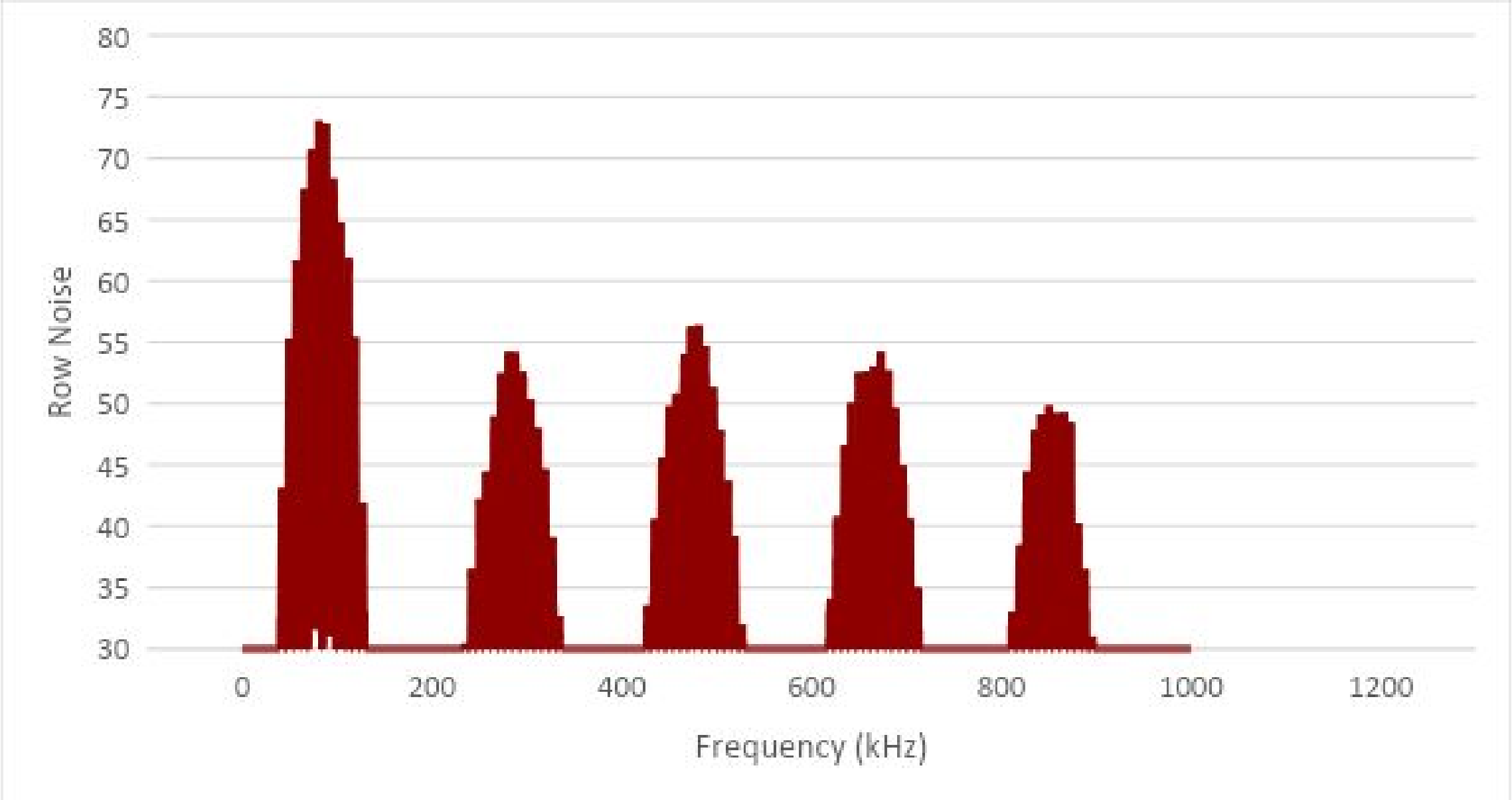

**Figure 37: Omnivision OV7955 Performance Characterisation**

The settings applied for the performance characterisation of the Omnivision OV7955 can be seen in Table 5.

**Table 5: Omnivision OV7955 Image Sensor Characterisation Settings**

| OV7955 Characterisation Settings | |
|---|---|
| **Format** | Raw |
| **Pixel** | Single |
| **Aspect Ratio** | 640x480 |
| **Frame Rate** | 30fps |
| **BLC** | Disabled – 0x5001 – Bit 5 |
| **PSU/Voltage** | AVDD/3.3V |
| **Noise Frequency** | 50Hz - 1MHz |
| **Noise Amplitude** | 1Vpp |

# Chapter 5 Analyses and Discussion

Analysis of the proposed characterisation method and of the characterisation results is carried out in this section of the dissertation as well as a further discussion of the decisions made in the design of the characterisation method and a discussion on the options for improving image sensor performance in response to power supply noise in automotive applications.

## 5.1. Calculating Row Noise

Row noise is not signal dependent and therefore its magnitude is unaffected by the amount of light captured by the image sensor. This allows testing to be carried out at 0 lux ensuring that all pixels are uniformly exposed to the same light intensity so that comparative analysis can be conducted to identify variation between rows in order to quantify row noise. As RAW images are taken of a uniformed 0 lux, the mean pixel value for each row should be the identical. The standard deviation of the row values represents the amount of variation recorded between rows for the same light intensity and therefore identifies any row noise present in the image. This is the founding principle of the row noise algorithm shown in Figure 38.

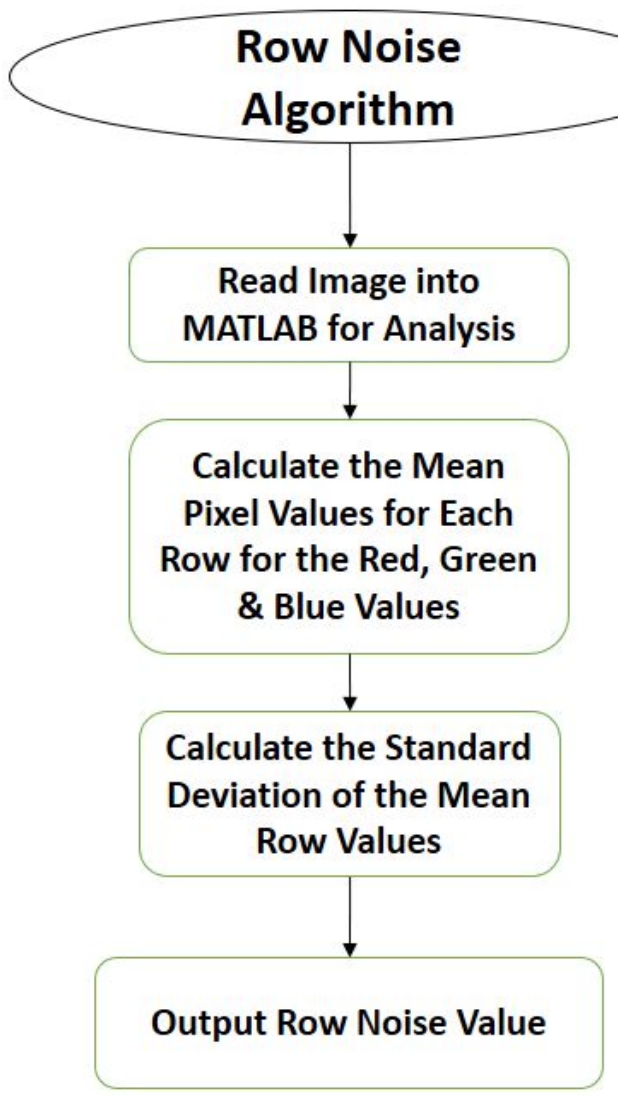

**Figure 38: Basic Row Noise Algorithm**

This basic row noise algorithm, based on a method proposed by Mikkonen 2014 (18), outputs a numerical value representing the magnitude of row noise present in an image.

Table 6 below illustrates how the row noise value visually corresponds to image noise for images with different quantities of row noise.

**Table 6: Row Noise Values for 0-100kHz Noise Input Images**

| **Input Noise Frequency** | **Captured Image** | **Row Noise** |
|---|---|---|
| No Noise | 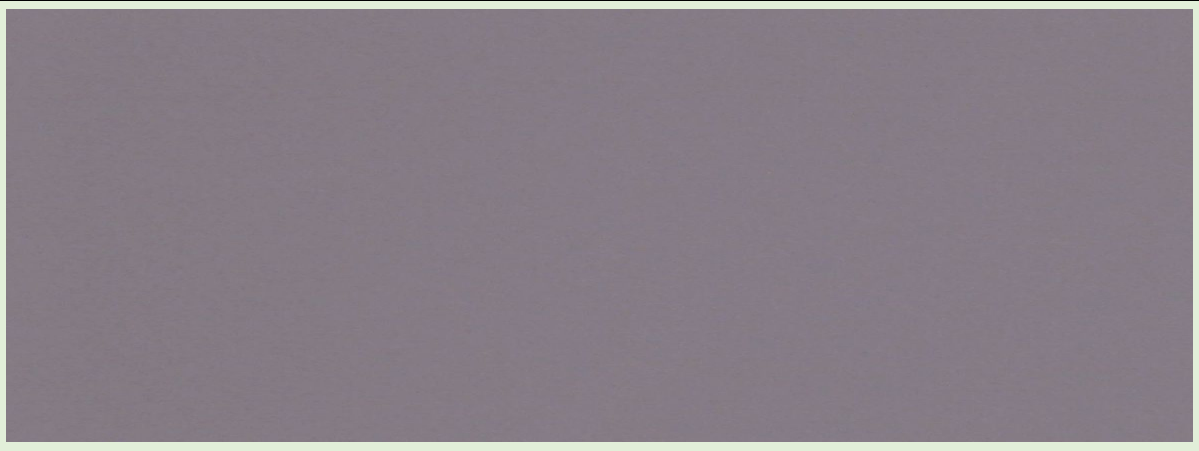 | **0.3805** |
| 25 kHz | 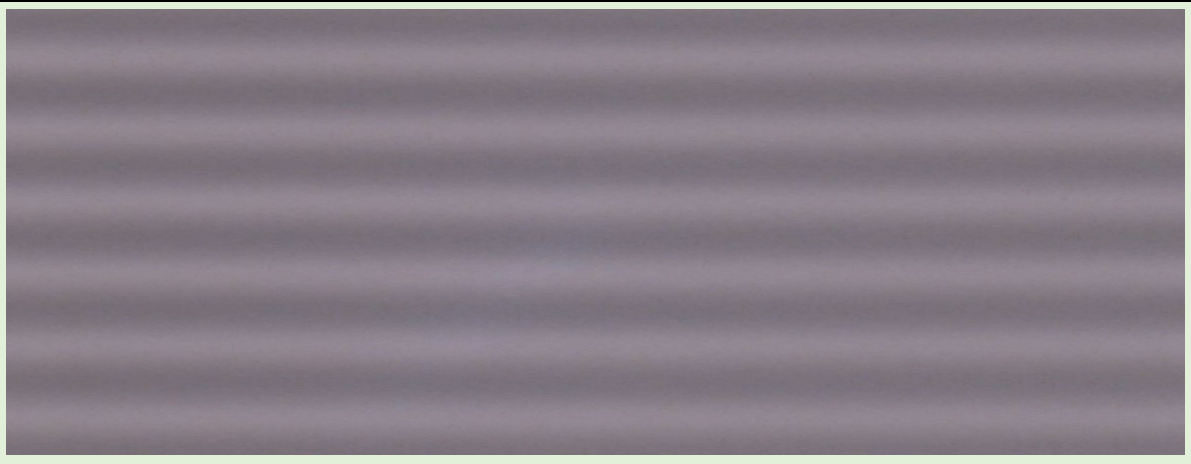 | **8.8069** |
| 50 kHz | 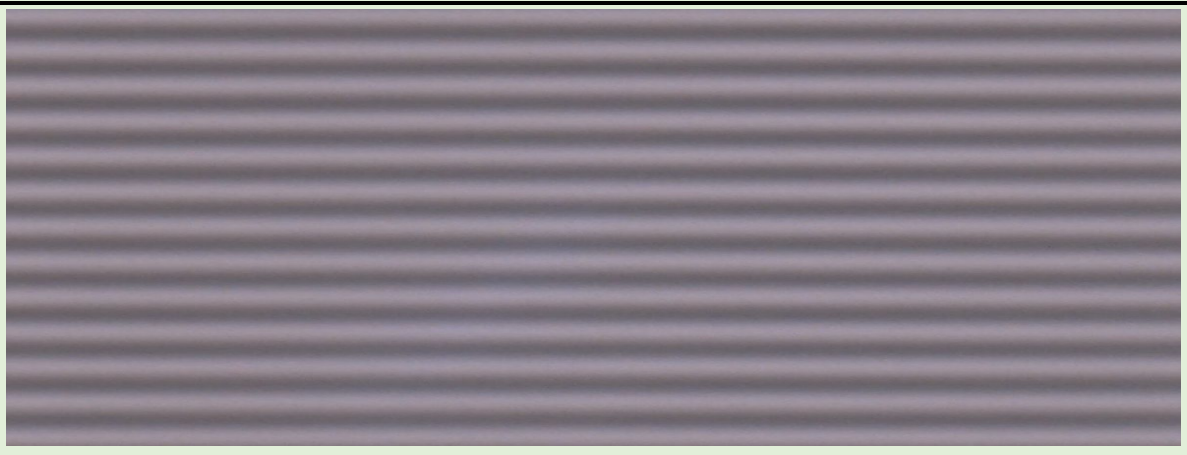 | **18.2273** |
| 75 kHz | 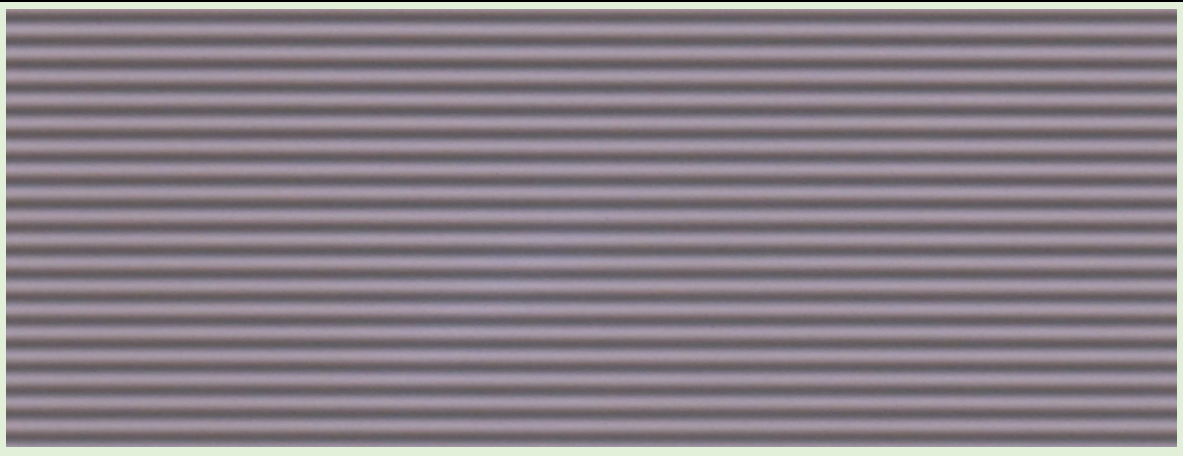 | **24.2720** |
| 100 kHz | 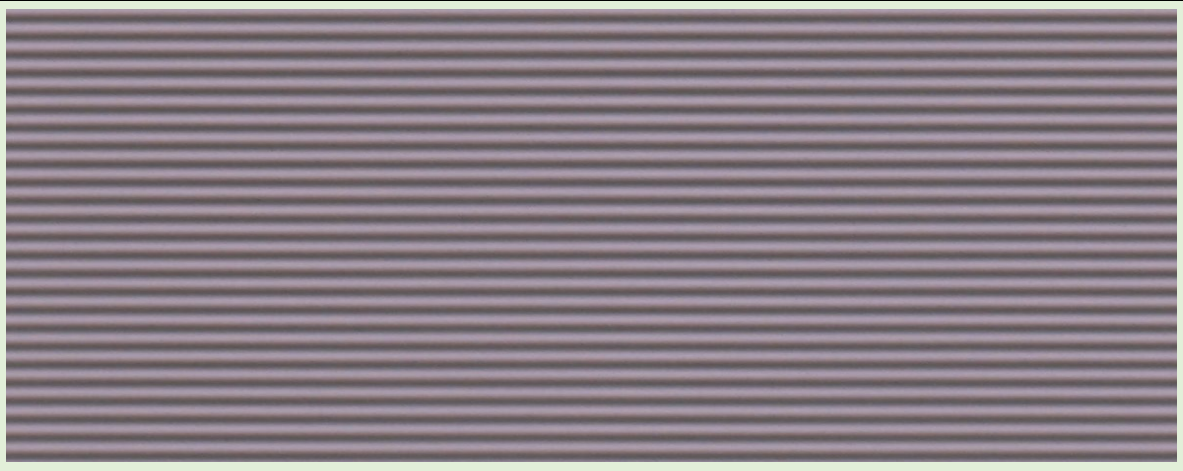 | **27.1401** |
| *Input noise amplitude 1Vpp, Images taken from OV10635 sensor, Short Pixel Raw 30fps, No BLC | | |

It can be seen from this example that the row noise algorithm successfully assigns an appropriate relative magnitude value based on the amount of visual row noise present in an image.

### 5.1.1. Reduction of Temporal Noise in Measurement

Temporal noise sources within an image sensor such as those discussed in the literature review, Section 2.6 can impact the precision of row noise measurement. Mikkonen 2014 attempts to resolve this issue by scaling down the RGB image before row noise analysis is carried out on the image data. Although this method can reduce the impact of temporal noise in the image, it also has the knock on effect of reducing narrow bands of row noise and only captures the wider band occurrences of row noise when analysis is carried out. Many image scaling techniques also require low pass filtering which can further reduce the impact of row noise on the image before analysis takes place. This method therefore reduces the precision and accuracy of row noise detection and is not suitable for thorough characterisation of automotive image sensors.

In order to reduce the impact of temporal noise on measurements without impacting the presence of row noise on the images, multiple images can be taken at each frequency step. Full analysis is carried out on all images for that frequency step and the average row noise measurement is the final row noise value. As temporal noise varies from frame to frame any temporal noise captured will not be consistent in multiple images and therefore will be filtered out. As row noise is a form of spatial noise and does not change over time and the row noise present in the image will be consistent for all images captured for a particular frequency step and therefore will be captured in the standard deviation of the row values for each image. The average of these standard deviations will therefore provide a more precise measurement of the row noise while reducing the impact of any temporal noise that has developed within the image sensor.

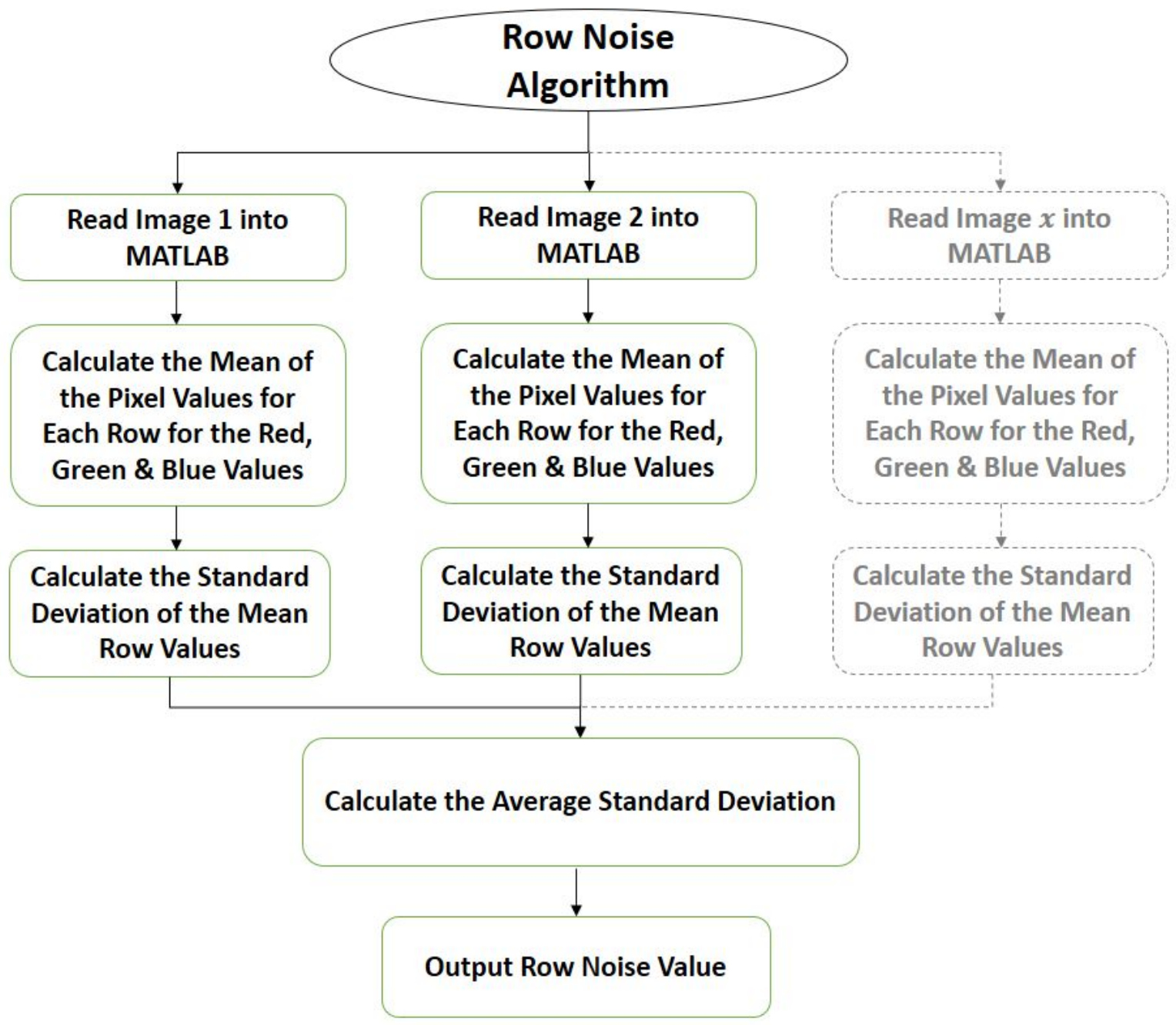

**Figure 39: Improved Row Noise Algorithm with Reduced Temporal Noise Impact**

The more images taken and analysed at each frequency step, the more precise the row noise measurement will be and the lower the impact that any temporal noise created within the image sensor will have. This however comes at the expense of time and processing power and therefore a compromise must be made particularly in instances where characterisation is carried out over a large frequency range.

### 5.1.2. Matlab Standalone Application

The row noise algorithm is implemented in Matlab. The program automatically identifies the aspect ratio of the image and analyses each image as a three dimensional array of the red, green and blue components of each pixel. The Matlab program developed can be seen in Figure 40 below:

```
%   ##########################################################
%   #  Row Noise Algorithm                                   #
%   #  Shane Gilroy                                          #
%   #  27th July 2016                                        #
%   #  "Impact of Power Supply Noise on Image Sensor         #
%   #   Performance in Automotive Applications"              #
%   #                                                        #
%   ##########################################################

A = cell(1,3); % Cell array to hold multiple image data

for i=1:3
s = ['C:\RowNoiseAnalyser\im' int2str(i) '.bmp']; %reads Image1-3 in .bmp format
A{i}= imread(s); % Read mulitple images from the directory above
end

means1 = mean(A{1},2); % Calculate the means for each row of each Pixel RGB of Image 1
means2 = mean(A{2},2); % Calculate the means for each row of each Pixel RGB of Image 2
means3 = mean(A{3},2); % Calculate the means for each row of each Pixel RGB of Image 3


STD1 = std(means1); % Standard Deviations of the image 1 means
STD2 = std(means2); % Standard Deviations of the image 2 means
STD3 = std(means3); % Standard Deviations of the image 3 means

RowNoise1 = ((STD1(1)+STD1(2)+STD1(3))/3); % Image 1 Row Noise Magnitude = Average of the 3
Pixel Standard Deviations(RGB)
RowNoise2 = ((STD2(1)+STD2(2)+STD2(3))/3); % Image 2 Row Noise Magnitude = Average of the 3
Pixel Standard Deviations(RGB)
RowNoise3 = ((STD3(1)+STD3(2)+STD3(3))/3); % Image 3 Row Noise Magnitude = Average of the 3
Pixel Standard Deviations(RGB)

AvgRowNoise = ((RowNoise1+RowNoise2+RowNoise3)/3); % Average Row Noise = Average of the 3
Image Standard Deviations

dlmwrite('RowNoiseOut.csv', AvgRowNoise, '-append');
```

**Figure 40: Row Noise Algorithm Matlab Code**

Traditionally NI TestStand Mathscript can be used in order to implement Matlab code as part of an automated sequence, however Mathscript is unable to handle three dimensional arrays and therefore an alternative method is required for the automation of the row noise algorithm. In order to circumvent this limitation, the Matlab program has been compressed into a standalone executable windows application which may be run on any PC regardless of whether Matlab is installed. This application can be called from within an automated test sequence or activated by double clicking the application icon. Once activated the application displays the following image on the user display while the program calculates the row noise value for the captured images:

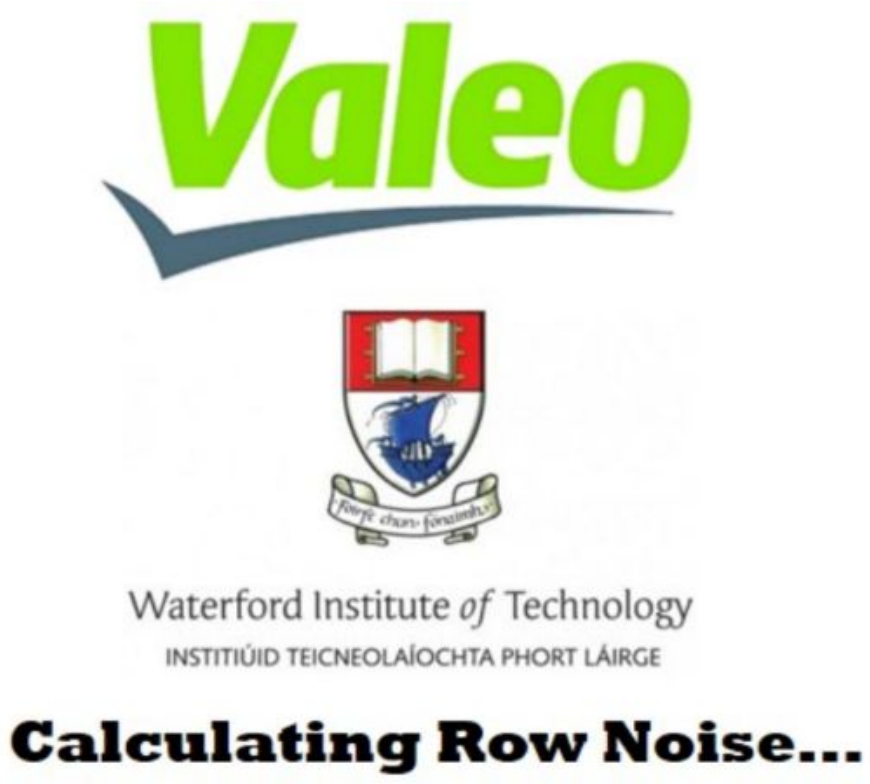

**Figure 41: Row Noise Application User Display**

Once complete the application produces a .csv file named “RowNoiseOut” in the same location as the application on the host PC. This file is updated accordingly each time the application is called. The use of a standalone application for calculating row noise, reduces the software requirements of a host PC for carrying out image sensor performance characterisation.

## 5.2. Automating Row Noise Analysis

In order to characterise image sensor performance in response to electrical noise, a large number of frequency steps must be analysed each one requiring the successful operation and data transfer between multiple software platforms in order to capture, process and run analysis on thousands of images.

LabVIEW and National Instruments TestStand automation is used in order to manage the operation of this analysis for the thousands of frequency steps required when conducting characterisation over a wide frequency spectrum.

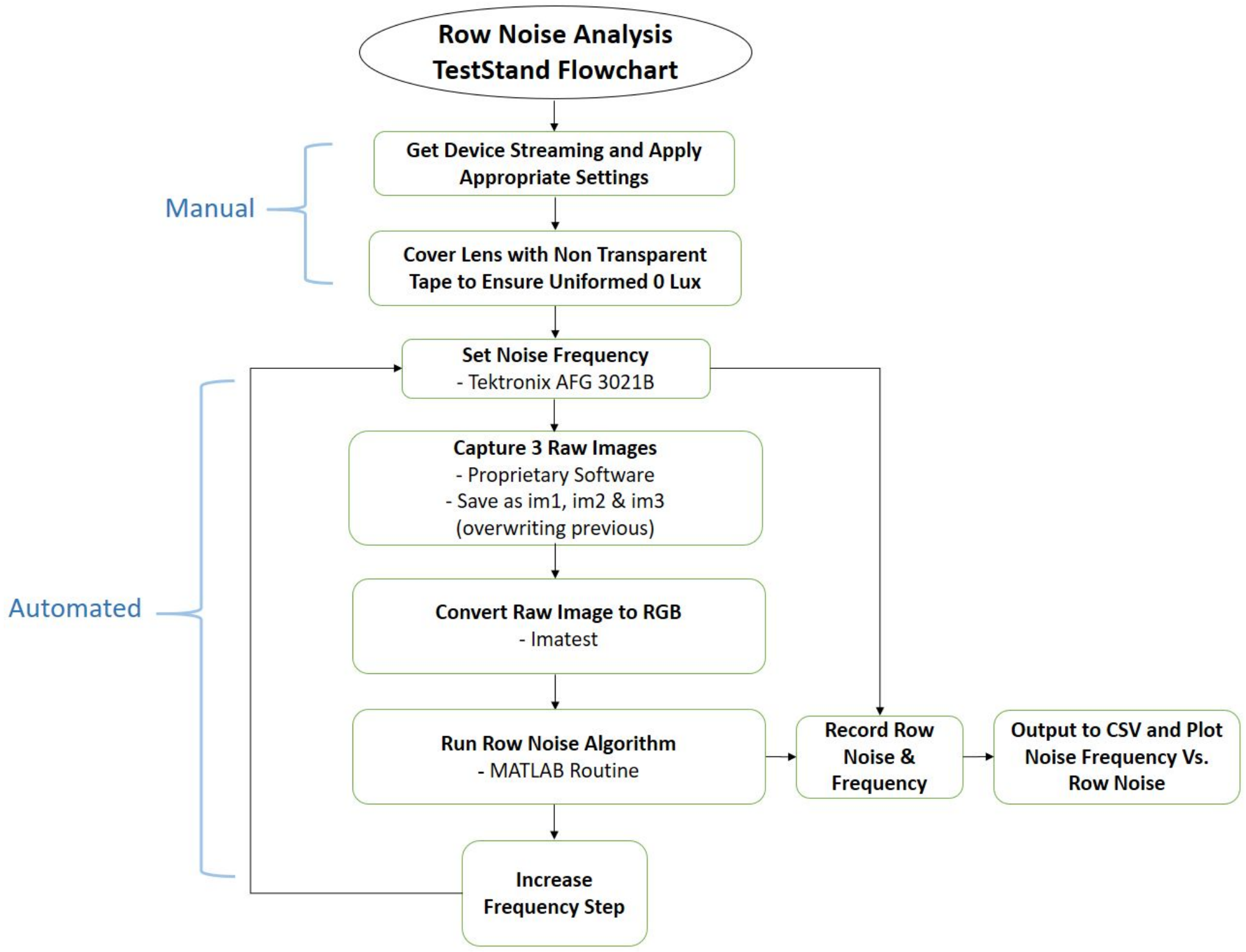

**Figure 42: Row Noise Analysis TestStand Flowchart**

A screenshot of the NI TestStand sequence can be seen in Figure 43 below.

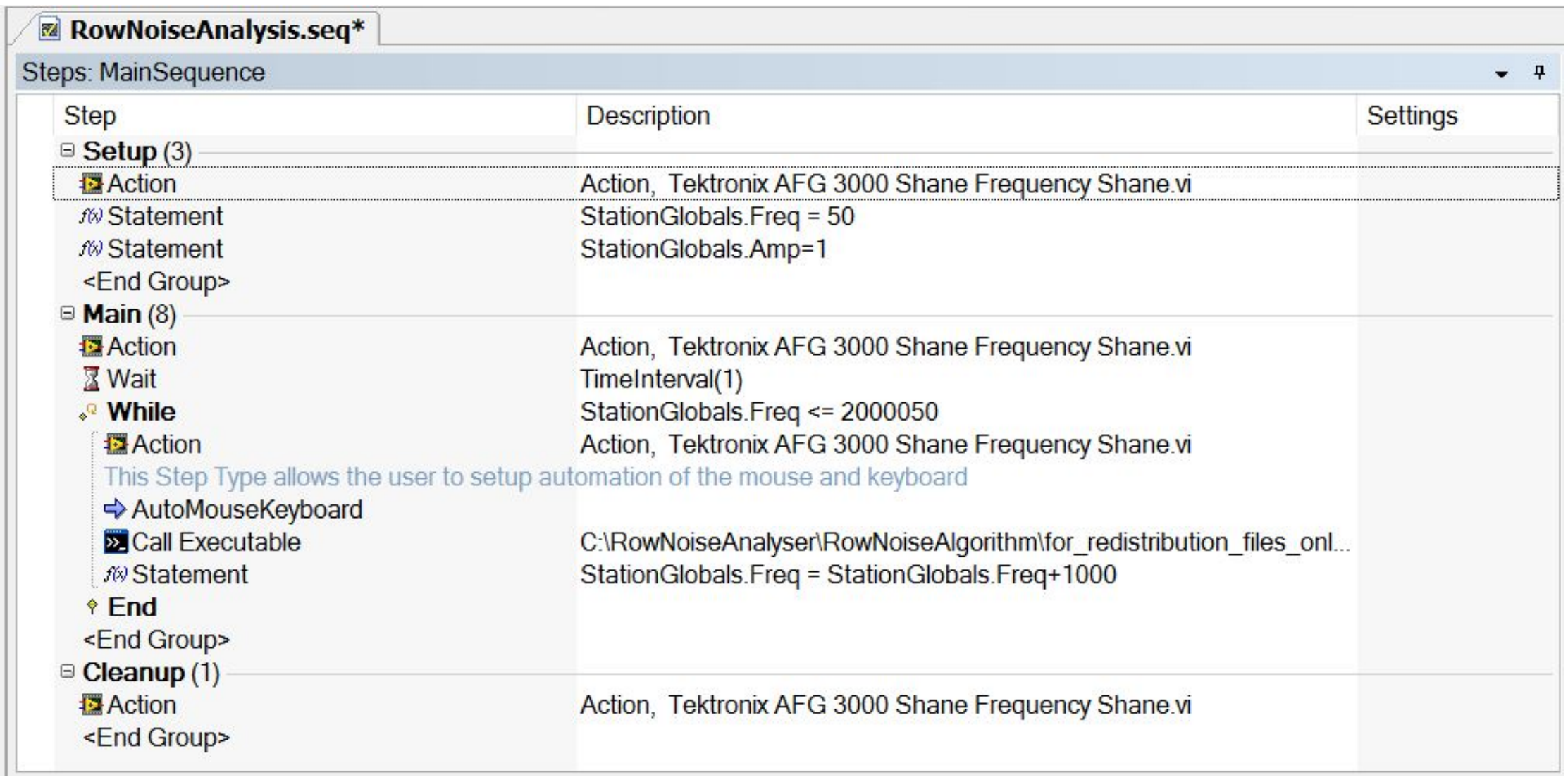

**Figure 43: NI TestStand Sequence**

Function generator communication and control is carried out through the use of a LabVIEW Virtual Instrument (VI) that is initialised, utilised and terminated within the TestStand sequence. Global Variables allow the user to specify the desired start frequency and amplitude range for custom performance characterisation. This feature allows the user to specify critical areas of concern for the image sensor under test. The frequency spectrum size is defined by the number of loops specified in the **While** loop declaration. The frequency step size is determined by the increment step value in Hertz (Hz) at the end of the **While** loop which can be seen in Figure 43 above as 1000Hz or 1kHz (StationGlobals.Freq= StationGlobals.Freq+1000).

RAW images are captured for analysis by the image sensor proprietary software, converted to RGB, titled and saved in the correct location through the use of the "AutoMouseKeyboard" application developed by the Valeo Vision Systems Test Engineering Department. AutoMouseKeyboard allows the user to record a sequence of mouse clicks and keyboard entries that is embedded into a macro which can be called from within NI TestStand. This feature provides a generic approach that allows the TestStand sequence to be used with any image sensor proprietary software installed on the host PC removing the need for a new characterisation sequence for each new image sensor.

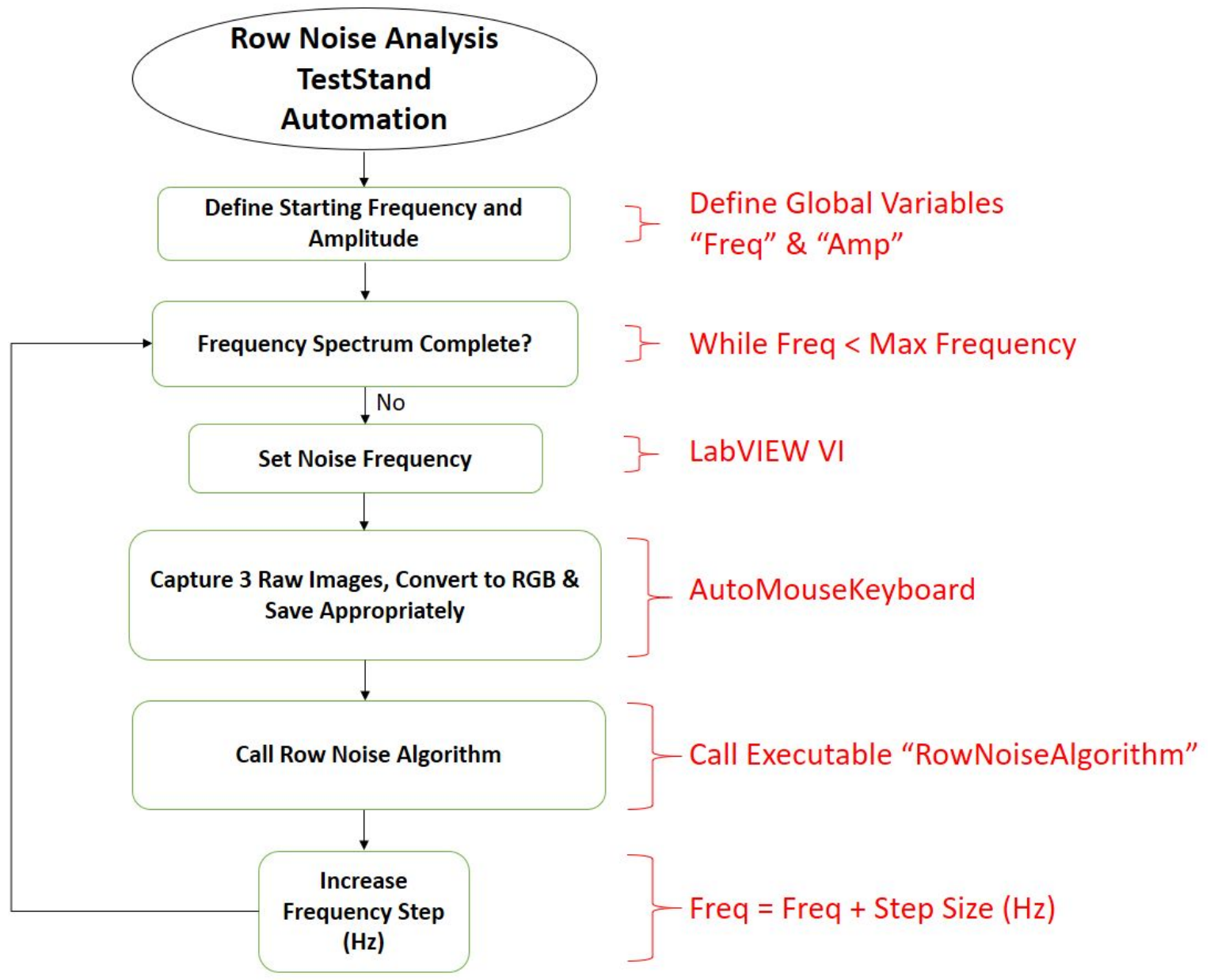

**Figure 44: TestStand Sequence Flowchart**

The row noise algorithm is implemented in Matlab and is carried out by calling the executable Matlab application “RowNoiseAlgorithm” specifically developed for image sensor performance characterisation. This application analyses RAW images and outputs the row noise value to a .csv file titled “RowNoiseOut” in the “C:\RowNoiseAnalyser\RowNoiseAlgorithm” location on the host PC. This file is updated with a row noise value for each frequency step in the **While** loop. This .csv file can then be used to plot Row Noise Magnitude vs. Input Frequency and for further analysis.

## 5.3. Image Sensor Settings

In the interest of future proofing the characterisation method and making it applicable for the widest range of image sensing technologies, minimal restrictions have been placed on the image sensor settings required for characterisation. The mandatory image sensor settings for performance characterisation can be seen in Table 7 below.

**Table 7: Image Sensor Settings for Characterisation**

| | |
|---|---|
| **Illumination Level** | 0 lux. Sensor lens should be covered with a non-transparent tape or cover |
| **Image / Data Format** | RAW Output / 8 Bit Data Format |
| **Black Level Control (BLC)** | Disabled |

In order to accurately carry out image sensor characterisation an identical input must be received by every pixel in the pixel array. As row noise is not signal dependent, the most accurate way of ensuring a uniformed input is to cover the sensor lens with a non-transparent material such as copper tape. This results in a uniformed 0 lux input to the image sensor. It is extremely important that the non-transparent material completely covers the lens and that no light is able to escape through and strike any part of the pixel array.

The image sensor must be configured for RAW output mode so that images can be captured as close as possible to the sensor core minimising the influence of any internal image processing such as auto white balance (AWB), de-noise etc. This is carried out to ensure that row noise is at its most visible when testing is completed in order to provide the most accurate characterisation of the image sensor under test. Configuring the device for an 8-bit data format will assign a value from 0-255 for each pixel in the pixel array.

### 5.3.1. Black Level Calibration

Modern image sensors contain a number of optical black rows, designated pixels which are electrically identical to all other pixels in the array however are not sensitive to light. These optical black rows can be used to carry out a processing technique known as Black Level Calibration (BLC). This process can detect errors and noise on the optical black rows and subtract them from the signal.

As row noise is not signal dependent it impacts the optical black rows in an identical manner to every other row in the pixel array. Therefore, when characterisation is carried out BLC must be disabled as it will attempt to suppress a portion of the row noise, leading to inaccurate characterisation results. This will manifest itself during testing by increasing and decreasing the magnitude of row noise periodically. This occurrence will alert the engineer that BLC has not been successfully disabled before commencing image sensor characterisation.

## 5.4. Characterisation Results

An experiment was conducted to determine an appropriate frequency sweep size for initial image sensor characterisation in response to electrical noise on the power supply lines, the results can be seen in Figure 35. It was concluded from this experiment that row noise is most likely to occur when electrical noise of less than 1MHz is present on the image sensor power supply lines. Therefore, a 50Hz to 1Mhz sweep was used for initial characterisation results with a sweep time of 5 Hours and 33 Minutes for each device. An experiment was undertaken using Omnivision OV7955 in order to confirm the repeatability of the characterisation method. Electrical noise in the range of 50Hz to 300kHz was conducted on the image sensor power supply lines with an amplitude of 1Vpp and a step size of 1kHz. The frequency sweep was repeated three times with the test setup dismantled and reinstalled between test 2 and 3 in order to rule out any dependency on setup. The results were then overlaid to allow accurate identification of test repeatability. The results of the repeatability testing provided in Figure 34 confirm that the image sensor performance characterisation method proposed in this document is highly repeatable.

Analysis of image sensor performance characterisation of Omnivision OV10635 and OV7955 image sensors can be viewed in Table 8 and Table 9 respectively. Individual image sensors are not directly comparable in terms of row noise magnitude as hardware differences, device features and offset variations exist between image sensor models, however the susceptibility and the immunity of each device to electrical noise on the power supply is clearly evident from the characterisation results. Information obtained from the characterisation plots can be used for the selection of suitable filter components in order to suppress the specific areas of concern or the selection of power supply regulators to ensure that critical frequencies of electrical noise are not produced by the surrounding circuitry.

Once characterisation is complete it is recommended that noise analysis of the vision system power supply circuit be carried out and overlaid with the characterisation results in order to assess the designs immunity to electrical noise on the power supply lines. This stage can be incorporated into schematic design simulation at the early stages of camera design in order to reduce the risk of row noise occurring in the final product.

Once peaks are identified through initial characterisation, a focused sweep of any areas of concern can be carried out and images of the row noise can be selected for subjective review by image quality engineers to determine if the row noise is critical for image viewing or analysis purposes as defined by the product application.

## 5.5. Omnivision OV10635

Analysis of Omnivision OV10635 performance characterisation is provided in Table 8.

**Table 8: OV10635 Characterisation Analysis**

<table>
<tr><th colspan="2">OV10635 Characterisation Analysis</th></tr>
<tr><td colspan="2">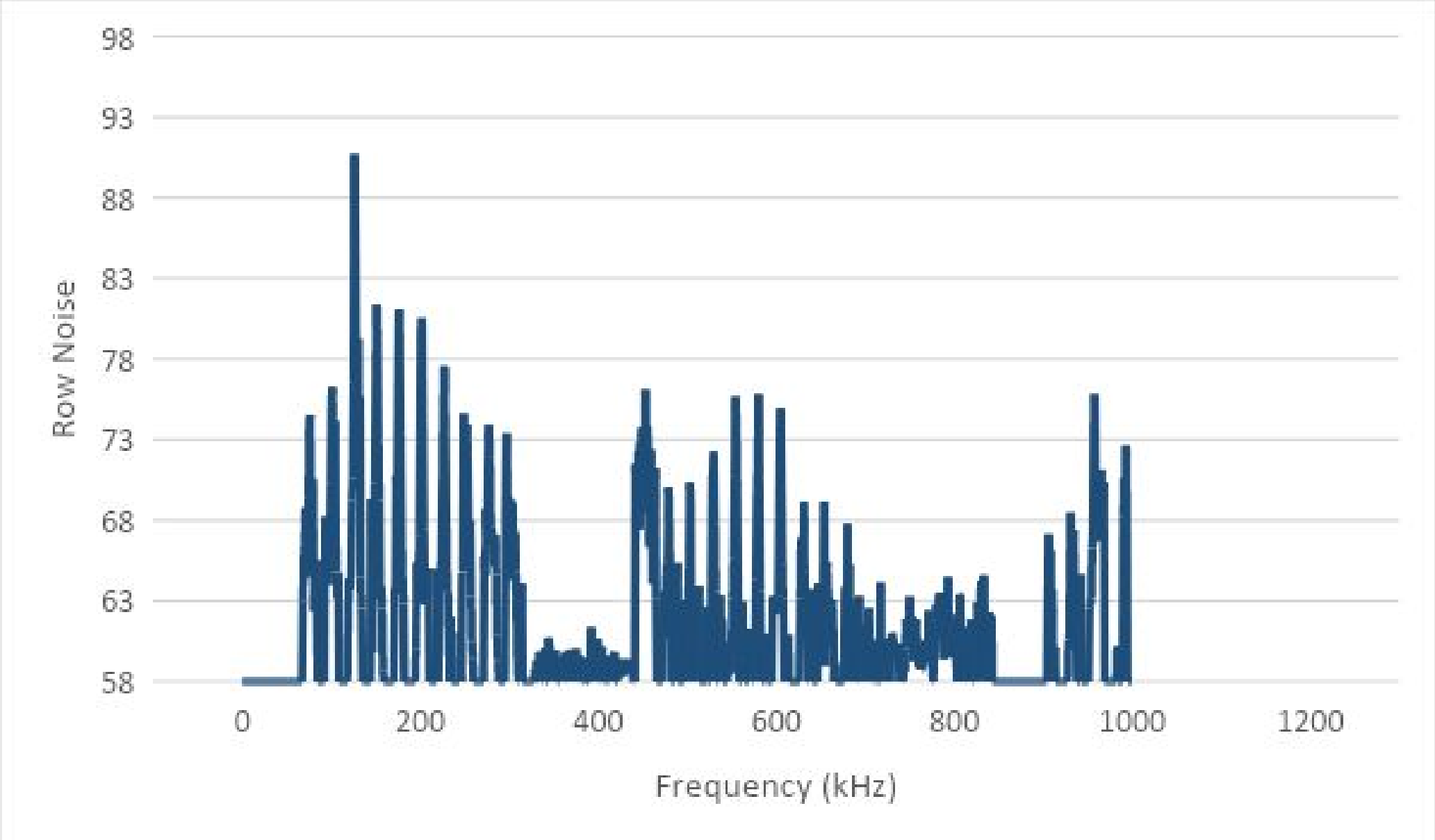
</td></tr>
<tr><th>Event</th><th>Electrical Noise Frequency</th></tr>
<tr><td>Row Noise Start</td><td>68 kHz</td></tr>
<tr><td>Peak Row Noise</td><td>126 kHz</td></tr>
<tr><td rowspan="5">Areas of Concern</td><td>76 kHz – 297 kHz</td></tr>
<tr><td>447kHz – 459 kHz</td></tr>
<tr><td>554 kHz</td></tr>
<tr><td>580 kHz</td></tr>
<tr><td>605 kHz</td></tr>
</table>

| | |
|---|---|
| | 958 kHz |

The performance characterisation of Omnivision OV10635 provided in Figure 36 provides detailed information on the image sensors immunity and susceptibility to electrical noise on the power supply lines. It can be seen from Figure 36 that the image sensor begins to become susceptible to row noise when electrical noise frequencies of over 68kHz are present on the power supply line. Peak row noise occurs in response to electrical noise of 126kHz and this is the most critical area to be considered in the design of a power supply circuit for the OV10635. Other particular areas of concern are highlighted in Table 8. These areas of concern can be reviewed and confirmed by the test engineer and provided to the hardware engineer as design input when designing a power supply circuit for the OV10635 for an automotive vision system.

## 5.6. Omnivision OV7955

Analysis of Omnivision OV7955 performance characterisation is provided in Table 9.

**Table 9: OV7955 Characterisation Analysis**

**OV7955 Characterisation Analysis**

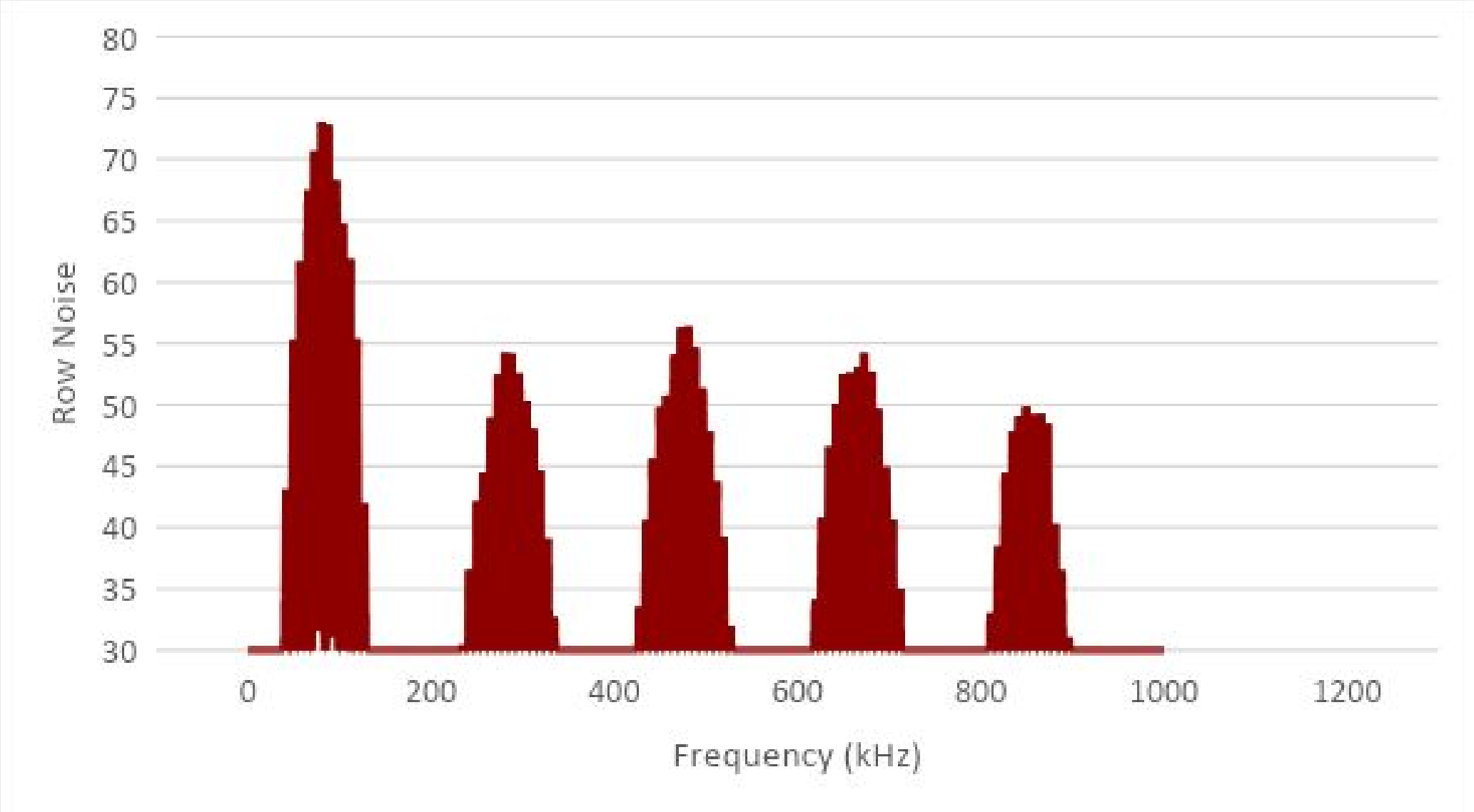

| Event | Electrical Noise Frequency |
|---|---|
| **Row Noise Start** | 42 kHz |
| **Peak Row Noise** | 81 kHz |
| **Areas of Concern** | 50 kHz – 119 kHz |
| | 265 kHz – 312 kHz |
| | 442 kHz – 503 kHz |
| | 635 kHz – 696 kHz |
| | 827 kHz – 873 kHz |

Performance characterisation of Omnivision OV7955 provided in Figure 37 provides detailed information on the image sensors rhythmic pattern of immunity and susceptibility to electrical noise on the image sensor power supply lines. Figure 37 indicates that OV7955 begins to become susceptible to row noise when electrical noise frequencies of over 42kHz are present on the power supply line. Peak row noise occurs in response to electrical noise of 81kHz and this is the most critical area to be considered in the design of a power supply circuit for the OV7955. Other particular areas of concern are highlighted in Table 9. These critical ranges can be reviewed and confirmed by the test engineer and provided to the hardware engineer as design input when designing an automotive camera using the Omnivision OV7955 image sensor.

## 5.7. Improving Image Sensor Performance

There are three main strategies for reducing the effect of row noise in a vision system:

1. Removing the noise source - removing the electrical noise on the power supply lines or the image sensor design characteristics which contribute to the occurrence of row noise.
2. Critical frequency matching or shifting to reduce the impact of the electrical noise on image sensor performance.
3. Removing row noise from images through post processing.

This section provides a brief summary each of these approaches.

### 5.7.1. Removing Noise Source

The primary methods of removing the occurrence of row noise at the source are to select an image sensing technology which does not use a column parallel readout architecture such as Digital Pixel Sensors (DPS) or to suppress the electrical noise on the power supply line before it impacts the image sensor.

#### 5.7.1.1. Digital Pixel Sensors

Row noise occurs as a result of the column parallel readout architecture used in image sensor devices which contain chip level or column level analogue processing technology. One method for preventing the occurrence of row noise as a result of fluctuations on the power supply lines is to use image sensors which carry out analogue processing at pixel level such as Digital Pixel Sensor (DPS).

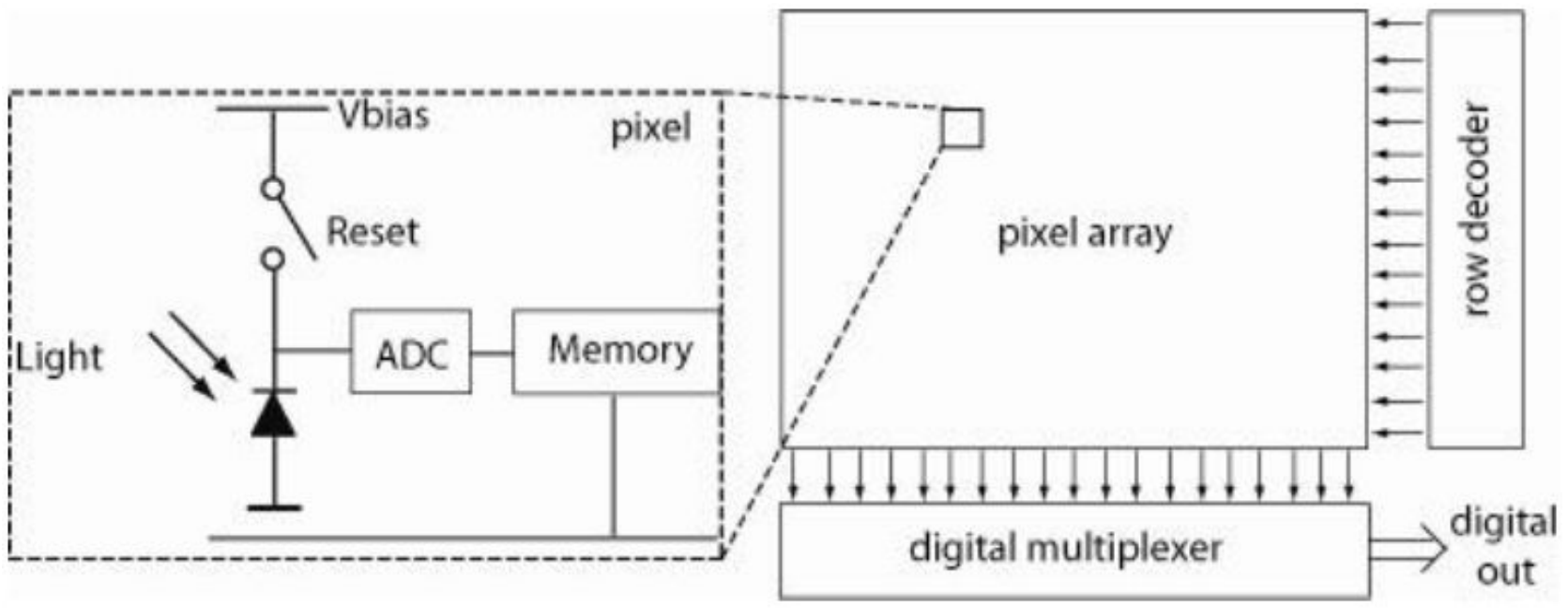

**Figure 45: Digital Pixel Sensor**

Digital Pixel Sensors conduct amplification and analogue to digital conversion at pixel level and as a result all pixels can be read out simultaneously resulting in every pixel having an identical bias voltage level when sampled. This removes the occurrence of row noise in the image as no variation occurs between rows. These type of sensors also remove the occurrence of column fixed pattern noise as every single pixel has its own ADC, rather than a shared ADC for each column, so any slight offset variations are much less visible. Pixel level FPN may still occur which is up to five times less visible to the human eye than column level FPN. Pixel readout speed is also significantly increased as all pixels can be read in parallel which can improve camera performance and greatly increase the maximum frame rates that can be achieved.

The primary drawback with this type of technology compared to chip level or column level processing is that the fill factor of each pixel is greatly reduced as a portion of the photosensitive area of the pixel is sacrificed in order to make room for components required for analogue processing. In order to maintain the photon capturing capability of other sensor architectures especially in low light applications, pixel size must be increased which directly results in reduced resolution or increased image sensor size which may restrict the suitability of Digital Pixel Sensors for some automotive applications.

#### 5.7.1.2. Filtering Capacitors

In most electronic applications, filtering capacitors could be used to reduce the noise source on the power supply lines before it enters the image sensor. Mikkonen 2014 (18) discusses the use of filtering capacitors attached to the power supply lines in order to increase the stability of the power supply voltage thereby reducing row noise. Experiments were conducted to determine the impact that the addition of 100nF, 1uF and 27uF capacitors on the image sensor power supply lines would have on the observed row noise for a similar cost and size sensitive application. The results of this experiment can be seen in Figure 46 below.

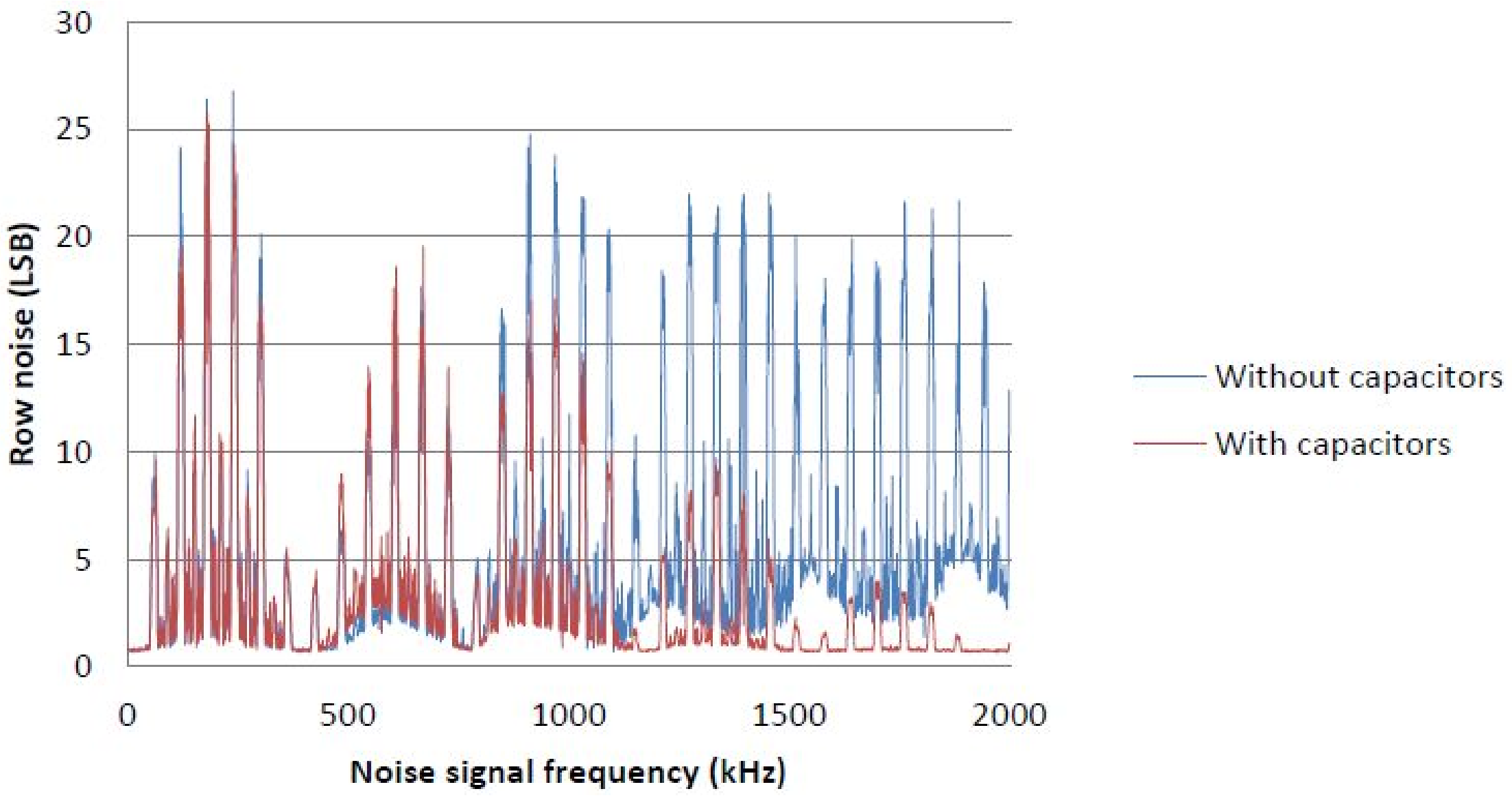

**Figure 46: Impact of Power Supply Filtering Capacitors on Reducing Row Noise**

Mikkonen 2014 concludes from this experiment that the impact of using filtering capacitors for reducing row noise only becomes visible for noise signals of over 750kHz (18) and is less effective for low frequencies of noise. Physically large capacitors would be required in practice in order to reduce the low frequencies of electrical noise which result in row noise in CMOS image sensors. This solution, although possible, is unacceptable for current automotive camera applications due to the significant increase in cost and PCB real estate that would be incurred as a result of adding very physically large capacitors to the image sensor PCB design.

### 5.7.2. Frequency Matching

Once the critical frequencies of power supply noise for a particular image sensor are identified through the use of the characterisation method outlined in this document, attempts can be made to synchronise or phase shift the noise frequency or line frequency as demonstrated in Figure 47 and Figure 48 below.

In cases where the frequencies of noise on the power supply line are in sync with the line frequency, the power supply is at the same level each time a row is sampled. This results in all rows having identical bias voltages when sampled and therefore no light intensity variations are observed between rows.

No Row Noise

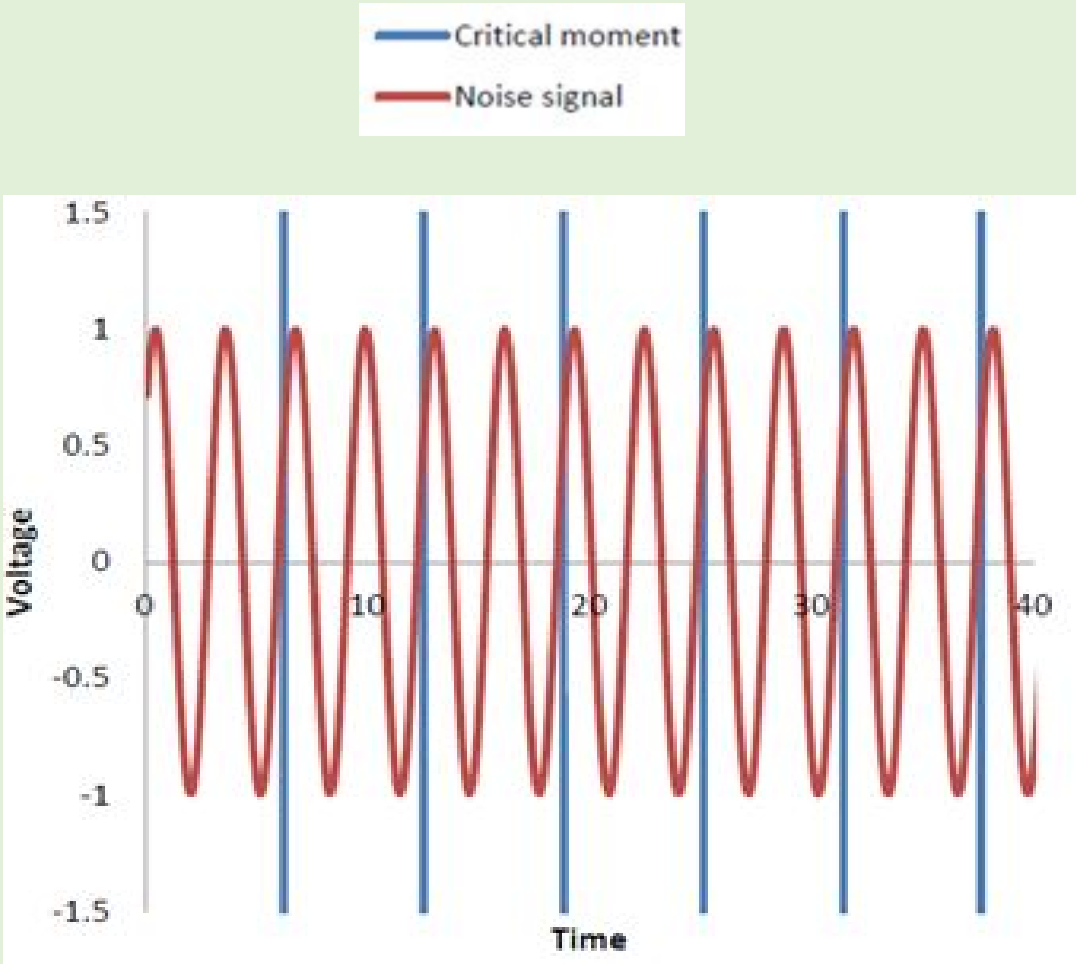

Figure 47: Synchronised Noise and Line Frequency, No Row Noise

Row Noise Occurs

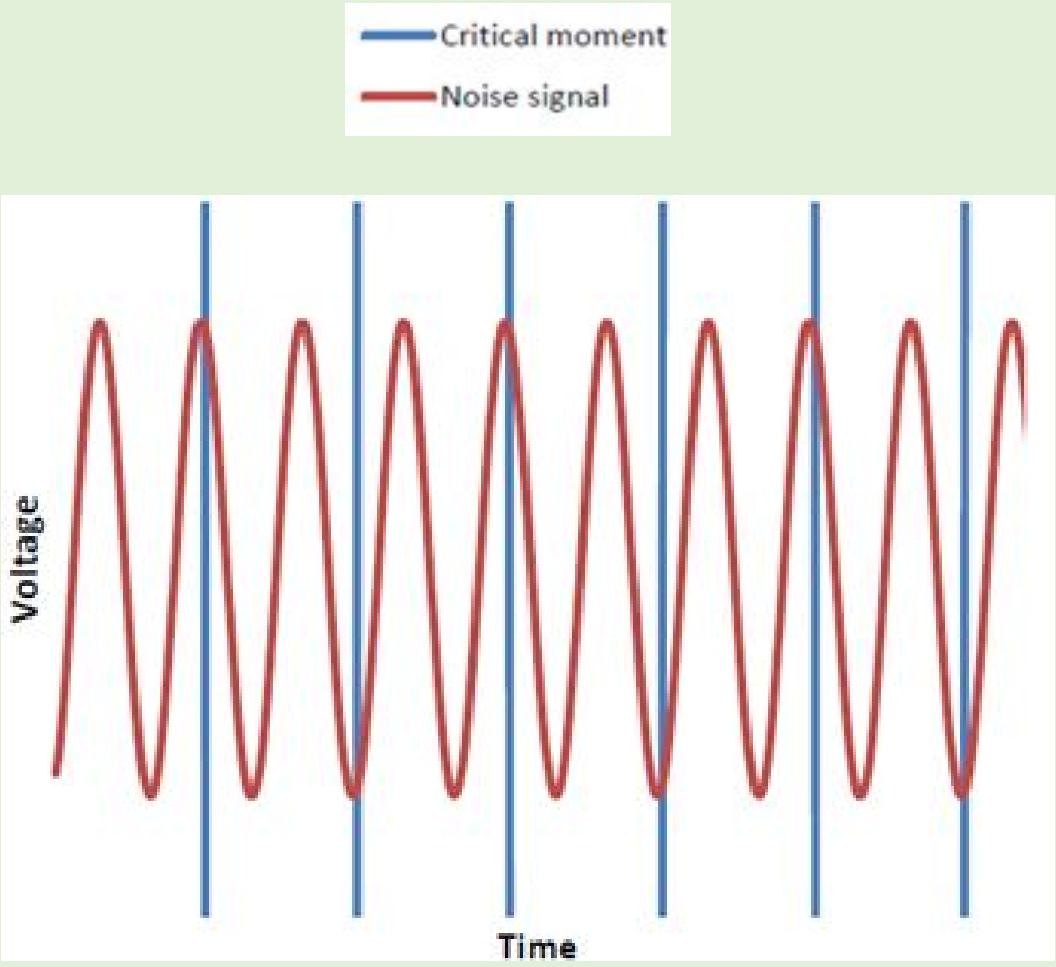

Figure 48: Unsynchronised Noise and Line Frequency, Row Noise Occurs

This process could be carried out pre-emptively by selecting power supply regulators and components which do not generate noise or harmonics in the specific critical ranges identified by image sensor characterisation. If this level of hardware selection or a hardware change is not suitable, the frequency relationship can be modified by changing the line frequency.

$$Line\ frequency = frames\ per\ second\ x\ frame\ length \qquad (2.12)$$

The line frequency can be manipulated by small changes in the frame length or the number of frames per second (FPS) of the camera. The risk in attempting to synchronise line frequency with electrical noise frequency is that if it is not accurately synchronised the effect of row noise can be exasperated on the image. This occurs as the widest banding of row noise can be seen when the line frequency is close to but not in sync with the power supply noise frequency or its harmonics (18).

Therefore, an alternative and perhaps more suitable option is to create the largest possible difference between noise frequency and line frequency so that the row noise variation is at its smallest. Row noise will be just one pixel high if the noise frequency is precisely in the middle of the line frequency harmonics as the noise signal is then in phase on every second row. This process of modifying the noise frequency using hardware selection or filtering such as inductors, ferrites and common mode chokes etc. or modifying the line frequency to create the largest difference can therefore reduce row noise height or banding to such a small scale that it becomes negligible and no longer visible to the human eye.

This process could be thought of as "tuning" the noise or line frequency in order to reduce the visible row noise of a camera in low light. Full characterisation should be repeated to validate any changes in order to reduce the risk of any negative side effects which may occur as a result of frequency matching.

### 5.7.3. Removing Row Noise from Image

A number of patents exist for various methods of reducing row noise from an image post exposure. A selection of these methods are outlined below.

Dark reference pixels can be used to estimate the amount of row noise on an image after exposure by assigning each row a number of optically black reference pixels which are identical in architecture to normal pixels but are not sensitive to light. This can be accomplished by permanently attaching the photodiode of a number of pixels to the power supply or covering pixels with a non-transparent material such as a black colour filter. The dark reference pixels for each row will be affected by power supply noise in an identical manner to the light sensitive pixels, therefore the estimated row noise can be calculated by averaging the dark reference pixels. This average row noise value can then be subtracted in digital form from the light sensitive pixels on the same row after analogue to digital conversion. The drawback from this method is that any temporal noise on the dark reference pixels can appear as row noise on the corresponding row after noise correction is carried out. This risk can be reduced by increasing the number of dark reference pixels used for each row (27, 28){Willassen, 2012 #20;Willassen, 2012 #20}.

Olsson *et* al 2014 (29) describe a process of low pass filtering an image using a non-linear, one dimensional digital finite impulse response filter and then creating a high pass filtered image by subtracting the low pass filtered image from the original image. Intermediate offset values are then created from the high pass filtered image on a row wise selection of pixel values. These offset values are then subtracted row by row from the original image in order to suppress row noise. A similar method outlined by Menikoff 2015 (30) discusses filtering columns of pixels through a low pass filter and calculating an offset value from average differences of rows which exceed a pre-set threshold and updating the original image with the row offset.

Another approach is to develop an algorithm which identifies the pixel values of a number of homogeneous pixels surrounding the subject pixel and comparing neighbouring values to a predetermined tolerance range. A pixel intensity correction value is then calculated by differencing the subject pixel and the average intensity value of neighbouring pixels within a pre-set tolerance. The correction value is then applied to the subject pixel in order to maintain uniformity between pixel rows (30).

# Chapter 6 Conclusions and Recommendations for Further Work

## 6.1. Conclusions

1. Image row noise occurs as a result of electrical noise on the power supply lines of a CMOS image sensor. Row noise is spatial, therefore consistent between frames, its visibility on an image can be related to the camera line frequency (frame rate x frames per second) and although it is not signal dependent, it is more apparent in lowlight applications when the signal to noise ratio of a camera system is greatly reduced.
2. The root cause of row noise in CMOS image sensors is found to be as a result of the fluctuation of the photodiode bias voltage in a pixel array which can occur when electrical noise is present on image sensor power supply line. These power supply fluctuations result in different light intensity measurements for a constant input over time. These varying measurements are then exasperated into a row formation by the column parallel readout mechanism used in CMOS image sensors that contain chip level and column level processing architectures.
3. A mathematical algorithm has been developed in order to assess the magnitude of row noise in an image, taking into account the impact of the temporal noise sources which occur in CMOS image sensors. An automated, systematic, repeatable characterisation method of image sensor performance in response to power supply noise has been developed around this mathematical algorithm which can be used to identify the noise frequency ranges that a specific model of image sensor is particularly immune or susceptible to across a wide frequency spectrum.
4. Characterisation of two state-of-the-art image sensor models currently used in the design of automotive cameras, Omnivision OV10635 and OV7955, have been carried out in order to demonstrate the effectiveness of the proposed method. The characterisation results indicate that image sensor performance in response to power supply noise can vary dramatically between image sensor models, even from the same manufacturer. This leads to the conclusion that a "one size fits all" solution for image row noise is not effective for the design of camera systems for automotive safety applications. Characterisation results should be used to as a design input for the component selection, schematic, PCB and software design of automotive cameras in order to improve low light performance for active automotive safety applications.

5. Three strategies have been presented for the improvement of image sensor performance in response to electrical noise on the power supply lines. The effectiveness and cost of implementation of each of these strategies varies for each specific application according to the design stage of the camera system. Therefore, is recommended to conduct performance characterisation of each new image sensor model and consider the results in a pre-emptive manner at the early design stages of automotive camera design in order to reduce the risk of poor low light performance in safety applications.

## 6.2. Recommendations for Further Work

The short timescale of the project placed a limitation on the project scope and there is a large potential for further work. The proposed characterisation method has focused on image sensor performance in response to electrical noise on the power supply lines only, however the method can be adapted in future to conduct image sensor characterisation in response to any noise source or stimulus as required by the vision system application. The structured, systematic nature of the proposed characterisation method allows the system to be updated by switching the row noise algorithm executable file with one defined by the new noise input allowing characterisation to be implemented by following the same steps. The method outlined could also be modified in future to allow the characterisation of complete camera modules for the purposes of debug or validation of automotive vision systems in response to electrical noise.